\documentclass[12pt]{article}
\usepackage{amssymb,amsmath,amsthm,amsfonts,amscd,amstext}
\usepackage[mathscr]{eucal}
\usepackage{array}
\usepackage{graphicx}
\date{{}}       
\input epsf  

\newtheorem{lem}{Lemma}[section]
\newtheorem{prop}[lem]{Proposition}
\newtheorem{thm}[lem]{Theorem}
\newtheorem{cor}[lem]{Corollary}
\newtheorem{df}{Definition}[section]

\newtheorem{claim}[lem]{Claim}

\newcommand{\Aut}{\mathrm{Aut}}
\def\aut{\operatorname {Aut}} 
\def\C{{\mathbb C}}  

\def\d  {{\rm d}}
\def\D{{\mathbb D}}

\def\der{\operatorname {Der}} 
\newcommand{\Der}{\mathrm{Der}}
\def\Det{\operatorname {Det}}
\def\Diff{\operatorname {Diff}}

\newcommand{\Tset}{\mathbb{T}}
\def\dim{\operatorname{dim}}  

\newcommand{\e}{{\mathrm e}} 
\def\E{{\mathbb E}}    
\def\End{\operatorname{End}} 
\def\eps {\epsilon}

\def\G{\textsf{G}}  
\def\H {\mathbb H} 
\def\HH {{\cal H}} 

\def\id{\operatorname{Id}}  
\newcommand{\interior}{\mathrm{int}}
\def\KK{{\cal K}}
\def\ker{\operatorname{ker}}  

\def\LL{\cal L}
\def\M {\mathbb M}   
\def\map{\operatorname{Map}}  
\def\met{\operatorname{Met}}  
\def\MM{\cal M}  
\def\OO{\cal O}  
 
\newcommand{\Obdle}{\mathfrak{L}} 
\def \Prob {{\bf P}}
\def\P{{\mathbb P}}  
\newcommand{\Pexp}{{{\cal P}\mathrm{exp}}}
\def\R{{\mathbb R}}    
\def\SS{{\Sigma}} 

\def\tr{\operatorname{tr}}  
\def\V{\textsf{V}}   
\def\vect{\operatorname{Vect}}
\newcommand{\Vir}{\mathrm{Vir}}
\def\vir{\operatorname{Vir}} 
\def\Vir{\operatorname{Vir}} 
\def\Weyl{\operatorname{Weyl}} 
\def\witt{\operatorname{Witt}}

\def\Z{{\mathbb Z}}    
\def\N{{\mathbb N}}    

\newcommand{\unit}{\mathbf{1}}
\newcommand{\Xpdet}{{|{\det}_{X,p}|}}
\newcommand{\Xdet}{{|{\det}_X|}}
\newcommand{\Xope}{\mathfrak{O}}
\newcommand{\be}{\begin{eqnarray}}
\newcommand{\ee}{\end{eqnarray}}

\begin{document}
\begin{titlepage}

\vspace*{7mm}

\begin{center}
{\bf \Large  On Connections of Conformal Field Theory and \\
Stochastic L{\oe}wner Evolution} \\
\vspace*{8mm}

{      R.~Friedrich} \\

\vspace*{3mm}

{\em Institute for Advanced Study} \\
{\em Princeton, NJ 08540, USA} \\
\vspace{2mm}

\vspace*{6mm}

\end{center}

\begin{abstract}
This manuscript explores the connections between a class of stochastic processes called ``Stochastic Loewner Evolution" (SLE) and conformal field theory (CFT). 

First some important results are recalled which we utilise in the sequel, in particular the notion of conformal restriction and of the ``restrcition martingale", originally introduced  in~\cite{LSWr}.

Then an explicit construction of a link between SLE and the representation theory of the Virasoro algebra is given. In particular, we interpret the Ward identities in terms of the restriction property and the central charge in terms of the density of Brownian bubbles. We then show that this interpretation permits to relate the $\kappa$ of the stochastic process with the central charge $c$ of the conformal field theory. This is achieved by a highest-weight representation which is degenerate at level two, of the Virasoro algebra.

Then we proceed by giving a derivation of the same relations, but from the theoretical physics point of view. In particular, we explore the relation between SLE and the geometry of the underlying moduli spaces. 

Finally we outline a general construction which allows to construct random curves on arbitrary Riemann surfaces. The key to this is to consider the canonical operator $\frac{\kappa}{2}L^2_{-1}-2L_{-2}$ in conjunction with a boundary field that is a degenerate highest-weight field $\psi$ as the generator of a diffusion on an appropriate moduli space.

\vfill

\begin{tabular}{ll}
{\em PACS 2003:} &
02.50.Ey, 
05.50.+q, 
11.25.Hf  
\\
{\em MSC 2000:}   &
60D05, 
58J52, 
58J65, 
81T40 
\\
{\em Keywords:}  & Probability Theory; Conformal Field Theory \\
 & \\
{\em Email:}    &  {\tt rolandf@ias.edu}
\end{tabular}
\end{abstract}
\end{titlepage}

\pagenumbering{Roman}
\setcounter{page}{1}

\include{preface}

\tableofcontents
\newpage
\pagenumbering{arabic}

\section{SLE and Restriction property}
\subsection{Introduction}
The following text, which is based on a couple of original publications~\cite{F, FW1, FW2, FK, KBonn} explores the relation between a class of stochastic processes, called ``Stochastic L{\oe}wner Evolution", abbreviated as SLE and (boundary) conformal field theory (CFT) from an analytic geometric point of view. The present manuscript shall evolve in the near future, as we will incorporate additionally worked out material, that is only sketched in this version.

The origin of SLE is a paper by O.~Schramm~\cite{S1} in which he revisited the notion of scaling limit and conformal invariance for the loop erased random walk and the uniform spanning tree. His ideas to use L{\oe}wner's differential equation in a stochastic context, turned out to be very fruitful in the general study of domain walls of two-dimensional critical systems in the continuum limit. The SLE process yields two-dimensional random curves, which can be obtained by iterating random conformal maps. In fact, these random curves are the only ones in two-dimensions which have a certain  Markov property, that is conjecturally satisfied by a number of random interfaces of critical planer models from statistical mechanics as e.g. the Ising model, percolation etc. Fundamental properties of SLE where covered in~\cite{RS} by S.~Rhode and O.~Schramm and  by G.~Lawler, O.~Schramm and W.~Werner in a series of articles~\cite{LSW1, LSW2, LSW3, LSWlesl, LSWsaw}, and they also added a new perspective~\cite{LSWr} that is important in our work.

CFT on the other hand, is by now a well established and very powerful tool in the investigation and understanding of two-dimensional statistical mechanics models. Its seminal foundations where introduced first by A.~Belavin, A.~Polyakov and  A.~Zamolodchikov~\cite{BPZ1,BPZ2} and then extend to the treatment of systems with boundaries (boundary conformal field theory BCFT), by J.~Cardy~\cite{Ca1,Cardy:ir}. CFT, which also plays a crucial role in string theory, is in general a very rich mathematical object. 

Therefore it is natural to assume,  that the two fields must be related with each other, even more as a series of early predictions could be proved, that have been previously derived by CFT methods. The most prominent among them is Cardy's formula~\cite{Ca2}, that can be derived easily by SLE methods
\cite {LSW1} (its actual validity for discrete 2d percolation models has been proved in \cite {Sm}).

The first published paper on a connection of SLE with CFT was provided by M.~Bauer and D.~Bernard~\cite{BB}. Over the past years, they also continued to explore the above mentioned relations for simply connected domains, in parallel and independently from the work presented here. In particular they covered the essential parts of SLE from the operator point of view in a comprehensive series of papers~\cite{BB, BB1, BB2,Bauer:2003kd}.

Let us now briefly summarise the content. 

In Section~1 we first recall the connection of two-dimensional discrete systems and the scaling limit, as described by a conformal field theory. In particular we explicit how a series of Gibbs measures defined on a discrete system should converge at criticality to a measure located at the unstable fixed point of the renormalisation group flow, conjecturally the SLE measure. This perspective is natural from the physics point of view.
Then we recall the main results of SLE, as needed in the sequel. These were mainly derived  by G.~Lawler, O.~Schramm and W.~Werner. The role of ``conformal invariance" and of the central ``restriction property" (cf.~\cite{LSWr}) are discussed.

We would like to emphasise the fact, commonly not very well known, that as SLE is thought today, it is the result of the subtle interplay of two directions, as started on one hand by Schramm in \cite{S1} and on the other hand by Lawler and Werner in \cite{LW1, LW2}. It is only the combination of the two viewpoints that brings the method to its full power.
The main ingredient in SLE, as the name states, is L{\oe}wner's differential equation, which is based on Hadamard's principle of boundary variation. In particular L{\oe}wner's equation results from a  singular deformation of the boundary. 

The motivation for Part~2 was the natural assumption, that if there is a conncetion between SLE and CFT, then one should be able to recover at least some of the physical objects and quantities from SLE.
The fundamental objects in CFT are the central charge, as-well the so-called Ward identities,  which capture symmetries at the level of correlation functions.
The relevant symmetry algebra  is the complex Virasoro algebra which is the central extension of the Witt algebra.

In this section, which is based on a publication with W.~Werner~\cite{FW1, FW2}, we study the basic but important example of the intersection probability of a random curve, that satisfies the restriction property, with slits at the boundary of a simply connected domain. In particular, we show that it satisfies the same recursive relations (Ward type identities), as the correlation functions of the stress-energy tensor with additional insertions of boundary operators, in a $c=0$ CFT, where the weight of the boundary field is given by the exponent for the restriction measure.  After this we provide for the central charge zero case a degenerate highest-weight representation at level two, on a Fock space.
In particular, this gives the relation between $\kappa$ and $c$, and between the associated exponential of the restriction measure and the highest weight of  a degenerate highest-weight representation of the Virasoro algebra with central charge $c$.

In Part~3 we first show how to obtain from a boundary CFT calculation the same results as derived in the previous section by purely probabilistic methods.

Now follows a part that is based on an article published with J.~Kalkkinen~\cite{FK}. So the main result of this chapter is to explain and to understand more abstractly and analytic-geometrically the restriction property and the so-called ``restriction martingale" $Y_t$.  More precisley, we investigate the relations between SLE and the modular geometry, which underlies CFT on non trivial surfaces more deeply.  The main idea to do so, is to explore models defined in arbitrary geometries by cutting open the surface or $n$-connected domain along the domain-wall and to calculate how the associated CFT correlators change.  

Part~4 is intended as a condensed and informal introduction to fundamental notions of quantum field theory. It has to be considered as an illustrated glossary and it is written with the only purpose to acquaint the reader with almost no prior knowledge of quantum field theory or CFT with the words and concepts, used by the physicists. A good example to illustrate the possible  linguistic difficulties that might arise in trying to read a physics paper is the word ``world-sheet"  that in mathematical  terminology would be referred to as a ``Riemann surface". We avoid deliberately the pseudo-rigorous treatment of quantum field theory and try not to drag the reader into so-called ``rigorous methods", which often lack the physical and mathematical beauty at the same time. Physics should serve as an inspiration for equally elegant mathematics! 

Section~5  prepares the setting for the final part. It contains material from complex-algebraic geometry. 
Further it illustrates how a concept from physics evolved and became in the hand of mathematicians a part of ``solid" mathematics. 

We start by recalling the basic definitions and properties of the moduli space of $n$-punctured curves and the decorated moduli space of curves with a formal parameter at the marked points.  Then we introduce the important concept of Virasoro uniformisation~\cite{ADKP, BS,K,TUY} and highlight its physical interpretation. In this context special emphasis is put on the role  of Schiffer variation (a special case of Kodaira-Spencer deformation theory); as it is an important tool to generalise the L{\oe}wner method or in the program of axiomatic CFT. We then continue by introducing the concept of determinant line bundles, as needed to model the partition function, as well to extend Virasoro uniformisation to the case, with non-trivial central charge.
Here we follow the excellent exposition of~\cite{BK}. This chapter is closed by discussing the ``Segal-Kontsevich style" axiomatic framework of CFT, which applies to arbitrary geometries. 

In the last chapter of this manuscript, which is based on a collaboration with M.~Kontsevich~\cite{KBonn} and J.~Kalkkinen, we embark on the program of connecting CFT and SLE in a very general setting, that unifies the previously introduced notions.

The fundamental object is the operator $\frac{\kappa}{2}L^2_{-1}-2L_{-2}$ as well as a degenerate highest-weight field $\psi$. This way we obtain a generator of a diffusion process on the moduli space of decorated surfaces.
Using the Virasoro uniformisation, one deduces, that the above operator  is hypo-elliptic. The stochastic process thus obtained explores, at least locally, the moduli space and should allow to construct the  ``restriction martingales" in arbitrary geometries. 

\subsection{Motivation}
Probably most of us came  along at one or the other occasion, the following problem. Given a sequence of numbers, simple figures or some other data one should continue it. 
This sort of problems are for example used to ``measure intelligence". But we  know, that sometimes without any further specifications one  can continue it  in different ways, depending on the knowledge  of the person. Nevertheless most often we think that we can extrapolate because   there should be something like a pattern hidden in the data. 
But it could also happen, that the sequence leads to some object outside the given set or class, and it is a priori not clear what it is.  
A simple experiment that reveals some of the features mentioned above is to use the spell-checker on the computer, since it has a finite search (=limit) space, that can be partitioned according to the chosen language (or multilingual) and additionally a limited correction algorithm. 

The archetypical example in this context is the notion of infinity and zero, that mankind had to add, following either the  ``infinite patern" of the natural numbers or to cope with ``nothing". In mathematics and  physics, one encounters this sort of difficulties constantly. They bear names like limit, compactification, renormalisation, phase transition etc. and constitute a beautiful and difficult conceptual world. So, let us just mention some other examples. E.g. in particle physics one has to deal with the  ideas concerning ``high-energy limit" and ``approximate symmetry". ``Energy" is a function on Minkowski momentum space. So ``high-energy limit" might mean, geometrically, passing to the boundary of a compactification of Minkowski space, say in the conformal compactification.  
Another important example shows up, as we will see later, when we consider the partition function for the Polyakov string.  Again, the partition function is thought of as ``living" on the compactification of the moduli space of algebraic curves of genus $h$. Although  the data for the interior of moduli space is the same, there  exist several solutions to the problem of what the compactified moduli space should be. Just let us mention the ones by Satake, Baily or Deligne-Mumford. Additionally, for the Teichm{\"u}ller space there is the one by Thurston. This shows that what the right answer might be, is a priori not clear, but is rather dependent  on how it fits in an existing  setting or how the concept is manageable from the computational point of view, i.e. how practical it is. 

To come back to the intended framework, we will discuss now the above mentioned ideas in the context of statistical mechanics of planar models defined on a lattice. The reason to introduce them is to simplify or substitute continuous problems. The advantage for working in the discrete setting is that one can avoid technicalities, do simulations or in general, the setup becomes easier to understand. 
But on the other hand the  price one has to pay for such simplifications is that inherent symmetries of the continuum model are lost or the discretisation process itself bears some arbitrariness in its formulation. However, there exists also the case, where the continuum is easier to analyse than the discrete world. 

Therefore  understanding the connections between grid-based models and continuum models is  of fundamental importance.  One reasonable way to proceed towards the continuous object  is by taking a {\bf scaling limit} of a series of grid approximations. This means making sense of the limit of a sequence of grids of finer and finer mesh. But most often it is not clear, what the answer should be.

However there are some prominent exceptions.  The classical  ones are the relation between simple random walk (SRW) and Brownian motion or the Ising model, which have been extensively studied and quite well understood. 

So let us now consider a domain $D$,  i.e. a nonempty open connected set in  $\C$ such that each component of $\partial D$ has positive diameter and further a  graph $\textsf{G}=\G(D,\delta)$, which is an approximation of the domain $D$ in a regular lattice $\Lambda$ with mesh size $\delta>0$. The set of vertices $\V$ is partitioned into two disjoint subsets. The interior vertices $\V_{\mbox{\tiny int}}(\G)$, of $\G$ are the vertices of $\V$ which are in $D$, and the boundary vertices, $\V_{\partial}(\G)$, are the intersections of edges of $G$ with $\partial D$. From the physical point of view the vertices constitute  the {\bf component subsystems} . So each state  of the subsystem $i$ is described mathematically as a point in a finite dimensional space $X_i$, that additionally comes equipped with a probability measure $d\mu_i$. For example $X_i=\R, \R^n, \Z_2$ or $SU(N)$. Most often the spaces $X_i$ are assumed to be identical.

The goal of statistical mechanics is to assign a probability measure on the product space of individual components. To account for the interactions, which also reflect the geometry of the index set (= vertices in the domain) these measures are not simple tensor product measure but generally have the form
\begin{displaymath}
d\mu^{(n)}=e^{-U}\cdot\bigotimes_{i=1}^{n} d\mu_i,
\end{displaymath}
where $U$ is the interaction energy. Hence these measures are essentially a density times a product of local measures.

For concreteness we continue  our discussion with a  two dimensional spin system, the celebrated Ising model.   A {\bf configuration of spins} on the set of vertices $\V$ is a map $\sigma: \V\rightarrow\{-1, +1\}$. There are $2^{|\V|}$ such configurations, where $|\V|$ denotes the cardinality of the set $\V$.  To construct the Gibbs measure, i.e. to account for the interaction we define an observable, the {\bf energy}, which is a functional on the space of configurations, as \begin{displaymath}
E[\sigma]:=-\frac{1}{2}\sum_{\{ij\}}\sigma(i)\cdot\sigma(j)\quad\qquad i,j\in\V
\end{displaymath}
where $\{ij\}$ denotes a pair of nearest neighbours. The energy functional takes its minimum on two specific configurations, namely when either all spins are up or equally down, i.e. $\forall i\in\V$: $\sigma(i)=+1$ resp. $\sigma(i)=-1$.   {\bf States} are  probability measures ${d\mu}=d\mu(\sigma)$ on the space of configurations $\{\sigma\}$. These measures form a convex set, i.e. with ${d\mu}_1$, ${d\mu}_2$ also any convex combination  
\begin{displaymath}
d\mu=\alpha_1\; {d\mu}_1+\alpha_2\; {d\mu}_2,\qquad\alpha_{1,2}\geq0\quad\mbox{and}\quad\alpha_1+\alpha_2=1
\end{displaymath}
is a (mixed) state. The thermodynamical equilibrium at the {\bf absolute  temperature} $T>0$ is described by the following {\bf Gibbs state} \begin{displaymath}
d\mu_G(\sigma):=\frac{1}{Z} e^{-\beta E[\sigma]},\qquad \beta:=\frac{1}{T}\quad\mbox{(inverse temperature)}
\end{displaymath} 
where the numerical pre-factor $Z>0$  gives the normalisation, such that the sum of all elementary events ads up to 1. It is defined as 
\begin{displaymath}
Z:=\sum_{\{\sigma\}} e^{-\beta E[\sigma]}
\end{displaymath}
and called the {\bf partition function}. 

The most interesting feature of statistical mechanics systems are phase transitions. However they exist only in the {\bf thermodynamic limit}. The reason is, that so far all expressions for the finite system $\V$ are polynomial in a fundamental energy scale. Therefore singularities can only occur after the delicate double limit,
\begin{displaymath}
|\V|\rightarrow\infty,\qquad \frac{|\V_{\partial}|}{|\V|}\rightarrow 0.
\end{displaymath}
has been taken. Only now thermodynamic quantities may become singular. Because of the role of conformal field theory, we will be primarily interested in continuous phase transitions. For the two-dimensional Ising model it has been shown by L. Onsager that it exhibits such a second order transition.  If the temperature  crosses a special value $T_c>0$, called the {\bf critical temperature}, then the behaviour of the system changes drastically. So  $T_c$   separates a high temperature disordered phase from a low temperature  ordered phase. In the high temperature phase the 2-point function of the order parameter (the spins)
\begin{displaymath}
\langle\sigma(x)\sigma(y)\rangle:=\sum_{\{\sigma\}}\sigma(x)\cdot \sigma(y)\, e^{-\beta E[\sigma]}
\end{displaymath}
will fall off exponentially, i.e. like
\begin{displaymath}
\langle\sigma(x) \sigma(y)\rangle\sim e^{-\frac{|x-y|}{\xi}}
\end{displaymath}
where the {\bf correlation length} $\xi=\xi(T)$ depends on the temperature (we see that $\xi^{-1}$ can be regarded as a mass for the theory). As the temperature  approaches its critical value, the correlation length increases towards infinity, like the inverse power of $T-T_c$:
\begin{displaymath}
\xi(T)\sim\frac{1}{|T-T_c|}
\end{displaymath}
and therefore the theory becomes mass-less.  As a consequence  the two-point function  falls off as a power law
\begin{displaymath}
\langle\sigma(x) \sigma(y)\rangle\sim -\frac{1}{|x-y|^{d-2+\eta}}
\end{displaymath}
where $d$ is the dimension of the system and this expression defines the {\bf critical exponent} $\eta$. The critical exponent calculated for the two dimensional Ising model is $\eta=1/4$. Therefore its correlator would behaves as
\begin{displaymath}
\langle\sigma(x) \sigma(y)\rangle\sim \frac{1}{|x-y|^{1/4}}.
\end{displaymath} 
So far we did not speak about Quantum Field Theory (QFT), although 
since the 1960s it has been realised  that the scaling limit of a general isotropic system near a continuous phase transition is / should be a {\bf Euclidean quantum field theory}. Moreover it was a real breakthrough around 1984, when Belavin, Polyakov and Zamolodchikov  \cite{BPZ1,BPZ2} showed, that the critical behaviour of two-dimensional systems is governed  by a so-called Conformal Field Theory (CFT). In many cases the CFT is known. For example the critical Ising model corresponds to a so-called {\bf minimal model} with central charge $c=1/2$. The alleged relation is the following.  If we take a near-critical lattice model, such that the correlation length $\xi\gg a$  where $a$ denotes the lattice spacing, the following limit exists:
\begin{equation}
\label{statmech2CFT}
\langle\phi(r_1)...\phi(r_N)\rangle_{\mbox{\tiny CFT}}=\lim_{\xi/a\rightarrow\infty} a^{-N \Delta_{\phi}}\langle \sigma(r_1)...\sigma(r_N)\rangle_{\mbox{\tiny lattice}}
\end{equation} 
where $\phi(r)$ is a local quantum field, and $\sigma(r)$ is the corresponding lattice quantity. The non-trivial power $\Delta_{\phi}\geq0$ is the {\bf scaling dimension} or {\bf conformal dimension} of the field $\phi$. The CFT correlator $\langle\phi(r_1)...\phi(r_N)\rangle_{\mbox{\tiny CFT}}\in\C$, which can be explicitly calculated, is a real analytic function on $\R^{2N}\setminus\{\mbox{all diagonals}\}$. The numbers $\Delta_{\phi}$ belong to a discrete subset of $\R_+$, the {\bf conformal spectrum} of the CFT. For the Ising model it looks like $\{0, 1/8, 1\}$. However the above correspondence, that is rooted in an emphasis on correlation functions of {\bf local} (or quasi-local) operators and their algebra encoded in the operator product expansion, has been proved in very few examples;  however  if assumed, it has many powerful consequences as we will see.

But what is the intuitive picture behind the correlation length that seems to play such a key role? Near the critical temperature the spin system is an aggregate of domains (droplets) with spin $-1$ and spin $-1$, such that droplets of all sizes up to the correlation length are present and droplets of different spin are in droplets etc. The {\bf domain walls} (phase boundaries) separating spin droplets form a complicated system of random, disjoint Jordan curves (arcs) covering the domain.

If   we restrict the setting further by considering a simply connected domain with two marked points $A$ and $B$ on the boundary and  a {\bf hexagonal lattice} approximating it, then the following boundary conditions:
\begin{figure}[ht]
\begin{center}
\includegraphics[scale=0.3]{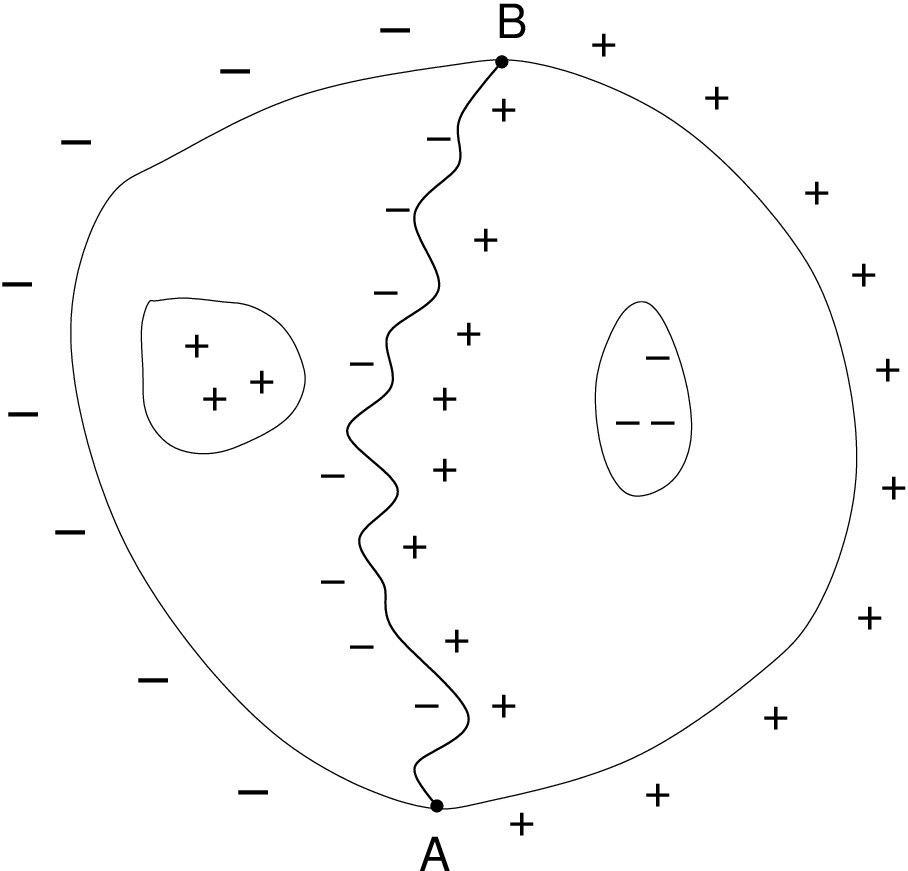}
\caption{The domain-wall with spin -1 to the left, $+1$ to the right, connecting $A$ and $B$.} 
\label{domainwall}
\end{center}
\end{figure}
clockwise spin $+1$ on the part of the boundary from $B$ to $A$ and  $-1$ from $A$ to $B$, give necessarily rise to  a long {\bf chordal} domain wall  from $A$  to $B$. It is a Jordan  curve (we call this also a {\bf simple curve}) connecting vertices such that on the left we always have spin $-1$ and to the right $+1$. Therefore with this  boundary conditions in each configuration there will be such  a line, that fluctuates according to the particular realisation of the spins in the interior.   

Before we proceed further  some words of caution are in place. Not all geometric objects, and more specifically the probabilities attributed to them, that are  observable in the discrete setting are suited to continue to exists in a nontrivial way if we proceed towards a ``limit". If for example we would consider the limit of the probability that in the set $\{1,2,...,N\}$ of natural numbers a particular one, say $n$ occurs, this would tend to zero under the assumption of equal probability as $N\rightarrow\infty$. On the other hand, if we ask for the probability of getting an even number, that would tend to $1/2$, and therefore would be a sensible event.

In the case of our model with the particular choice of boundary conditions, we  have an object that  persists  on all finite scales  and should therefore  be macroscopically observable.

So we can introduce on the set of configurations $\{\sigma\}$  an equivalence relation by declaring two  realisations $\sigma_1$ and $\sigma_2$ as equivalent if they have exactly  the same chordal domain wall, that connects the  two marked  boundary points, cf. Fig.~(\ref{domainwall}). The probability measure on the quotient space is now the image of the  Gibbs measure $d\mu_G$ under the natural projection $\pi$, i.e.
\[
\begin{CD}
\{\sigma\}\\
@V \pi VV\\
\{\sigma\}\big/{\sim} 
\end{CD}
\]
We note that the $\sigma$-algebra on the set of equivalence classes is coarser than the original one.  Hence 
the set of observables is smaller and therefore not all can be reconstructed  from the knowledge of the chordal lines; one looses information.
\begin{figure}[ht]
\begin{center}
\includegraphics[scale=0.4]{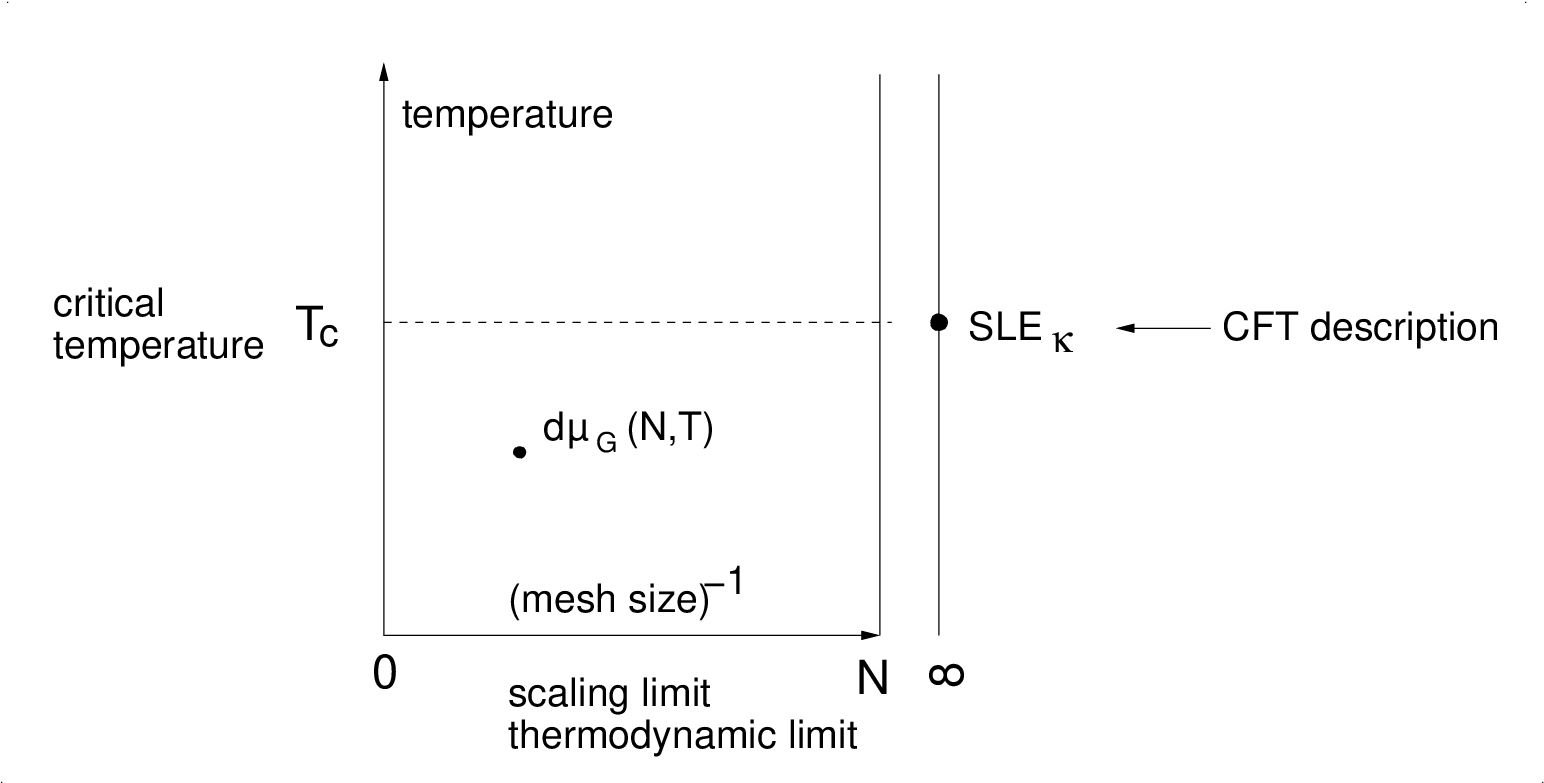}
\caption{The $\mbox{SLE}_{\kappa}$ claim: $\mbox{SLE}_{\kappa}$ is the unstable fixed point of the renormalization group flow.} 
\label{phase_diagram}
\end{center}
\end{figure}

The  family of measures $\mu_{\delta, T}:=\pi_*\,d\mu_G$ we introduced, depends on two parameters, namely the mesh size $\delta$ and the temperature $T$; cf. Fig.~(\ref{phase_diagram}). Further they are supported on the compact sets of the underlying domain. On general grounds we expect that as we approach along the critical temperature $T_c$ the thermodynamic limit, the limiting measure  should have some interesting properties. From the physical point of view one might understand this as follows. Since the scaling limit measure is described as an unstable fixed point of the renormalisation group (RG) flow there should be an associated  conformal field theory, with some central charge. The measure should therefore be conformally covariant.

This was first stated by  M. Aizenman~\cite{Ai} for critical percolation, and his conjecture then led  to  Cardy's  beautiful formula \cite{Ca2}.
Indeed, as we will see, probability theory  can say much more about the special point in Fig.~(\ref{phase_diagram}), that is denoted $\text{SLE}_{\kappa}$, and corresponds to the RG fixed point.
 
We close this section with two remarks. 
\begin{itemize}
  \item The procedure just described, is not particular to the Ising model.  We can choose other  models, e.g. $Q$-states Potts model (i.e. different Boltzmann-weights) as well as other lattices (not necessarily hexagonal) with appropriate boundary conditions (e.g. wired and free). Then it is conjectured that in the scaling limit at the critical temperature $T_c$ the choices made should be irrelevant for so-called universality classes; however this has not been proved generally. 
  \item For lattice models there are  other  ``good" probabilistic events. In the Ising model slightly below the critical temperature $T_c$ we observe, as mentioned  above, the sea of nested Jordan curves of domain boundaries.
Heuristically, by passing to $T_c$ and simultaneously rescaling we obtain a dense collection of closed, non-intersecting loops, carrying a scale invariant probability distribution. Again, it would be derived from the series of Gibbs measures as we explained in the case of chordal lines.
\end{itemize}

\subsection{Stochastic L{\oe}wner Evolution}

We shall now rephrase in more specific  terms the above discussion and show, what is known at a detailed and not only conceptual physical level. In fact, to a large extent, the rigourous part is what SLE is about. It is the confluence of two perspectives on the same problem, namely to understand the scaling limit of two-dimensional critical discrete systems under the hypothesis of  conformal covariance. One of the approaches is  O. Schramm's \cite{S1} treatment of   the {\bf loop erased random walk} (LERW) on the square lattice.  His treatment of the problem, in particular the question of a scaling limit and its dynamical description via the L{\oe}wner equation, led to  fundamental new insights and results. The other direction originated in G. Lawler and W. Werner's~\cite{LW1, LW2} attempt to classify  a class of Brownian measures on compact sets, with respect to their behaviour under conformal maps. This one, what we might call the static approach, was geared towards a better understanding of aspects of CFT. In principle it is the subtle interplay between the static and dynamic point of view that gives this rich theory.  To some extent it is as in ordinary statistical mechanics where one might approach the problems by averaging over configurations or equally well, given the ergodic hypothesis holds,  by taking the time average of an observable. The interplay as it is described, can be found  in a series of papers by G. Lawler, O. Schramm and W. Werner \cite{LSW1,LSW2,LSW3,LSWlesl,LSWsaw}  and also S. Rohde and O. Schramm \cite{RS}.  The prospective usefulness of the SLE method for physics, as so often, was recognised by J.~Cardy~\cite{Ca3}.

We are now going to recall briefly  some of the main results about SLE that we will need in the sequel. 

The fundamental task when studying a {\bf scaling limit} is to set up a conceptual foundation for it, which  means answering the following two questions:
\begin{enumerate}
  \item What kind of object is the scaling limit?
  \item What does it mean to be the scaling limit?
\end{enumerate}
For models defined on a lattice there is often more than one ``right" answer to the first question. But even if the conceptual framework is fixed, the next problem that emerges is to prove the existence of the scaling limit in the pre-specified category. 

So let us consider the complex plane $\C=\R^2$, but viewed as  a subset of the Riemann sphere $\hat{\C}:=\C\cup\{\infty\}$, which gives  compactness. Further we endow $\hat{\C}$ with the  spherical metric  $d_{\mbox{\tiny sp}}$.  $D$ shall denote a domain, i.e. a nonempty open connected set in $\C$. We need 
\begin{df}
Let $(\Omega,{\cal F})$ and $(E,\mathscr{E})$ be measurable spaces. A mapping $Y:\Omega\rightarrow E$,  $\omega\mapsto Y(\omega)\in E$ is a {\bf random set}, if $\{\omega : Y(\omega)\in B\}\in{\cal F}$ for every $B\in\mathscr{E}$, (i.e. ${\cal F}/{\mathscr{E}}$-measurable) where $E$ is a set of sets and $\mathscr{E}$ is a $\sigma$-algebra defined on $E$.
\end{df}
To make sense of the concept of the {\bf scaling limit} for the  random curves in $D$, as previously considered, we think of them as a {\bf random set} in $\overline{D}$. The Borel $\sigma$-algebra we are going to specify is induced by the Hausdorff metric.

So if  $(X,d)$ is a metric space and if $A\subset X$ and $\epsilon>0$, let $U(A,\epsilon)$ be the $\epsilon$-neighbourhood of $A$. Let ${\cal K}={\cal K}(X)$ be the collection of all nonempty {\bf closed}, {\bf bounded} subsets of $X$. If $A,B\in{\cal K}$, define 
\begin{displaymath}
d_H(A,B):=\inf\{\epsilon | A\subset U(B,\epsilon)\;\,\mbox{and}\;\,B\subset U(A,\varepsilon)\}.
\end{displaymath}
Then $d_H$ is a metric on ${\cal K}(X)$ called the {\bf Hausdorff metric}. If $(X,d)$ is complete, so is $({\cal K}, d_H)$ and further if $(X,d)$ is totally bounded, so is $({\cal K}, d_H)$. The following is known to hold:

\begin{thm} If $X$ is compact in the metric $d$, then the space ${\cal K}$ is compact in the Hausdorff metric $d_H$.
\end{thm}

Now on the collection ${\cal K}(D)$ of closed subsets of $D\subset\hat{\C}$, we use the metric 
\begin{displaymath}
d_{{\cal K}(D)}:=d_H(A\cup\partial D, B\cup\partial D)
\end{displaymath}
and therefore ${\cal K}(D)$ is compact with this metric.
For the planar discrete models as described previously we have the following
\begin{claim}
The chordal domain wall in $D$ or the set of loops in $D$ is  a (are) random element(s) in ${\cal K}(D)$, and its distribution $\mu_{\delta}=\mu_{\delta,D}$, derived from the Boltzmann
weights,  is a probability measure on the associated Borel $\sigma$-algebra $\mathscr{B}({\cal K}(D))$.
\end{claim}
Because the space of Borel probability measures on a compact space is compact in the weak topology, there is at least a sequence $\delta_j\rightarrow0$ such that the {\bf weak limit}
\begin{equation}
\label{weak_limit}
\mu_0:=\lim_{j\rightarrow\infty}\mu_{\delta_j}
\end{equation}
exists. 
\begin{df} The measure $\mu_0$ as  in (\ref{weak_limit}) will be called a {\bf subsequential scaling limit} measure of the chordal domain wall connecting $A,B\in\partial D$  or of the set of Jordan loops in $D$. If $\mu_0=\lim_{\delta\rightarrow0}\mu_{\delta}$, then we say that $\mu_0$ is the {\bf scaling limit measure} of the chordal line or the loops in the domain $D$.
\end{df}

In physics this would be called somewhat vaguely: ``to have the same statistics as...". Actually, there are not many known discrete models where the existence of the scaling limit measure has been established.
The few exeptions are the loop-erased random walk (LERW)~\cite{S1}, the uniform spanning tree (UST)~\cite{S1} and critical site percolation on the triangular lattice~\cite{Sm}.

So-far we did not discuss the support of the scaling limit measure. So what happens to simple curves on the lattice? Will they remain simple in the limit?

We say that the measure is supported on simple path if there is a collection of simple paths whose complement has zero measure. 
Again, the following is one of the few completely understood cases.

\begin{thm}[\cite{S1}]
Let $D$ be a domain in $\hat{\C}$ such that each connected component of $\partial D$ has positive diameter, and let $p\in D$. Then every sub-sequential scaling limit measure of LERW from $p$ to $\partial D$ is supported on simple paths.\\
Similarly, if $p,q$ are distinct points in $\hat{\C}$, then every sub-sequential scaling limit of the LERW from $p$ to $q$ on $\delta\Z^2$ is  supported on simple paths. 
\end{thm}
\begin{figure}[ht]
\begin{center}
\includegraphics[scale=0.4]{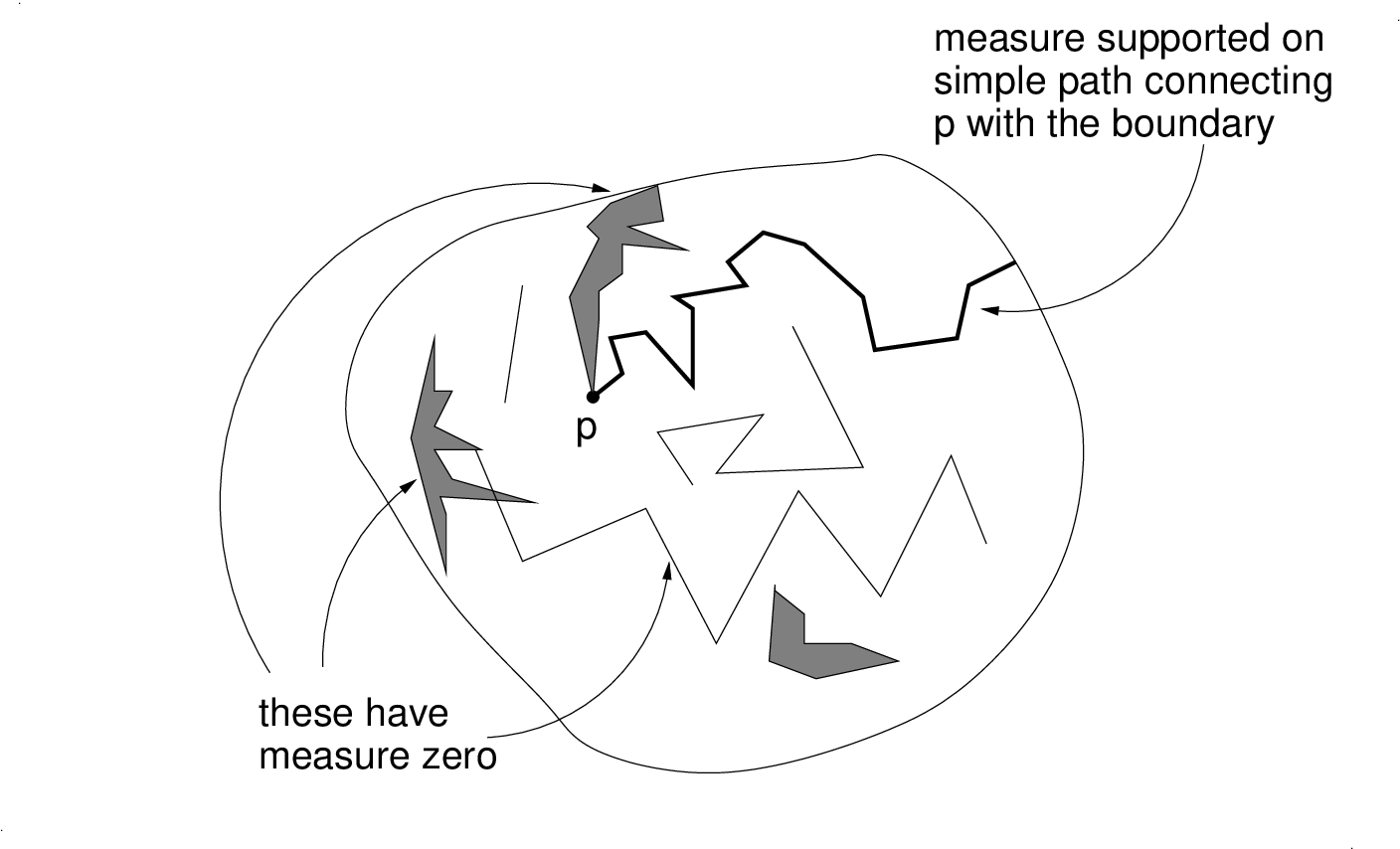}
\caption{The support of the scaling limit measure in the radial case.} 
\label{cd}
\end{center}
\end{figure}

The results so-far only concerned the existence and the nature of the scaling limit measure that was associated to an individual domain. But 
if we consider two domains $D,D'\subset\hat{\C}$, then every homeomorphism $f: D\rightarrow D'$ induces a homeomorphism $f_*:{\cal K}(D) \rightarrow {\cal K}(D')$. Consequently, if $\mu$ is a probability measure on $\mathscr{B}({\cal K}(D))$, then there is an induced probability measure $f_*\mu$ on $\mathscr{B}({\cal K}(D'))$. But in general the two measure will not coincide. However it is believed (proofed)  that there is stronger statement,  usually called \cite{Ai, LPS, S1}: 
\begin{center}
``the assumption of conformal invariance".
\end{center}
The task therefore consists in understanding how the genuinely assigned probability spaces, as derived from lattice systems, couple to the geometry of the underlying domain, and what are the iso-measurable homeomorphisms. 

So let $D\subsetneq\hat{\C}$ be a simply connected domain and $({\cal K}(D),\mathscr{B}({\cal K}(D)))$ the measurable space as introduced previously. 

\begin{claim} There exists a class of probability measures $\mu_D$, obtained as scaling limits, such that the diagram 
\[
\begin{CD}
D@>f>> D'\\
@V F VV  @V FVV\\
({\cal K}(D), \mathscr{B}({\cal K}(D)),\mu_D) @>f_*>> ({\cal K}(D'), \mathscr{B}({\cal K}(D')),\mu_{D'})
\end{CD}
\]
commutes, if $f$ is a biholomorphic map $f$ and $f_*$ denotes the induced measurable mapping.
\end{claim}
One should observe first that the measures are  constructed completely by  intrinsic means on each individual domain.  Then  we compare the ``indigenous" measure with the ``immigrant" measure, i.e. with the image measure (push-forward) constructed on the other domain. The result is that if the measure ``traveled" via a conformal map that respects some boundary conditions like mapping (ordered) marked points into (ordered) marked points, then after arrival the measure is the same, as the one, that is already there (and hence has the same support). This property is what is called {\bf conformal covariance}.  We can phrase it also this way. Immagine two observes performing experiments to determin the measures in two different frames. Then, if they relate their observations by a biholomorphic map, the results should agree. The whole thing would be pointless if we just would define the new measures by simply pushing them forward.  However, if we know that conformal covariance holds then we only need one production plant for the scaling limit measures (standard domain) and then we can send the product by conformal homeomorphisms. This is what is done for SLE! 
 
The above discussion is not void, in particular because we have
\begin{thm}[\cite{S1}]
Let $D\subset\C$ be a simply connected domain in the complex plane with $p\in D$. Suppose that $f: D\rightarrow D'$ is a conformal homeomorphism onto another (necessarily) simply connected domain $D'\subset\C$. Then 
\begin{displaymath}
f_*\,\mu_{p,D}=\mu_{f(p),D'}
\end{displaymath}
where $\mu_{p,D}$ is the scaling limit measure of LERW from $p$ to $\partial D$, and $\mu_{f(p),D'}$ is the scaling limit measure of LERW from $f(p)$ to $\partial D'$.
\end{thm}

In fact there is more since   there is  a continuous one-parameter family  of probability measures, called {\bf SLE}-measures that are supported on the trace of (simple) paths  \cite{S1}  as well a family of measures supported on ``{\bf Brownian loops}" \cite{LW3} such that the above diagram commutes. The crucial difference however is the nature of their respective support.

In the case of (simple) path the assumption of conformal covariance leads to the afore mentioned explicit description of the scaling limit measure for random curves in terms of solutions of L{\oe}wner's differential equation with a Brownian motion parameter. For a detailed derivation that starts directly  with K. L\"owner's approach one may consult \cite{A2,D} and for the stochastic case \cite{S1,Wstf,Lin}. 
The version we are recalling  is the definition of chordal SLE${}_{\kappa}$  in the
upper half-plane $\H$ that starts from $0$ and continues to
infinity \cite{Lin, Wstf}.

So let $(\Omega,{\cal F}, ({\cal F}_t),P)$ denote a standard filtered probability space that is complete and continuous from the right.
Then the  chordal SLE$_\kappa$ curve $\gamma$
is characterised as follows: 
\begin{df}[L{\oe}wner Equation]
\label{loewner_eq}
For $z\in\H, t\geq0$ define $g_t(z)$ by $g_0(z)=z$ and 
\begin{equation}
\frac{\partial g_t(z)}{\partial t}  = \frac{2}{ g_t(z) - W_t}
\label{lowner}.
\end{equation}
\end{df}
The maps $g_t$ are normalised 
such that $g_t (z) = z + o(1) $ when $z \to \infty$
and  $W_t:= \sqrt{\kappa}\,B_t$ where  $B_t(\omega)$ is the standard
one-dimensional Brownian motion defined on
$\R_+\times\Omega$ starting in 0 and with variance
$\kappa>0$. Given the initial point $g_0(z)=z$, the ordinary differential
equation (\ref{lowner}) is well defined until a random time
$\tau_z$ when the right-hand side in (\ref{lowner}) has a pole.
There are two sets of points that are of interest, namely the preimage of infinity $\tau^{-1}(\infty)$ and its complement. For those in the complement we define:
\begin{equation}
\label{Khull}
K_{t}:=\overline{\{z\in\H: \tau(z)<t\}}
\end{equation}
The family $(K_t)_{t\geq0}$, called  {\bf hulls}, is an increasing family of compact sets in $\overline{\H}$ where $g_t$
is the uniformising map from $\H\setminus K_t$ onto $\H$. It
has been shown in \cite{RS, LSW2} that there exists a continuous process
$(\gamma_t)_{t\geq0}$ with values in $\overline{\H}$ such that
$\H\setminus K_t$ is the unbounded connected component of
$\H\setminus\gamma[0,t]$ with probability one. This process is
the {\bf trace} of the SLE${}_{\kappa}$ and it can be recovered from
$g_t$, and therefore from $W_t$, by
\begin{equation}
\gamma_t  =  \lim_{z\rightarrow W_t, z\in\H} g_t^{-1}(z)~.
\end{equation}
Now, for another simply connected domain $D$ with two boundary points $A,B\in\partial D$ the chordal $SLE_{\kappa}$ in $D$ from $A$ to $B$ is defined as
\begin{displaymath}
K_t(D, A,B):=h^{-1}(K_t(\H,0,\infty))
\end{displaymath}
where $K_t(\H,0,\infty)$ is the hull as in (\ref{Khull}) and $h$ is the conformal map (cf. diffeomorphism covariance) from $D$ onto $\H$ with $h(A)=0$ and $h(B)=\infty$.
The value of the constant $\kappa$ characterises the nature of the resulting
curves 
\begin{prop}[\cite{RS}] Let $\gamma$ be the trace of an $SLE_{\kappa}$. Then a.s.:
\begin{itemize} 
  \item if\, $0\leq\kappa\leq 4$, the curve $\gamma$ is simple,
  \item if\, $4 < \kappa < 8 $ the trace is a self-touching curve (curve with double points, but without
crossing its past) and
  \item if\, $8\leq \kappa $ the curve $\gamma$ becomes space filling (Peano curve).
\end{itemize}
\end{prop}
Further we know by
\begin{prop}[\cite{Be3}] The Hausdorff dimension  of the $SLE_{\kappa}$ trace is:
\begin{equation}
\label{Hausdorff_Dim}
\min\,(1+\frac{8}{\kappa},2)\quad\text{a.s.}
\end{equation}
\end{prop}

The results that follow now where derived in \cite{LSWr}. They inspired a lot of the research presented in this work, especially the ``mysterious form"  of Proposition~\ref{Y_t} played a crucial role. At this point we recommend to consult \cite{LSWr} for the detailed description and further information.

Let us consider chordal $\mbox{SLE}_{\kappa}$ for $\kappa\leq4$ which produces a simple curve $\gamma:\R_+\rightarrow\overline{\H}$ with the following properties: $\gamma(0)=0$, $\gamma(0,\infty)\subset\H$ and $\gamma(t)\rightarrow\infty$ as $t\rightarrow\infty$. We define a {\bf hull} as a bounded set $A\subset\overline{\H}$ such that $A=\overline{A\cap\H}$, $\{0,\infty\}\notin A$ and $\H\setminus A$ is simply connected. Let  $A$ now be a compact hull and 
\begin{displaymath}
\phi_A: \H\setminus A\rightarrow\H
\end{displaymath}
the conformal homeomorphism with $\phi_A(0)=0$, $\phi_A(\infty)=\infty$ and $\phi'_A(z)\sim z$ as $z\rightarrow\infty$. 
Then the composition of these conformal mappings defines a pseudo-semi-group on hulls:
\begin{displaymath}
\phi_{A.B}=\phi_B\circ\phi_{A}.
\end{displaymath}
\begin{figure}[ht]
\begin{center}
\includegraphics[scale=0.5]{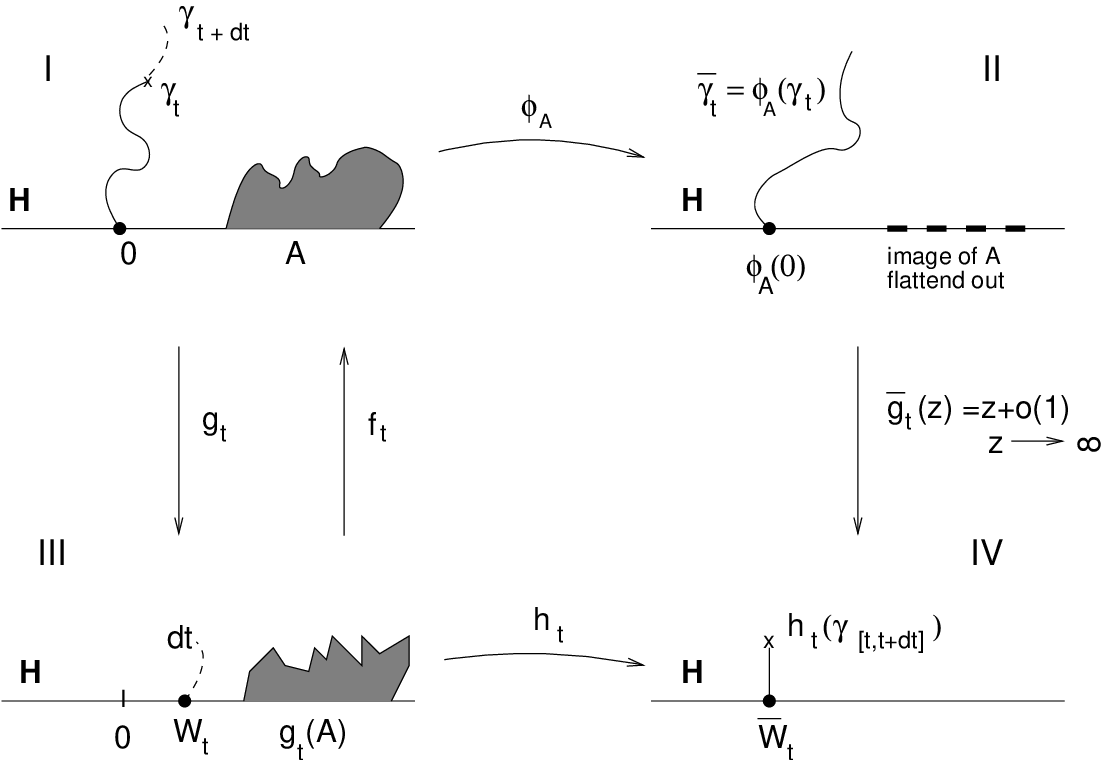}
\caption{The commutative diagram from~\cite{LSWr}.} 
\label{cdr}
\end{center}
\end{figure}
To ease the understanding of the following steps compare with  Fig.~\ref{cdr}. Let $(g_t)$ be a L{\oe}wner chain with driving process $W_t$, and $A$ a hull. For $A\subset g^{-1}(\H)$ define $h_t:=\phi_{g_t(A)}$ and also $\overline{W}_t:=h_t(W_t)$. Then $\bar{g}_t:=\phi_{g_t(A)}\circ g_t\circ\phi^{-1}_A$ is itself a L{\oe}wner chain, given $t$ is small enough.

Suppose now that the driving process $(W_t)$ of the chain is a semimartingale with
\begin{displaymath}
dW_t=\sqrt{\kappa}\,dB_t+b_t dt
\end{displaymath}
where $B_t$ is a standard Brownian motion and $b_t$ some bounded progressive process. Let $z$ be a point in $\H\setminus g_t(A)$ or in a punctured neighbourhood of $W_t$ in $\R$. Then the following formul{\ae}  hold 
\begin{eqnarray}
\partial_t h_t(z)& = & -\frac{2h'_t(W_t)^2}{h_t(z)-\overline{W}_t}+\frac{2 h'_t(z)}{z-W_t}\\
\partial_t h'_t(z) & = & -\frac{2h'_t(W_t)^2h'_t(z)}{(h_t(z)-\overline{W}_t)^2} +\frac{2 h'_t(z)}{(z-W_t)^2}-\frac{2h''_t(z)}{z-W_t}\\
{[\partial_t h_t]}(W_t)& = & \lim_{z\to W_t}\left(\frac{2h'_t(W_t)^2}{h_t(z)-\overline{W}_t}-\frac{2h'_t(z)}{z-W_t}\right)=-3h''_t(W_t)\\
{[\partial_t h'_t]}(W_t)& = &\lim_{z\to W_t}\partial_t h'_t(z)=\frac{h''_t(W_t)^2}{2h'_t(W_t)}-\frac{4h'''_t(W_t)}{3}.
\end{eqnarray}
At this step one has to use a rather refined version of It\^o's formula, as it can be found in [Revuz-Yor], to derive the following stochastic differential equations (SDEs):
\begin{eqnarray}
d\overline{W}_t  &=&  h'_t(W_t)dW_t+\left(\frac{\kappa}{2}-3\right)h''_t(W_t)dt \\
dh'_t(W_t)  &=& h''_t(W_t) dW_t+\left(\frac{h''_t(W_t)^2}{2h'_t(W_t)}+\left(\frac{\kappa}{2}-\frac{4}{3}\right)h'''_t(W_t)\right)dt 
\end{eqnarray}
Let us now briefly describe the conformal restriction property~\cite{LSWr} which we shall need in the sequel. So let us consider a simply connected domain in the complex plane $\C$, e.g. the upper half-plane $\H$. If we suppose that two boundary points are given, say $0$ and $\infty$ then we are interested in closed random subsets $K\subset\H$ such that: 
\begin{itemize}
  \item $K\cap\R = \{0\}$, $K$ is unbounded and $\H\setminus K$ has two connected components.
  \item For all simply connected subsets $H\subset\H$ such that $\H \setminus H$ is bounded and bounded away from the origin, the law of $K$ conditioned on $K\subset H$ is equal to the law of $\phi(K)$, where
$\phi$ is a conformal map from $H$ onto $\H$ that preserves the boundary points $0$ and $\infty$.
\end{itemize}
The law of such a set $K$ is called a {\bf (chordal) restriction measure}. One can show that there exists only a one-parameter family ${\bf P}_{\alpha}$ of such probability measures, where $\alpha$ is a positive number, the {\bf restriction exponent}, and that 
\begin{equation}
\label{resmeas}
{\bf P}_{\alpha}[K \subset H] = \phi' (0)^{\alpha}
\end{equation}
when $\phi : H\rightarrow\H$ is chosen in such a way that $\phi(z)/z\rightarrow1$ as $z \rightarrow\infty$. 

Let us define the set 
\begin{displaymath}
V_A:=\{\omega\; |\; \gamma[0,\infty)\cap A=\emptyset\}
\end{displaymath} 
which is measurable and has positive probability. On the event $V_A$ we can consider  the path $\phi_A\circ\gamma(t)$.
\begin{df}
$\mbox{SLE}_{\kappa}$ satisfies the {\bf chordal restriction property} if the distribution of $\phi_A\circ\gamma(t)$ conditioned on $V_A$ is the same as (a time change of) $\mbox{SLE}_{\kappa}$
\end{df}
Then the series of the above formul{\ae} yield
\begin{prop} \label{rest5/8}Chordal $\mbox{SLE}_{\kappa}$ satisfies the restriction property for $\kappa=8/3$ and for no other value of $\kappa\leq 4$. If   $A$ is a hull, $\kappa=8/3$ and $\phi_A$ as above, then 
\begin{displaymath}
{\bf P}\{\gamma[0,\infty)\cap A=\emptyset\}=\phi'_A(0)^{5/8}.
\end{displaymath}
\end{prop}
For a proof see  \cite {LSWr}.

Recall that the {\bf Schwarzian derivative} $Sf$ of $f\in C^3(\C)$ is defined as
\begin{equation}
\label{schwarzian}
Sf(z):=\frac{f'''(z)}{f'(z)}-\frac{3}{2}\left[\frac{f''(z)}{f'(z)}\right]^2.
\end{equation} 
Let 
\begin{eqnarray}
\label{magic_c1}
\lambda & := & \frac{(3\kappa-8)(\kappa-6)}{2\kappa}, \\
\alpha & := & \frac{6-\kappa}{2\kappa}.
\label{magic_c2}
\end{eqnarray}
Then we have from~\cite {LSWr},~Proposition 5.3:
\begin{prop}[\cite{LSWr}]
\label{Y_t}
Suppose $\kappa\leq 4$ and $\alpha$, $\lambda$ are defined as in (\ref{magic_c1},\ref{magic_c2}). Set $W_t:=\sqrt{\kappa} B_t$ and let
\begin{equation}
\label{martingaleY}
Y_t:=h'_t(W_t)^{\alpha}\cdot \exp\left(\frac{\lambda}{6}\int_0^t S h_s(W_s)ds\right).
\end{equation}
Then $Y_t$ is a local martingale for $t<\tau_A:=\inf\{t: K_t\cap A\neq\emptyset\}$. If $\kappa\leq 8/3$, then $Y_t$ is a bounded martingale.
\end{prop}
Note that by It\^o's formula we get for $Y_t$ as defined above in~(\ref{martingaleY}):
\begin{displaymath}
\frac{dY_t}{Y_t}=\alpha\,\frac{h''_t(W_t)}{h'_t(W_t)}\,dW_t.
\end{displaymath}
 
\section{Operator--algebraic aspects of SLE}
\label{OA}

We are now going to derive several algebraic identities which resemble  objects found in CFT  very much. In fact we will explain later, how they can be interpreted in the language of conformal field  theory. Nevertheless the discussion here shows, that we have a fair chance to build a mathematically rigorous  model of CFT, based on probabilistic considerations. That indeed reveals the existence of  deep connections to other  mathematical frameworks like vertex operator algebras.

\subsection{Correlation functions and Ward type identities}

Let us consider chordal SLE for $\kappa\leq4$ such that its trace is a simple curve. If we define for $z\in\overline{\H}$
\begin{displaymath}
f_t(z):=g_t(z)-\sqrt{\kappa} B_t,
\end{displaymath}
where $B_t$ is a standard real-valued Brownian motion with $B_0\equiv0$ and $g_t$ as in (\ref{lowner}). It follows from the Markov property of $B_t$, that the law of $(f_{t_0+t}\circ f^{-1}_{t_0}, t\geq0)$ is identical to that of $(f_t, t\geq0)$. We can now write the stochastic L{\oe}wner equation (\ref{lowner}) in It\^{o} form as
\begin{equation}
\label{loewner_itoform}
df_t(z)=-\sqrt{\kappa}\,dB_t+\frac{2}{f_t(z)}dt.
\end{equation}
It\^o's formula shows that 
for any set of points $\{x_1, \ldots, x_n\}\subset\R$ and any smooth
function $F: \R^n \to \R$ 
\begin {eqnarray*}
\lefteqn{ dF (f_t (x_1) , \ldots, f_t (x_n))
= 
-\sqrt{\kappa}\; dB_t \sum_{j=1}^n \partial_{j} F (f_t (x_1), \ldots ,f_t ( x_n))
}\\
&&+ 
dt 
\left\{ \frac {\kappa}{2}
(\sum_{j=1}^n \partial_j)^2 + (\sum_{j=1}^n \frac {2}{f_t (x_j)} 
\partial_j ) \right\} F (f_t (x_1), \ldots, f_t( x_n)).
\end {eqnarray*}

If one defines the operators 
\begin{displaymath}
L_N := - \sum_{j=1}^n x_j^{1+N} \partial_j,
\end{displaymath}
and the value  
\begin{displaymath}
F_t := F(f_t(x_1), \ldots, f_t(x_n)),
\end{displaymath}
then we get the important  
\begin{lem}[SLE-Martingale]
$$
dF_t = -\sqrt{\kappa}\; dB_t\; L_{-1}F_t + dt (\frac{\kappa}{2} L_{-1}^2 - 2 L_{-2} ) F_t
.$$
\end{lem}
We note that 
the chordal crossing probabilities \cite {LSW1,LSW3} are 
identified using the fact that the drift term vanishes iff $F_t$
is a martingale i.e.  
\begin{equation}
\label{singular_2}
(\frac{\kappa}{2} L_{-1}^2 - 2 L_{-2} )F_t=0,
\end{equation}
and that the operators $L_N$ satisfy the commutation relations of the Witt algebra.

Suppose now that $\gamma$ is a random simple curve in $\overline{\H}$ that satisfies the restriction property (in this paragraph however we do not use the fact that this implies $\gamma$ is chordal $\mbox{SLE}_{8/3}$ and that $\alpha=5/8$). Then by Prop.~\ref{rest5/8} we know that the probability that it avoids the slit $[x, x+i\epsilon\sqrt{2}]$, $x\in\R$, must be of the form $\Phi'(0)^{\alpha}$ for some $\alpha>0$ where
\begin{displaymath}
\Phi(z):=\sqrt{(z-x)^2+2\epsilon^2}-\sqrt{x^2+2\epsilon^2}
\end{displaymath} 
is the conformal map from $\overline{\H}\setminus[x, x+i\epsilon\sqrt{2}]$ onto $\overline{\H}$ that preserves $0$ and $\infty$ with $\Phi(z)\sim z$ as $z\rightarrow\infty$. For $\epsilon>0$ and $x\in\R^*$ we define the event 
$$
E_\eps (x) := \{ \gamma \cap [x, x + i \eps \sqrt {2} ] \not= \emptyset \},
$$
i.e. that the curve intersects the given slit. If we fix the real numbers $x_1, x_2,...,x_n$, then again by the restriction property we have for the event that the path intersects all slits (but of variable size $\epsilon_i$)  
$${\bf P} [ E_{\eps_1} (x_1) \cup \ldots \cup E_{\eps_n} (x_n)]
= 1 - \Phi_{\overline{\H} \setminus \cup_{j=1}^n [x_j, x_j + i \eps_j \sqrt {2}]}'(0)^\alpha.
$$
In principle these derivatives can be calculated by the Schwarz-Christoffel transformation; cf. \cite{A1}. The application of an inclusion-exclusion formula yields the values of the probabilities
$$
p(x_1, \eps_1, \ldots, x_n , \eps_n)
:={\bf P} [ E_{\eps_1}(x_1) \cap \ldots \cap E_{\eps_n} (x_n) ]
$$
as a function of the $x_i$'s and $\eps_i$'s. For example for $n=1$,
\begin{displaymath}
p(x,\eps)={\bf P}[E_{\eps}(x)]=1-\left(\frac{x}{\sqrt{x^2+2\eps^2}}\right)^{\alpha}.
\end{displaymath}
In particular we get the asymptotic behaviour for $\eps\rightarrow0$
\begin{displaymath}
p(x,\eps)\sim\frac{\eps^2\alpha}{x^2}.
\end{displaymath}
We then can define 
\begin{equation}
\label{ }
B^{(\alpha)}_1(x):=\frac{\alpha}{x^2}=\lim_{\eps\rightarrow 0}\eps^{-2} p(x,\eps).
\end{equation}
This enables to define and to compute the functions $B_n = B_n^{(\alpha)}$
as
\begin{equation}
\label{fw1}
B_n ( x_1, \ldots, x_n)
:=
\lim_{\eps_1 \to 0, \ldots, \eps_n \to 0}
\eps_1^{-2} \ldots \eps_n^{-2}\;
p(x_1, \eps_1, \ldots, x_n, \eps_n ).
\end{equation}
An indirect way to justify the existence of the limit in (\ref{fw1}) is the following.
Note that when $\alpha=1$, then the description of
$\gamma$ as the right-boundary of a Brownian excursion (see 
\cite {LSWr}) yields 
the following explicit expression for $B_n$:
$$
B_n^{(1)}
(x_1, \ldots, x_n )
= \sum_{s \in {\mathfrak S}_n}
\prod_{j=1}^{n-1}
(x_{s(j)} - x_{s(j-1)})^2
,$$
where $\mathfrak{S}_n$ denotes the group of permutations of $\{1, \ldots, n\}$
and by convention $x_{s(0)} = 0$.

This is due to the fact  that $\gamma$ intersects all slits if and only if the Brownian excursion itself intersects all these slits. One then decomposes this event according to the order with which the excursion actually hits them, and one uses its strong Markov property.

Similarly, an analogous reasoning using the Brownian motion reflected on the negative half-axis, and conditioned not to hit the positive half-axis (and its strong Markov property), yields the existence of the limit in (\ref{fw1}) for all $\alpha<1$. 

Also, since the right-boundary of the union $K_1\cup...\cup K_N$ of $N$ independent sets satisfying the restriction property with exponents $\alpha_1,...,\alpha_N$ satisfies the (one-sided) restriction property with exponent $\alpha_1+\cdots+\alpha_N$, we get the existence of the limit in (\ref{fw1}) for all $\alpha$ (using the existence when $\alpha_1,...,\alpha_N\leq1$, and the following property of the functions $B$: 
For all $R: \{1, \ldots, n \} \to \{1, \ldots, N\}$,
write
$r(j) = \hbox {card} ( R^{-1} \{ j\})$. Then,
\begin {equation}
\label {sg}
B_n^{(\alpha_1  + \cdots +\alpha_N)} (x_1, \ldots, x_n)
= \sum_R \prod_{j=1}^N B_{r(j)}^{(\alpha_j)} ( x_{R^{-1} (j)})
,\end {equation}
where $B_0 = 1$ and $x_I$ denotes the vector with coordinates
$x_k$ for $k \in I$.
This yields a simple explicit formula for $B^{(n)}$ when $n$ is a 
positive integer.

In the general case, one way to compute $B_n^{(\alpha)}$
is to use the following inductive relation (together with $B^{(\alpha)}_0
\equiv1$):

\begin {prop}[Boundary Ward Identities]
For all $x, x_1, \ldots, x_n\in\R_+^{*}$ and  $n\in\N$
\begin {eqnarray}
\nonumber
\lefteqn{ B_{n+1}^{(\alpha)} ( x, x_1, x_2, \ldots, x_n )
\ = \  \frac {\alpha}{x^2} B_n^{(\alpha)} (x_1, \ldots , x_n)} \\
&&
-\sum_{j=1}^n \left\{ (\frac {1}{x_j-x} + \frac 1x ) \partial_{x_j}
  - \frac {2}{(x_j-x)^2} \right\}
 B_n^{(\alpha)} (x_1, \ldots , x_n).
\label {ward1}
\end {eqnarray}
\end {prop}
Those who are familiar with CFT will recognise the similarity of the above relation (\ref{ward1}) with the usual Ward identities. Indeed, as we shall see later, the above corresponds precisely to the insertion of the stress tensor at positions $x_i$ in the central charge zero case. 
\medbreak
\noindent
{\bf Proof.}
Suppose now that the real numbers $x_1, \ldots, x_n$ are fixed and
let us focus on the event 
$E:=E_{\eps_1}(x_1)\cap...\cap E_{\eps_n}( x_n)$.
Let us also choose another point $x \in \R$ and a small $\delta$.
Now, either the curve $\gamma$
avoids $[x , x+ i \delta \sqrt {2}]$ or it does hit it.
This additional  slit is hit (as well as the $n$ other ones)
with a probability $A$ comparable
to
$$\eps^{2n} \delta^2 B_{n+1} (x_1, \ldots, x_n, x)$$
when both $\delta$ and $\eps$ vanish.
On the other hand, 
the image 
of $\gamma$ conditioned to avoid $[x, x+i \delta \sqrt {2}]$ 
under the map
$$\varphi 
(z) := \Phi_{\overline{\H} \setminus [x, x+ i \delta \sqrt {2}]}
= \sqrt { (z-x)^2 + 2 \delta^2 } - \sqrt { x^2 + 2 \delta^2}
$$
 has the same law as $\gamma$.
In particular, we get  that 
\begin {eqnarray}
\label{SLE-Tensor}
\nonumber
A' &:= &
\Prob [  E
\mid \gamma \cap [x, x + i \delta \sqrt {2} ] = \emptyset ]
 \\ 
&\sim& 
\eps^{2n} \prod_{j=1}^n |\varphi'(x_j)|^2\;
B ( \varphi(x_1), \ldots, \varphi (x_n) )
\end {eqnarray}
when $\eps \to 0$ (this square for the derivative can be interpreted as the fact that the boundary exponent for the restriction measures is always 2). But for small $\delta$, 
$$
\varphi (z) = z +  {\delta^2} \left(
\frac 1 {z-x} + \frac 1 x \right) + o (\delta^2)
$$
and
$$
\varphi'(z) = 1 - \frac {\delta^2}{(z-x)^2} + o (\delta^2).
$$
On the other hand,
\begin {equation}
\label {e.exp}
\Prob [ E ] 
= 
A + A'\cdot \Prob [ \gamma \cap [x, x+i \delta \sqrt {2} ] = \emptyset ]
\end {equation}
is independent of $\delta$
and
$$
\Prob [ \gamma \cap [x, x+ i \delta \sqrt {2} ] = \emptyset ]
= \varphi'(0)^{\alpha}
= 1 - \frac {\alpha \delta^2}{x^2} + o (\delta^2)
$$
when $\delta \to 0$.
Looking at the $\delta^2$ term in the $\delta$-expansion of
(\ref {e.exp}), we get (\ref {ward1}). 
\qed

\subsection{Highest-weight representations}
\label{HWR}

We start by recalling  some  basic facts from the theory of highest weight representations of the Virasoro algebra. This particular infinite Lie algebra plays an important role in string theory and as we will see, in SLE. (cf. e.g.~\cite{DiFrancesco:nk, FbZ, KR})

The Witt algebra is usually defined as the Lie-algebra of complex (polynomial) vector fields on the punctured unit disc $\D^{\times}$.  These can be seen as derivations.
\begin{displaymath}
\witt\subset\der_{\C}\,\C((t))=\C((t))\partial_t.
\end{displaymath}
The elements
\begin{displaymath}
\ell_n:=-t^{n+1}\partial_t,\;n\in\Z
\end{displaymath}
yield a $\C$-basis of $\witt$, i.e.
\begin{displaymath}
\witt=\bigoplus_{n\in\Z}\C\,\ell_n
\end{displaymath}
and the Lie bracket is given by 
\begin{displaymath}
[\ell_n,\ell_m]=(n-m)\ell_{n+m}
\end{displaymath}
as a calculation shows.

A Lie algebra ${\mathfrak{a}}$ is called {\bf abelian} if the Lie bracket of ${\mathfrak{a}}$ is trivial, i.e. $[X,Y]=0$ for all $X,Y\in{\mathfrak{a}}$.
\begin{df}
Let ${\mathfrak{a}}$ be an abelian $\C$-Lie-algebra and ${\mathfrak{g}}$ a (possibly $\infty$-dimensional) Lie-algebra over $\C$. An  exact sequence of Lie-algebra homomorphisms 
\begin{displaymath}
0\rightarrow{\mathfrak{a}}\rightarrow{\mathfrak{h}}\rightarrow{\mathfrak{g}}\rightarrow0
\end{displaymath}
is called a {\bf central extension} of ${\mathfrak{g}}$ by ${\mathfrak{a}}$, if $[{\mathfrak{a}},{\mathfrak{h}}]=0$, i.e. $[X,Y]=0\;\;\forall X\in{\mathfrak{a}}$ and $Y\in{\mathfrak{h}}$, where we have identified ${\mathfrak{a}}$ with the corresponding sub-algebra of ${\mathfrak{h}}$.
\end{df}

\begin{df} The {\bf Virasoro algebra} $\vir$ is the one-dimensional universal central extension of $\C((t))\partial_t$ by $\C$. It is a complex vector space with bases $L_n\,\, n\in\Z$ and ${\bf c}$, i.e.
\begin{displaymath}
\vir=\C((t))\partial_t\oplus\,\C\,{\bf c}
\end{displaymath} 
where the multiplication is  given $\forall n,m\in\Z$ by
\begin{eqnarray}
{[L_n, {\bf c}]} & =&  0 \\  \label{Vir_com}
{[L_n, L_m]} & = & (n-m)\cdot L_{n+m}+\delta_{n,-m}
\cdot\frac{n^3-n}{12}\, {\bf c}
\end{eqnarray}
\end{df}
We are mainly interested in such representations $\rho$ of $\vir$ on a complex vector space $V$, where ${\bf c}$ acts as a complex multiple of the identity, i.e. $\rho({\bf c})= c\cdot\id_V$. Therefore $c$ is called the {\bf central charge}.
A priori  there is no preferred representation of the Virasoro algebra. But one class does specially well serve physical purposes, the so called (unitary) highest weight representations. These are common in the theory of angular momentum.

Let $V$ be a vector space over $\C$.
\begin{df} A representation  $\rho:\vir\rightarrow\End_{\C}(V)$ (i.e. a $\vir$ module) is a {\bf highest-weight representation} if there is a pair $(h,c)\in\C^2$  and a vector $v\in V$ called {\bf highest weight vector} such that:
\begin{enumerate}
  \item $\rho(L_n)v=0\qquad\forall n>0,\,n\in\Z$,
  \item $\rho(L_0)=h\cdot v$\quad and\quad $\rho({\bf c})v=c\cdot v$,
  \item the vector $v$ is cyclic, i.e. the set $\{\rho(L)v: L\in{\cal U}(\vir)\}$ spans $V$.
\end{enumerate}
\end{df}

The notation used in physics is $|h\rangle$ instead of $v$ and $\hat{L}_n|h\rangle$ instead of $\rho(L_n)v$. We will switch freely between the different notations and even mix it without any further notice. 

A {\bf Verma module} ${\cal V}(c,h)$ is a special form of a highest weight representation.  The additional requirement is that the so-called ``{\bf descendant states}" of the  highest weight vector, i.e.
\begin{equation}
\label{descendants}
\hat{L}_{-k_1}\hat{L}_{-k_2}...\hat{L}_{-k_n}|h\rangle\quad\quad1\leq k_1\leq...\leq k_n,\, k_i\in\N
\end{equation}
and $|h\rangle$ itself form a vector space basis of ${\cal V}(c,h)$. We note that a descendant state as defined in (\ref{descendants}) is an eigenstate of the operator $\hat{L}_0$ with eigenvalue 
\begin{displaymath}
h'=h+k_1+k_2+...+k_n=h+N,
\end{displaymath}
where $N$ is called the {\cal level} of the state. So a state at level 2 is spanned by the operators $\hat{L}^2_{-1}$ or $\hat{L}_{-2}$, acting on $|h\rangle$. We will need the following existence statement
\begin{lem}For every pair $(h,c)\in\C^2$ there is a Verma module ${\cal V}(c,h)$.
\end{lem}

\begin{df} We call a vector 
 $|w\rangle$  in a HW representation {\bf singular}, if it is different from the highest weight vector  and if $\hat{L}_n |w\rangle=0$ for all $n>0,\,\,n\in\N$.
\end{df}
Therefore it generates its own representation. Consider  in a Verma module ${\cal V}(c,h)$ the following state at level 2:
\begin{displaymath}
|\chi\rangle:=(\frac{\kappa}{2}\hat{L}^2_{-1}-2\hat{L}_{-2})|h\rangle.
\end{displaymath}
The goal is to tune $\kappa$ and $h$ in such a way that $|\chi\rangle$ is a {\bf null state} (or singular vector). It follows from the commutation relations (\ref{Vir_com}) of the Virasoro algebra that the conditions $\hat{L}_1|\chi\rangle=\hat{L}_2|\chi\rangle=0$ are sufficient, since then $\hat{L}_n|\chi\rangle=0$ for all $n\geq 3$. A calculation shows
\begin{eqnarray*}
\hat{L}_1|\chi\rangle & = & (\kappa+2\kappa h-6)\hat{L}_{-1}|h\rangle,\\
\hat{L}_2 |\chi\rangle & = &  (3\kappa h-8h-c)|h\rangle.
\end{eqnarray*}
Therefore the conditions on $c$ and $h$ for $|\chi\rangle$ to be singular, as a function of $\kappa$, are 
\begin{eqnarray}
\label{ckappa}
c & = &\frac{(3\kappa-8)(6-\kappa)}{2\kappa} , \\
 h & = &\frac{6-\kappa}{2\kappa}.
\label{hkappa}
\end{eqnarray}
We notice that the above relations are the same as (\ref{magic_c1}), (\ref{magic_c2}) for $\lambda$ and $\alpha$.

After this reminder of representation theory we proceed now further. So let us  define, for all $N \in \Z$,
the operators
$$
{\cal L}_N := \sum_j
\{ - x_j^{1+N} \partial_{x_j} - 2 (N+1) x_j^N \}
$$
acting on functions of the
real variables $x_1, x_2 , \ldots $.
In fact, one should 
in principle (but we will omit this) make precise the range of $j$ i.e.
define ${\cal L}_N$ on the product over $n$ of the spaces $V_n$ of functions
of $n$ variables $x_1, \ldots, x_n$.

Note that these operators satisfy the commutation
relation
$$[ {\cal L}_N, {\cal L}_M ]
= (N-M) {\cal L}_{N+M}
$$
just as the operators $L_N$ do.
In other words, the vector space generated by these operators is
(isomorphic to) the Witt algebra, i.e. 
the Lie algebra of polynomial vector fields on the unit circle.

Note also that one can rewrite the Ward identity in terms of these
operators as:
\begin {equation}
\label {ward2}
B_{n+1}^{(\alpha)} (x, x_1, \ldots, x_n)
= \frac {\alpha}{x^2} B_n^{(\alpha)} (x_1, \ldots, x_n)
+ \sum_{N \ge 1} x^{N-2} {\cal L}_{-N} B_n^{(\alpha)}  (x_1, \ldots, x_n).
\end {equation}
We are now going to consider vectors $w = (w_0, w_1, w_2, \ldots)$
such that the $n$-th component $w_n$ is in fact a function of
$n$ variables $x_1, \ldots, x_n$.
An example of such a vector is 
\begin{displaymath}
B=B^{(\alpha)}=
(B_0, B_1, B_2, \ldots )
\end{displaymath}
where we put $B^{(\alpha)}_0 \equiv 1$ (we will now fix $\alpha$ and not always 
write the $(\alpha)$ supersript).

For such a vector $w$, we define for all $N \in \Z$ the
operator $l_N$
in such a way that
$$
w_{n+1}
(x, x_1, \ldots, x_n)
= \sum_{N \in \Z} x^{N-2} ( l_{-N}(w))_n (x_1, \ldots, x_n)
.$$
In other words, the $n$-variable component $(l_N (w))_n$ of $l_N(w)$
is the $x^{-N-2}$ term in the Laurent expansion of $w_{n+1} (x, x_1, \ldots, x_n)$
with respect to $x$.

For example, the Ward identity (\ref {ward2}) gives the values of $l_N(B)$
\begin {equation}
\label {hwv}
l_N (B) = 
\left\{ 
 \begin {array}{l@{\hbox {\quad if }}l}
 (0, 0, \ldots ) & \quad N>0 \\
(\alpha B_0, \alpha B_1, \ldots )  &\quad N=0 \\
({\cal L}_N B_0, {\cal L}_N B_1, \ldots)  &\quad N<0
\end {array}
\right.
\end {equation}

More generally:
\begin {lem}
\label {llemma}
For all $k \ge 1$ and  negative $N_1, \ldots, N_k$,
\begin {equation}
\label {elemma}
 ( l_{N_1} \cdots l_{N_k} B )_n
= {\cal L}_{N_1} \ldots {\cal L}_{N_k} B_n.
\end {equation}
\end {lem}
Note that this only holds when the $N$'s are negative. For instance,
$$
{\cal L}_0 ( B_1) = 0 \not= \alpha B_1 = (l_0 B)_1.
$$
\medbreak
\noindent
{\bf Proof of the Lemma.}
This is a rather straightforward consequence of (\ref {ward2}).
We have just seen that it holds for $k=1$.
Assume that (\ref {elemma}) holds for some integer $k \ge 1$.
Then, 
\begin {eqnarray*}
\lefteqn{
({\cal L}_{N_2} \cdots {\cal L}_{N_k} B)_{n+1}
(x, x_1, \ldots, x_n)
}\\
&=&
u + \sum_{N \le -1} x^{-N-2} {\cal L}_{N} {\cal L}_{N_2} \ldots 
{\cal L}_{N_k} B_n (x_1, \ldots, x_n)
\end {eqnarray*}
where $u$ is a Laurent series in $x$
such that $u(x, x_1, \ldots, x_n) = O (x^{-2})$
when $x \to \infty$.
We then apply ${\cal L}_{N_1}$ (viewed as acting on the space of functions of the $n+1$ variables $x, x_1,...,x_n$) to this equation, where $N_1<0$.
There are two $x^{-N-2}$ terms in the expansion of the right-hand side:
The first one is simply
$$
x^{-N-2}
{\cal L}_{N_1} {\cal L}_{N} {\cal L}_{N_2} \ldots 
{\cal L}_{N_k} B_n (x_1, \ldots, x_n)
.$$
The second one comes from the term 
\begin {eqnarray*}
\lefteqn{
({\cal L}_{N_1} x^{-N-N_1-2}) {\cal L}_{N+N_1} {\cal L}_{N_2} 
\ldots {\cal L}_{N_k} B_n (x_1, \ldots, x_n)
} \\
&=&
 (N-N_1) x^{-N-2} {\cal L}_{N+N_1} {\cal L}_{N_2} 
\ldots {\cal L}_{N_k} B_n (x_1, \ldots, x_n)
.
\end {eqnarray*}
The sum of these two contributions is indeed
$$
x^{-N-2}
{\cal L}_N {\cal L}_{N_1} \ldots {\cal L}_{N_k} B_n (x_1, \ldots, x_n)$$
because of the commutation relation of the ${\cal L}$'s.
This proves (\ref {elemma}) for $k+1$.
\qed

\medbreak
We now define, the vector space $V$
generated by the vector $B$ and all vectors
$l_{N_1} \ldots l_{N_k} B$
for negative
$N_1, \ldots, N_k$ and positive $k$
(we will refer to  these vectors 
as the generating vectors of $V$).
Then:
\begin {prop} For all $v \in V$, for all $M,R  \in \Z$,
$$
l_M (v) \in V\quad\mbox{and}\quad[ l_M , l_R ] v = (M-R) l_{M+R}\; v
.$$
\end {prop}

We insist on the fact that $l_N$ only coincides with ${\cal L}_N$
for negative $N$,
and that the commutation relation for the $l_N$'s
does not hold for a general vector. But, the above statement
shows that it is valid on this special vector space $V$.

\medbreak
\noindent
{\bf Proof.}
Note that the commutation relation holds
for negative $R$ and $M$'s because
of Lemma \ref {llemma}.

Suppose now that 
$N_1, \ldots, N_k$ are negative.
Then,
\begin {eqnarray*}
{\cal L}_{N_1} \ldots {\cal L}_{N_k}
B_{n+1}
&=&
\sum_{N \le 0, I}
{\cal L}_{N_{i_1}} \ldots {\cal L}_{N_{i_r}} (x^{-2-N}) \\
&& \hskip 1cm
\times
{\cal L}_{N_{j_1}} \ldots {\cal L}_{N_{j_s}} (l_N B)_n (x_1, \ldots, x_n)
\end {eqnarray*}
where the sum is over all $I:=\{ i_1 , \ldots, i_r \}
\subset \{1, \ldots, k \}$. One then writes
$\{ j_1, \ldots j_s\} = \{1, \ldots, k \} \setminus 
\{ i_1, \ldots, i_r \}$ (and the $i$'s and $j$'s are increasing).
We use $l_N (B)_n$ instead of ${\cal L}_N B_n$
 to simplify the expression (otherwise
the case $N=0$ would have to be treated separately).

Since 
\begin {eqnarray*}
\lefteqn{
{\cal L}_{N_{i_1}} \ldots {\cal L}_{N_{i_k}}
(x^{-2-N})
}\\
&=& 
(N - 2 N_{i_r}) (N - N_{i_r} -
2N_{i_{r-1}}) \ldots \\
&& \ldots (N - N_{i_r} - \ldots 
-N_{i_2} - 2 N_{i_1}) x^{-2-N+N_{i_1} + \cdots +N_{i_k}},
\end {eqnarray*}
it follows immediately  that for all integer $M$,
\begin {eqnarray}
\nonumber
\lefteqn {(l_M l_{N_1} \ldots l_{N_k} B)_n
}\\
&=&
\sum_{I : \ M+N_{i_1} + \cdots +N_{i_r} \le 0}
(M+N_{i_1} + \ldots + N_{i_{r-1}} - N_{i_r}) \ldots (M - N_{i_1}) 
\nonumber \\
&& \hskip 1cm
\times 
 {\cal L}_{N_{j_1}} \ldots {\cal L}_{N_{j_s}} (l_{M+N_{i_1} + \ldots + N_{i_r}}
  B)_n
\label {compl}
\end {eqnarray}
This implies that indeed, $l_M (V) \subset V$.
When $M \le 0$, then for any $i_1, \ldots, {i_r}$, $M +N_{i_1}
+ \ldots + N_{i_r} \le 0$, so that 
the sum is over all $I$.

Suppose now that $M \ge 0$, $R < 0$, and consider 
$v = l_{N_1} \ldots l_{N_k}$ for some fixed negative
$N_1, \ldots, N_k$.
We can apply (\ref {compl}) to get the expression of 
$l_{R+M} v$, of $l_M l_R v$ and of $l_M v$.
Furthermore, we can use the Lemma to deduce the following
expression for $l_R l_M v$:
\begin {eqnarray*}
\lefteqn {(l_R l_M v)_n}\\
&=&
\sum_{I : \ M+N_{i_1} + \cdots +N_{i_r} \le 0}
(M+N_{i_1} + \ldots + N_{i_{r-1}} - N_{i_r}) \ldots (M - N_{i_1}) \\
&& \hskip 1cm
\times 
 {\cal L}_R{\cal L}_{N_{j_1}} \ldots {\cal L}_{N_{j_s}} (l_{M+N_{i_1} + \ldots + N_{i_r}}
  B)_n
  \end {eqnarray*}
On the other hand,
\begin {eqnarray*}
\lefteqn{
(l_M l_R v)_n
}
\\
&=&
\sum_{I_0 : \ M+N_{i_0} + \cdots +N_{i_r} \le 0}
(M+N_{i_0} + \ldots + N_{i_{r-1}} - N_{i_r}) \ldots (M - N_{i_0}) \\
&& \hskip 1cm
\times 
 {\cal L}_{N_{j_1}} \ldots {\cal L}_{N_{j_s}} (l_{M+N_{i_0} + \ldots + N_{i_r}}
  B)_n,
  \end {eqnarray*}
where this time, the sum is over 
$\{ i_0, \ldots, i_r \} \subset \{ 0, \ldots, k \}$, and
we put $R= N_0$.
The difference between these two expressions is due to the 
terms (in the latter) where $i_0 =0$:
\begin {eqnarray*}
\lefteqn { [ l_M, l_R] v } \\
&=&
(M-R) 
\sum_{I : \ M+N_{i_1} + \cdots +N_{i_r} \le 0}
(M+R+N_{i_1} + \ldots + N_{i_{r-1}} - N_{i_r}) \ldots\\
&& \ldots  (M+R - N_{i_1}) 
 {\cal L}_{N_{j_1}} \ldots {\cal L}_{N_{j_s}} (l_{M+R+N_{i_1} + \ldots + N_{i_r}}
  B)_n\\
  &=& 
  (M-R) l_{M+R}.
\end {eqnarray*}
This proves the commutation relation for negative $R$
and arbitrary $M$.

Finally, to prove the commutation relation when both $R$ and 
$M$ are negative and
$v = l_{N_1} \ldots l_{N_k}$ 
as before, it suffices to use the previously proved
commutation relations to write 
$l_M v$, $l_R v$ and $l_{M+R} v$
as linear combination of the generating vectors of $V$.
Then, one can iterate this procedure to 
express $[l_M, l_R] v$ as a linear combination of 
the generating vectors of $V$. Since this formal algebraic calculation
is identical to that one would do in the Witt algebra, 
one gets that indeed 
$[l_M, l_R] v = (M-R) l_{M+R}$, 
which therefore also holds for any $v \in V$. 
\qed

\medbreak

This shows that to each
(one-sided) restriction measure, one can simply
associate a highest-weight representation of the Witt algebra acting on a
certain space of function-valued vectors.
The value of the highest weight is the exponent 
of the restriction measure.

Note (cf. \cite{LSWr}) that the right-sided boundary of a simply connected
set $K$ satisfying the two-sided  restriction
property satisfies the one-sided restriction property
(so that one can also
associate a representation to it).
In this case,
the function $B_n$ also represents the limiting value
of
$$
\eps^{-2n}\, 
{\bf P} [ K \hbox { intersects all slits } [x_j, x_j + 2 i \eps \sqrt {2}] ,\,
j =1 , \ldots , n ]
$$
even for negative values of some $x_j$'s.

\subsection{Evolution and degeneracy for  SLE$_{8/3}$}

We are now going to see how to combine the previous 
considerations with a Markovian property. For instance, 
does there exist a value of $\kappa$ such that SLE$_\kappa$
satisfies the restriction property? We know from \cite {LSWr}
that the answer is yes, that the value of $\kappa$ is $8/3$ 
and that the corresponding exponent is $5/8$. 
This  
``boundary exponent'' for SLE$_{8/3}$ has appeared before in the 
theoretical physics literature (see  \cite {Ca1})
as the boundary exponent for 
long self-avoiding walks (which is consistent with the conjecture 
\cite {LSWsaw} that this SLE is the scaling limit of the 
half-plane self-avoiding walk). 
This exponent was identified as the only
possible highest-weight of a highest-weight representation 
of the Witt algebra that is {\em degenerate} at level two.

We are now going to see that indeed, the Markovian property of SLE
is just a way of saying that the two vectors
$l_{-2} (B)$ and $l_{-1}^2 (B)$
 are not 
independent. This shows (without using the computations in \cite {LSWr})
why the values $\kappa=8/3$, $\alpha=5/8$ pop out.

Let $\gamma$ be an $\mbox{SLE}_{\kappa}$.
Consider the event 
$E:=
E_{\eps_1}(x_1)\cap\ldots\cap E_{\eps_n}(x_n)$ 
as in the definition of $B_n^{(\alpha)}$.
If one considers the 
conditional probability of $E$ given $\gamma$ up to time $t$, 
then it is the probability that an (independent) SLE 
$\tilde \gamma$ hits the (curved) slits 
$f_t ([x_j, x_j + i \eps_j \sqrt {2}])$.
At first order, this is equivalent to hitting 
the straight slits
$$[f_t(x_j), f_t(x_j) + i \eps_j \sqrt {2} f_t'(x_j)].$$

If the SLE satisfies the restriction property
with exponent $\alpha$, then this means that 
$$
f_t'(x_1)^{-2} \ldots f_t'(x_n)^{-2}\; B_n^{(\alpha)} 
( f_t (x_1), \ldots, f_t (x_n) )
$$
is a local martingale. 
Recall that
$$
\partial_t f_t (x) =- \sqrt {\kappa} dB_t + \frac {2 }{f_t (x)}\;\hbox { and }\;
\partial_t f_t' (x) = \frac {-2 f_t'(x)}{f_t(x)^2},\, \mbox{where}\,\, B_t\,\,\mbox{is Brownian motion}.
$$
Hence, since the drift term of the previous local martingale
vanishes,
It\^o's formula yields
$$
\frac {\kappa}{2} {\cal L}_{-1}^2 B_n - 2 {\cal L}_{-2} B_n 
= 0 
$$
for all $n \ge 1$.
Note that the operators are 
 ${\cal L}$'s and not $L$'s (as in the 
 crossing probabilities formulas)
 because of the local scaling properties 
 of the functions $B^{(\alpha)}$.
 
 In other words, $l_{-2}(B)$ and $l_{-1}^2(B)$
 are collinear and the previously described highest-weight representation of 
the polynomial vector fields on the unit circle  must be degenerate at level two.
One can now deduce the values of $\alpha$ and $\kappa$,
using the fact that
 $$
 l_2 (\frac {\kappa}{2} l_{-1}^2 - 2 l_{-2})B
 = (3 \kappa - 8) l_0 B = 0
 $$
 which implies that $\kappa=8/3$
 and
 $$
 l_1 (\frac {\kappa}{2} l_{-1}^2 - 2 l_{-2} ) B
 = \frac {\kappa}{2} (4 l_{-1} l_0 B + 2l_{-1} B)
 -6 l_{-1} B =
 (2 \kappa \alpha + \kappa - 6 ) l_{-1} B =0
 $$
 which then implies that $\alpha = 5/8$.
\subsection{The cloud of bubbles}
In this section we are going to introduce another important construction related to SLE. At the same time it seems  to make the bridge to  the Coulomb gas formalism. 
 
In general, it would be interesting to pursue more systematically this directions.  

We are now going to use the description of the ``restriction paths'' $\beta$ via SLE curves to which one adds a Poissonian cloud of Brownian bubbles, as explained in \cite{LSWr}. Let us briefly recall how it goes. Consider an ${\rm SLE}_{\kappa}$ for $\kappa < 8/3$. As we have just seen, it does not satisfy the restriction property. However, if one adds to this curve an appropriate random cloud of Brownian loops, then the obtained set satisfies the two-sided restriction property for a certain exponent $h > 5/8$ (and its right-boundary $\beta$ satisfies the one-sided restriction property). More details and properties of the Brownian loop-soup and the procedure of adding loops can be found in \cite{LSWr, LW3}.

Intuitively this phenomenon can be understood from the case, where $\kappa = 2 : {\rm SLE}_2$ is the scaling limit of the loop-erased random walk excursion (see \cite{LSWlesl}). Adding Brownian loops to it, one should (in principle) recover the Brownian excursion that satisfies the restriction property with parameter $h=1$.

More generally, let $\kappa < 8/3$ be fixed, and consider an ${\rm SLE}_{\kappa}$ curve $\gamma$, with its usual time-parametrisation. There exists a natural (infinite) measure on Brownian bubbles in ${\mathbb H}$ rooted at the origin. This is a measure supported on Brownian paths of finite length in ${\mathbb H}$ that start and end at the origin (more generally, we say that a bubble in $\H$ rooted at $x \in \partial \H$ is a path $\eta$ of finite length $T$ such that $\eta (0,T) \in \H$ and $\eta (0) = \eta (T) = x$). Consider a Poisson point process of these Brownian bubbles in ${\mathbb H}$, with intensity $\lambda$ (more precisely, $\lambda$ times the measure on Brownian bubbles). A realisation of this point process is a family ($\hat\eta_t$, $t \geq 0$) such that for all but a random countable set $\{ t_j \}$ of times, $\hat\eta_t = \emptyset$ and for the times $t_j$, $\hat\eta_{t_j}$ is a (Brownian) bubble in ${\mathbb H}$ rooted at the origin. We then define for all $t$, $\eta_t = f_t^{-1} (\hat\eta_t)$, so that $\eta_t$ is empty if $t \notin \{ t_j \}$ and is a bubble in ${\mathbb H} \backslash \gamma [0 , t_j]$ rooted at $\gamma (t_j)$ if $t = t_j$. Another equivalent way to define this random family ($\eta_t$, $t \geq 0$) via a certain Brownian loop-soup is described in \cite{LW3}.

Define the union $\Gamma$ of $\gamma$ and the bubbles $\eta_t$, i.e.
$$
\Gamma = \bigcup_{t \geq 0}\,\left (\{ \gamma_t \} \cup \eta_t\right) \, .
$$
We let ${\mathcal F}_t$ denote the $\sigma$-field generated by ($\gamma_s$, $\eta_s$, $s \leq t$).

The right outer-boundary $\beta$ (see \cite{LSWr, LW3}) of $\Gamma$ then satisfies the restriction property (actually $\Gamma$ satisfies the two-sided restriction property). This is proved in \cite{LSWr} studying the conditional probabilities that $\Gamma$ avoids a given set $A$ with respect to the filtration generated by $\gamma$ alone. As observed in \cite{LSWr}, the relation between the density $\lambda (\kappa)$ of the loops that one has to add to the ${\rm SLE}_{\kappa}$ and the exponent $h(\kappa)$ of the corresponding restriction measure (i.e. $h = (6 - \kappa) / 2\kappa$ and $\lambda = (8 - 3\kappa) h$) recalls the relation between the central charge and the highest-weight representations of the Virasoro algebra. We shall try in this subsection to give one way to explain the relation to representations, via the functions $B_n^{(h)}$, and therefore recover these values of $h$ and $\lambda$, just assuming that if one adds the cloud of bubbles with intensity some $\lambda$, one obtains a restriction measure.

It is worthwhile emphasising that in this context, the functions $B_n^{(h)}$ are only indirectly related to the SLE curve via this Poissonian cloud of loops. They do for instance not represent the probabilities that the SLE itself does visit the infinitesimal slits, but the probability that some loops that have been attached to this SLE curve do visits the infinitesimal slits.

Recall that the functions $B_n^{(h)}$ are related to a highest-weight representation of the Witt algebra, as discussed in the previous section. As in the $\kappa = 8/3$ case, we will try to obtain an additional information on this representation, using the evolution of the SLE curve. More precisely: How does the (conditional) probability with respect to ${\mathcal F}_t$ of the event $E$ that $\beta$ intersects the $n$ slits $[x_j , x_j + i \varepsilon_j \sqrt 2]$ for infinitesimal $\varepsilon_j$'s evolve with time? Here is a heuristic discussion, that can easily be made rigorous:

Consider an infinitesimal time $\Delta$. Let $\tilde\Gamma_{\Delta}$ denote the union of $\gamma [\Delta , \infty)$ and the loops that it does intersect. More precisely,
$$
\tilde\Gamma_{\Delta} = \bigcup_{t > \Delta}\, (\{ \gamma_t \} \cup \eta_t) \, .
$$
Typically (for every small $\Delta$), there is no bubble $\eta_t$ for $t \in [0,\Delta]$ that does intersect one of these $n$ slits. In this case, the conditional probability of the event $E$ given ${\mathcal F}_{\Delta}$ is simply the probability that $\tilde\Gamma_{\Delta}$ does intersect these $n$ slits (given ${\mathcal F}_{\Delta}$). The definition of $\gamma$ and of the bubbles show that the conditional law of $f_{\Delta} (\tilde\Gamma_{\Delta})$ given ${\mathcal F}_{\Delta}$ is independent of $\Delta$ (in particular, it is the same as for $\Delta = 0$ i.e. the law of $\Gamma$). This shows that (exactly as in the $\kappa = 8/3$ case), the conditional probability of $E$ has a drift term due to the distorsion of space induced by the SLE (i.e. by $f_{\Delta}$) of the type
$$
\left( \frac{\kappa}{2} \, {\mathcal L}_{-1}^2 B_n - 2 {\mathcal L}_{-2} B_n \right) \Delta \, .
$$

But there is an additional term due to the fact that one might in the small time-interval $[0,\Delta]$, have added a Brownian loop $\eta_t$ to the curve that precisely goes through one or several of the $n$ slits $[x_j , x_j + i \varepsilon_j \sqrt 2]$. The probability that one has added a loop that goes through the $j$-th slit is of order $\lambda \varepsilon_j^2 \Delta / x_j^4$. This fact is due to scale-invariance. Here $\lambda$ is the (constant) density of loops that is added on top of the SLE curve (we use this definition for this density $\lambda$ in this paper, as in \cite{LSWr}; in other contexts, replacing $\lambda$ by $\lambda / 6$ can be more natural). One way to understand the $\varepsilon_j^2 / x_j^4$ term is that the Brownian bubble has to go from $0$ to the slit, which contributes a factor $\varepsilon_j^2 / x_j^2$, and then back to the origin, which contributes also $1/x_j^2$. If such a loop has been added, the conditional probability of $E$ is (at first order) the probability that the SLE $+$ loops hits the remaining $n-1$ slits, i.e. $f_{n-1} (x_{\{ 1 , \ldots n \} \backslash \{ j \}}) \prod_{l \ne j} \varepsilon_l^2$ (here and in the sequel $x_J$ stands for $(x_{j_1} , \ldots , x_{j_p})$ when $J = \{ j_1 , \ldots , j_p \}$). More generally, define $T_0 = 0$, $T_1 (x) = 1/x^4$, and for $p \geq 2$,
$$
T_p (x_1 , \ldots , x_p) = \sum_{s \in \sigma_p} \frac{1}{x_{s(1)}^2 (x_{s(2)} - x_{s(1)})^2 \ldots (x_{s(p)} - x_{s(p-1)})^2 x_{s(p)}^2} \, .
$$
Each $s$ corresponds intuitively to an order of visits of the infinitesimal slits by the loop. For $J = \{ j_1 , \ldots , j_p \} \subset \{ 1 , \ldots , n \}$ with $\vert J \vert = p \geq 1$, the probability to add a loop that goes precisely through the slits near $x_j$ for $j \in J$ is of the order of
$$
\varepsilon_{j_1}^2 \ldots \varepsilon_{j_p}^2 T_p (x_J) \lambda \Delta \, .
$$
We are therefore naturally led to define the operator $U$ by
$$
(Uf)_n (x_1 , \ldots , x_n) = \sum_{J \subset \{ 1 , \ldots , n \}} T_p (x_J) \times f_{n-p} (x_{\{ 1 , \ldots, n \} \backslash J}) \, .
$$
Then, the fact that $P(E \mid {\mathcal F}_t)$ is a martingale, shows that the drift term vanishes i.e. that
\begin{equation}
\label{FW9}
\left\{ \frac{\kappa}{2} \, l_{-1}^2 - 2l_{-2} + \lambda U \right\} B = 0 \, .
\end{equation}

Note that the definitions of $l_N$ and $U$ show easily that for any $w = (w_0 , w_1 , \ldots)$ (not only in $V$),
$$
([l_N , U] w)_n (x_1 , \ldots , x_n) = \sum_{J \subset \{ 1 , \ldots , n \}} (l_N (T))_p (x_J) \times w_{n-p} (x_{\{ 1 , \ldots, n \} \backslash J}) \, .
$$
In order to compute $l_N (T)_p$, one has to look at the Laurent expansion (when $x \to 0$) of $T_{p+1} (x,x_1 , \ldots , x_p)$. Recall that $T_1 (x) = 1/x^4$ and note that for $p \geq 1$,
\begin{equation}
\label{FW10}
T_{p+1} (x,x_1 , \ldots , x_p) = 2x^{-2} T_p (x_1 , \ldots , x_p) + o (x^{-2})
\end{equation}
(the only terms in the sum that contribute to the leading term are these corresponding to $x$ being visited first or last by the loop). It follows that $l_N T = 0$ if $N > 2$ and if $N=1$ (there are no $x^{-N-2}$ terms in the expansion). Also, $l_2 T = (1,0,0, \ldots)$ (the only case where there is an $x^{-4}$ term is $p+1 = 1$). Finally, $l_0 T = 2T$ because of (\ref{FW10}). Hence,

$$
[ l_n, U ] =
\begin{cases}
     0 & \text{if }\;\; N>2, \\
     \id & \text{if}\;\; N=2,\\
     0 & \text{if}\;\; N=1,\\
     2U & \text{if}\;\; N=0.
\end{cases}
$$
This enables as before to relate $\lambda$ to $\kappa$ and $h$:
$$
l_2 (\kappa l_{-1}^2 / 2 - 2l_{-2}) B = l_2 (-\lambda UB) = -\lambda B - \lambda U l_2 B = -\lambda B
$$
and
$$
l_1 (\kappa l_{-1}^2 / 2 - 2l_{-2}) B = l_1 (-\lambda UB) = -\lambda U l_1 B = 0 \, .
$$
This last relation implies that
$$
h = \frac{6-\kappa}{2\kappa}
$$
and the first one then shows that
$$
\lambda = (8- 3\kappa) h = \frac{(8-3 \kappa) (6-\kappa)}{2\kappa} \, ,
$$
which are the formulae appearing in \cite{LSWr}.

This relation between $h$ and $-\lambda$ is indeed that between the highest-weight and the central charge for a representation of the Virasoro algebra that is degenerate at level two. Recall (cf. relation (\ref{Vir_com})) that if $L_n$'s are the generators of the Virasoro Algebra and $\bf c$ its central element, then $[L_2 , L_{-2}] = 4 L_0 + {\bf c}/2$, so the little two by two linear system leading to the determination of $\kappa$ and $h$ for a degenerate highest-weight representation of the Virasoro algebra is the same (and therefore leads to the same expression); roughtly speaking, $l_{-2} - \lambda U / 2$ plays the role of $L_{-2}$.

Note that the previous considerations involving the Brownian bubbles is valid only in the range $\kappa \in (0,8/3]$ and therefore for $c \leq 0$. This corresponds to the fact that two-sided restriction measures exist only for $h \geq 5/8$. In this case all functions $B_n^{(h)}$ are positive for all (real) values of $x_1 , \ldots , x_n$.

\subsection{Analytic continuation}

In the representations that we have just been looking at, we considered simple operators acting on simple rational functions. All the results depend analytically on $\kappa$ (or $h$). In other words, for all real $\kappa$ (even negative!), if one defines the functions $B_n^{(h)}$ recursively, the operators $l_n$, the vector $B^{(h)}$ and the vector space $V = V^{(h)}$ as before, then one obtains a highest-weight representation of the Witt algebra with highest weight $h$. The values of $\kappa$, $\lambda$ and $h$ are still related by the same formulae, but do not correspond necessarily to a quantity that is directly relevant to the SLE curve or the restriction measures.

When $h \in (0,5/8)$, the functions $B_n^{(h)}$ can still be interpreted as renormalised probabilities for one-sided restriction measures. They are therefore positive for all positive $x_1 , \ldots , x_n$ but they can become negative for some negative values of the arguments. The ``SLE $+$ bubbles'' interpretation of the degeneracy (i.e. of the relation (\ref{FW9})) is no longer valid since the ``density of bubbles'' becomes negative (i.e. the corresponding central charge is positive). In this case, the local martingales measuring the effect of boundary perturbations are no longer bounded (and do not correspond to conditional probabilities anymore).

For negative $h$, the functions $B_n^{(h)}$ can still be defined. This time, the functions $B_n^{(h)}$ are not (all) positive, even when restricted on $(0,\infty)^n$ and they do not correspond to any restriction measure. These facts correspond to ``negative probabilities'' that are often implicit in the physics literature.

Note that $c$ (i.e. $-\lambda$) cannot take any value: For positive $\kappa$, $c$ varies in $(-\infty , 1)$ and for negative $\kappa$, it varies in $[25 , \infty)$. The transformation $\kappa \leftrightarrow -\kappa$ corresponds to the well-known $c \leftrightarrow 26-c$ duality (e.g. \cite{Ne}).

In other words, the $B_n^{(h)}$'s provide the highest-weight representations of the Virasoro algebra with highest-weight $h$ and central charge zero. Each one is related to a highest-weight representation of the Virasoro algebra that is degenerate at level 2. Furthermore, all $B_n^{(h)}$'s are related by (\ref{sg}).

\section{Discussion of the results so far obtained}
\label{CFTresults}
In this section we are going to explain, how the results from SLE, as reviewed in the previous part,  can  be explained from the perspective of conformal field theory. 

As we already mentioned, in the scaling limit a critical lattice model can be related to a CFT. But to generate the chordal domain walls, that are supposed to converge to SLE, we need to specify appropriate boundary conditions. 
So by fixing them  we force the phase boundaries to
include in particular a path that connects the points on the
boundary, where the boundary conditions are changed. Since the partition function with free boundary conditions, $Z_f$, provides a measure for the number of states in the physical system, the partition function
$Z_{\alpha\beta}$ with fixed boundary conditions $(\alpha\beta)$
accounts for those configurations with such a chordal path. The fraction
\be
\frac{Z_{\alpha\beta}}{Z_f}
\ee
should be  the fraction of phase boundaries that include the paths forced
by these changes of boundary conditions among all possible phase
boundaries included in the full partition function. It can be
heuristically considered to be the probability that
some path connects these specified boundary points in the
unconstrained theory.

On physical grounds it seems reasonable to expect that the (positive) ratio 
$Z_{\alpha_1\ldots\alpha_n}/Z_f$ is bounded by $1$; this issue should nevertheless be settled by carefully studying candidate CFTs case by case.  

In CFT we cannot restrict configurations to phase boundaries as in
the case of classical statistical mechanics models,  because
configurations in a quantum field theory do not lend themselves to a
classical treatment. The only well-defined restrictions are indeed
boundary conditions.

To see which CFT we are exactly dealing with, we have to determine its central charge. But  the condition on being an SLE-martingale gives the  link between the degenerate highest-weight representations and the value $\kappa$ of the SLE process, as well as with the parameters for $Y_t$ cf.~(\ref{martingaleY}) to be a bounded local martingale. As we recall it is:
\begin{eqnarray}
\label{42}
-\lambda=c & = & \frac{(6-\kappa)(3\kappa-8)}{2\kappa}, \\
\label{43}
\alpha=h & = & \frac{6-\kappa}{2\kappa}.
\end{eqnarray}
\subsection{Boundary correlators of the Stress Tensor }
In the case of $\text{SLE}_{8/3}$ the above formulae (\ref{42},\ref{43}) yield, that the theory has as  central charge $0$, what is in agreement with the physics literature, assuming the $\text{SLE}_{8/3}$ process is the scaling limit of long self-avoiding walks. Further the scaling dimension of the boundary field at $0$ is $5/8$, again as it can be found in the  literature. 

Looking at the expression in (\ref{SLE-Tensor}) suggests that we have a weight two field. But we know that correlators in a domain whose boundary is infinitesimally  deformed are related to those in the unperturbed one, by insertions of the stress tensor, which is an anomalous  field of conformal dimension 2.

\begin{figure}[ht]
\begin{center}
\includegraphics[scale=0.5]{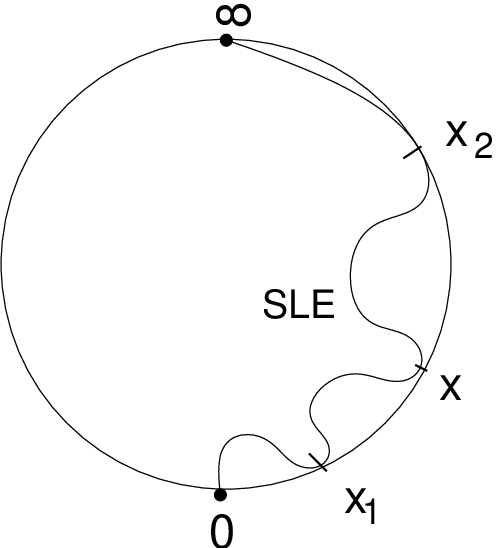}
\caption{The boundary correlator for SLE} 
\label{cd}
\end{center}
\end{figure}
Therefore the general structure of the $\mbox{SLE}_{\kappa}$ correlator we are calculating is 
\begin{equation}
\label{SLE-CFT-correlator}
\langle \Phi(0)\; T(x_1)\; T(x_2)...T(x)...T(x_n)\; \Phi(\infty)\rangle
\end{equation}
where $\Phi$ is a primary boundary field of weight $h_{1,2}=5/8$ in the Verma module 
${\cal V}_{1,2}$, inserted at zero and infinity, i.e.  where the boundary conditions change discontinuously. $T(x_i)$ denotes insertions of the stress tensor at  points $x_i\in\R\setminus\{0\}$.

Then we have
\begin{equation}
\label{ }
B'_0:=\langle\Phi(0)\Phi(\infty)\rangle
\end{equation} for the unperturbed domain.
After normalising  $B'_0$ we can set $B_0\equiv 1$ and continue with the sequence of normalised expressions. Then
\begin{displaymath}
B_1=\langle\Phi(0) T(x)\Phi(\infty)\rangle
\end{displaymath}
and using the OPE (\ref{Stress-OPE}) we get
\begin{displaymath}
\langle\frac{h}{x^2}\Phi(0)+\frac{1}{x}\partial_x\Phi(0),\Phi(\infty)\rangle=\langle\frac{h}{x^2}\Phi(0)\Phi(\infty)\rangle+\underbrace{\langle\frac{1}{x}L_{-1}\Phi(0),\Phi(\infty)\rangle}_{=0}=\frac{h}{x^2} B_0.
\end{displaymath}
Here we have used the fact, that the correlator of a primary field with its descendant is zero. {\bf Descendant fields}  arise in general as the fields associated to descendant states through the state-field correspondence. The descendant field corresponding to $\hat{L}_{-n}|h\rangle$, where $|h\rangle$ is a primary field of weight $h$, is denoted $\Phi^{(-n)}(w)$ resp. for several $\hat{L}_{-n}$'s as $\Phi^{(-n_1,...,-n_k)}(w)$. The natural definition of the descendant field associated with the state $\hat{L}_{-n}|h\rangle$ is 
\begin{equation}
\label{descendant_OPE}
\Phi^{(-n)}(w):=\hat{L}_{-n}|h\rangle(w)=\frac{1}{2\pi}\oint_w dz\,\frac{1}{(z-w)^{n-1}} T(z)\Phi(w).
\end{equation}
In particular $\Phi^{(0)}(w)=h\Phi(w)$ and $\Phi^{(-1)}=\partial_w\Phi(w)$. The physical properties of these fields (i.e. their correlation functions) may be derived from those of the ``ancestor"  primary field.  Therefore let us consider the correlator 
\begin{displaymath}
\langle (L_{-n}\Phi)(w){\mathfrak O}\rangle,
\end{displaymath}
where ${\mathfrak O}:=\Phi_1(w_1)...\Phi_N(w_N)$ is an $N$-tuple of primary fields with conformal dimensions $h_i$. This correlator may be calculated by using the Ward identity for $T$, c.f.~(\ref{Stress-OPE}). This way we get
\begin{eqnarray}
\langle (L_{-n}\Phi)(w){\mathfrak O}\rangle & = & \frac{1}{2\pi i}\oint_w dz\, \frac{1}{(z-w)^{n-1}}\langle T(z)\Phi(w){\mathfrak O}\rangle \\
 & = & - \frac{1}{2\pi i}\oint_{\{w_i\}} dz\,\frac{1}{(z-w)^{n-1}}\sum_{i=1}^N [\frac{1}{z-w_i}\partial_{w_i}\langle \Phi(w){\mathfrak O}\rangle\\
 &  &+\frac{h_i}{(z-w_i)^2}\langle \Phi(w){\mathfrak O}\rangle]\\
 &\equiv&{\LL}_{-n}\langle\Phi(w){\mathfrak O}\rangle\qquad(n\geq1)
\end{eqnarray}
wherein we defined the differential operator 
\begin{equation}
\label{ }
{\LL}_{-n}:=\sum_{i=1}^N\left[\frac{(n-1)h_i}{(w_i-w)^n}-\frac{1}{(w_i-w)^{n-1}}\partial_{w_i}\right].
\end{equation}
Let us note that in the above calculation the integration contour encircles only the pole at $w$ and that  by applying the ``inside-outside" theorem, which assumes that we are on the Riemann sphere,  we can reverse the contour and sum over the contributions from the poles at $w_i$.
We thus reduce the evaluation of a correlator containing a descendant field to that of a correlator of primary fields, on which we must apply a differential operator ${\LL}_{-n}$. In the boundary case ${\LL}_{-1}$ is not trivial and equal to 
\begin{equation}
\label{ }
-\sum_{i=1}^N\frac{\partial}{\partial w_i}.
\end{equation}
Now we can resume again and calculate the  next correlator by just writing it as  a sum over its principal parts. Therefore we bring the new insertion close to the old ones, i.e. close to the diagonal and decompose it according to the singular parts which leads us  to consider the pairs $(0, x)$, $(x_1, x)$ and $(x_2, x)$. 
Now we can evaluate the correlator $B_2$ by making use of the results  about correlators containing descendant fields . Since  for $\mbox{SLE}_{8/3}$ the central charge is zero, the corresponding stress-energy tensor transforms as a weight 2 primary field and therefore  the above discussion applies. Hence
\begin{eqnarray}
B_2=
\langle\Phi(0) T(x_1) T(x) \Phi(\infty)\rangle = \langle(\frac{h}{x^2}\Phi(0)-\frac{1}{x}\partial_w\Phi(0))T(x_1)\Phi(\infty)\rangle\\
  +  \langle\Phi(0)(\frac{c/2}{(x-x_1)^4}+\frac{2T(x_1)}{(x-x_1)^2}+\frac{\partial_{x_1}}{(x-x_1)})\Phi(\infty)\rangle\\
 =\frac{h}{x^2}B_1-\frac{1}{x}\partial_{x_1} B_1+\frac{2}{(x-x_1)^2}B_1+\frac{1}{(x-x_1)}\partial_{x_1}B_1\\
 =\frac{h}{x^2}B_1-\left[(\frac{1}{x}+\frac{1}{x_1-x})\partial_{x_1}-\frac{2}{(x_1-x)^2}\right]B_1
\end{eqnarray}

At infinity the expression is regular, since the points of insertion are finite. We note, that however, on the disk we should take also the second point where the boundary conditions change, into account.  But from our previous considerations we know, that in the case of $\mbox{SLE}_{8/3}$ the corresponding central charge is zero. Therefore the first term in the OPE~(\ref{Stress-Stress-OPE}) will not contribute. 

The general result follows now by induction over $N$.

\subsection{Moduli under L{\oe}wner process and infinitesimal deformations}
\label{moduli}
We continue our discussion with two objectives. The first is to prepare the setting for the more general treatment of SLE on ``arbitrary" surfaces and its connection with moduli spaces, that are the proper objects from the CFT point of view. Second, in the process of developing this perspective, we will see, how the restriction property and in particular the martingale $Y_t$~(\ref{martingaleY}) have a natural explication in this framework.

We note, that the question of how to define
statistical mechanics models on general Riemann surfaces properly, is still an open problem. Although there are several propositions, e.g. via triangulations, we prefer for the present purpose to translate the problem directly to a question in CFT, i.e. to the continuum theory.

A Riemann surface $X$, i.e.~the complex structure on the two-dimensional real manifold $X$, corresponds to a conformal class of metrics, which locally can be expressed in the form 
$\d s^2 = \e^\sigma|\d z|^2$, where $z$ is a local isothermal  coordinate. In general, given a two dimensional manifold with a metric we can deform the metric  by local reparametrisations given by a global vector
field $v \in \Gamma(TX)$ on the surface, local Weyl rescalings given by a global function $\varphi$ on $X$, and Teichm\"uller
reparametrisations given by the Beltrami differentials 
\be
\mu \in
\Omega^{(-1,1)}(X) := T^{(1,0)}X \otimes \Omega^{(0,1)}(X) 
\ee
and $\bar\mu\in \Omega^{(1,-1)}(X)$. Under these transformations a
correlator
\be
\langle \Xope \rangle := \langle \prod_i \Phi_i(z_i) \rangle
\label{xope}
\ee
of (holomorphic) primary fields -- of given spins $s_{i}$ and scaling dimensions $\Delta_i = 2 h_i - s_i$ -- inserted at points $z_i \in X$ transforms as \cite{DiFrancesco:nk}
\be 
\langle \delta \Xope \rangle 
&=& \frac{1}{2\pi i} \int_X
\d^2z \Big[  (\nabla_{\bar{z}} v^{z} + \mu) \langle T(z) \Xope
\rangle + (\nabla_{{z}} v^{\bar{z}} + \bar\mu) \langle
\bar{T}(\bar{z}) \Xope \rangle  \Big] \nonumber \\ 
&& - \sum_i
\Delta_i \varphi(z_i) \langle   \Xope \rangle \label{varX}~. 
\ee 
The last term here arises from 
delta-functions due to insertions of $T_{z\bar{z}}+T_{\bar{z}z}$, though only when we choose to change the representative of the conformal class $\sigma \longrightarrow \sigma + \varphi$ as well. We use here a conformally flat reference metric proportional to $|\d z|^{2}$. Choosing the transformations judiciously, this identity implies also the standard conformal Ward identity.

On manifolds with boundary, the diffeomorphisms 
generated by
the vector field $(v^z, v^{\bar{z}})$ are required to preserve the
boundary; this necessitates also that only one independent copy of
$\Vir$ and $\overline{\Vir}$ is preserved so that at the boundary
the stress-energy tensors coincide, i.e. $T(x) = \bar{T}(x)$ for $x \in \partial X$.

As the above deformations are all of the deformations we can
perform in two dimensions (conformal and complex structures being
equivalent), then the Beltrami differentials are the only true
deformations of the moduli of the theory, and can be thought of as
(anti-)holomorphic vector fields on the tangent space of the
moduli space $(\mu, \bar{\mu}) \in  T_X{\cal M}$. 
In the case of the moduli space of Riemann
surfaces with boundary the moduli space has, again, only real
analytic structure.

The partition function $Z$ depends, first of all, on the moduli of
the Riemann surface $X$ and the details of the CFT defined on that
surface \cite{Friedan:1986ua}. In particular,
locally it is therefore a function of the coordinates $m,\bar{m}$ of the moduli
space. In a nontrivial CFT $c\neq 0$ there is a trace anomaly
and the partition function depends on the choice of a representative
of the conformal class. In defining the partition function in this
way we need to specify, therefore, that the partition function
$Z(m,\bar{m})$ is evaluated, for instance, in the constant
curvature background metric $g_{\tiny const.}$ in the conformal equivalence
class of metrics we are interested in. The dependence on the
representative of the equivalence class arises through the
Liouville action $S_L[\sigma,g_{\tiny const.}]$ so that if we know the
partition function in the constant curvature background metric
$Z(m,\bar{m})$, on general conformally equivalent backgrounds, i.e. for metrics of the form $g = \e^{\sigma} g_{\tiny const.}$ the
physical partition function becomes
\be
Z[g] &=& \e^{{c}S_L[\sigma,g_{\tiny const.}]} ~Z(m,\bar{m}) ~. \label{weyl}
\ee
Neither is the partition function $Z(m,\bar{m})$ 
in general a well-defined
function on the moduli space of Riemann surfaces, but rather a
section of a (projective) line bundle on it. 
In the case of closed surfaces this line bundle factorises $E_c
\otimes \bar{E}_c$ to holomorphic and anti-holomorphic parts. The
line bundle $E_c$ comes equipped with a 
connection, with respect to which the partition function is
covariantly constant. This can be
seen also form Eq.~(\ref{varX}) by choosing $\Xope=\unit$ so that
$\delta \Xope = \delta \unit = 0$, which implies that 
\be
\nabla_\mu  Z \equiv
\delta_\mu Z - \frac{1}{2\pi i} \int_X \d^2 z ~ T(z)
\mu(z,\bar{z}) ~ Z = 0 ~. \label{covaT}
\ee
This tells us that the partition function is parallel transported
along the vector field $\mu$ on the moduli space with respect to the 
connection $\d + T$.
The holomorphic part can be recognised as a
tensor power of the standard determinant bundle 
\be 
E_c &=&
{\det}_X^{\otimes c/2} 
\ee 
otherwise known as the inverse Hodge bundle $\Det_1 = {\det}_X^{-1}$.

In the case of Riemann surfaces with boundary components the
holomorphic and the anti-holomorphic sectors are related by
complex conjugation. This means that the partition function 
$Z(m,\bar{m})$ with $m=\bar{m}^*$  is
actually a section of the emerging real-analytic 
bundle 
\be
\label{realanalytic}
{\Xdet}^{\otimes c} \longrightarrow {\cal M}~.
\ee
Boundary operators can be described, at least in the unitary case,  in terms of bulk operators on the Schottky double by insertions of operators and their complex conjugates on the original surface, and its mirror image, respectively. This means in particular that if in the expectation values of bulk operators inserted at the point $z \in X$ transform as elements of $(T^{*(1,0)}_{z}X)^{\otimes h'}$ for some conformal weights $h'$, then the corresponding \cite{DiFrancesco:nk} expectation values of  boundary operators at $z \longrightarrow x \in \partial X$ transform as elements of $|T^{*}_{x}\partial X|^{\otimes h}$ for some conformal weights $h$ that depend on the structure of the theory at hand.

Eq.~(\ref{covaT}) above is an example of the pairing given by integration over the Riemann surface $X$ of Beltrami differentials $\mu \in  T_X{^{(1,0)}\cal M}$ and holomorphic quadratic differentials $\nu \in \Omega_X^{(2,0)}(X)$ 
\be
(\nu,\mu) &:=& \int_X \nu \wedge \mu ~, 
\ee 
as the stress-energy tensor $\nu = T(z) (\d z)^2 \in \Omega^{(2,0)}(X)$ is  a locally defined quadratic differential. Holomorphicity is required here for guaranteeing independence of the choice of the representative of the Beltrami differential $\mu \sim \mu + \bar\partial v$. On manifolds with boundary we must restrict to vector fields $v$ that generate flows that leave the boundary invariant. In this sense we can identify $\Omega_X^{(1,0)}{\cal M} \simeq \Omega^{(2,0)}(X)$.  

Let us consider the generalised L{\oe}wner process associated to a parameterised path $\gamma \subset X$, i.e. the sequence of unique harmonic maps (for general surfaces) or biholomorphic maps (for simply connected domains)  that uniformise the cut surface,  and suppose that we have subdivided it in ``sufficiently" small parts $\gamma = \bigcup_{i} \gamma_{i}$. By sufficiently small we  mean that the pertinent measure on each path $\gamma_{i}$ -- Hausdorff or Lebesgue --  is arbitrarily small. In what follows we denote this measure by $\d t$, given a specific parameterisation of the original curve.  

Consider now  the generalised   L{\oe}wner procedure for the  first of these infinitesimal instalments $\gamma_{0}$; the idea is to  iterate the procedure over all of the infinitesimal contributions to get a finite result, as we shall see: 
The Beltrami differential associated to cutting the surface along the  ``infinitesimal" path $\gamma_{0}$ can be thought of as an infinitesimal vector $\mu_t \in T_X {\cal M}$. Its length is proportional to the volume of the cut-out set, i.e.~$\d t$. It has support only on the path $\gamma_0$, if anywhere. Given, as a test function, a cotangent vector from $T_X^*{\cal M}$ represented by a smooth quadratic differential $\nu \in \Omega^{(2,0)}(X)$ the  pairing is, therefore, of the form 
\be
\Big(\mu_t,\nu(z) (\d z)^2 \Big) \sim \d t ~ \nu(\gamma_0)~.
\ee
Here $\nu(z)$ is assumed constant across the infinitesimal set $\gamma_0$. This means that the Beltrami differential can be put formally in the form
\be
\mu_t &:=& \mu_t(z) \frac{\d \bar{z}}{\d z} = - 2\pi i ~ \delta({\gamma_0}) \frac{\d t}{(\d {z})^2} ~,   \label{formalMu}
\ee 
where the formal current 
\be
- 2\pi i ~ \delta({\gamma_0}) {\d t} &=& \mu_t(z) \d^{{2}} {z} \label{msures}
\ee
has distributional support on the cut-out path $\gamma_0$. The details of the distribution will depend of how we choose to regularise it. 

Under the generalised L{\oe}wner procedure the conformal class of the original Riemann surface $X$ changes as a function of the parameter $t$ along the path $\gamma \subset X$. The resulting family of Riemann surfaces $X_{t} \subset {\cal M}$ traces similarly over a certain path $\Gamma \subset {\cal M}$ in the moduli space. The choice of  regularisation of the formal current above amounts then to a choice of parameterisation of this curve as we shall presently see. 

To make the discussion indeed somewhat more concrete, let us look at the situation on the upper half-plane. Let $\ell$ be  a slit of length $t\alpha^{\alpha} /(1-\alpha)^{\alpha}$, extending in from $0$ at an angle $\pi\alpha\in(0,\pi)$ on $\H$. Then the inverse of the uniformising map is given by (cf. the discussion after Def.~\ref{loewner_eq} )
\be
g_t^{-1} : \H \longrightarrow \H \setminus\ell ~;~ z \mapsto
\left(z + t \right)^{1-\alpha} \cdot 
\left(z - \frac{\alpha}{1-\alpha} t \right)^{\alpha} ~. \label{HL}
\ee
The points $-t$ and $+\alpha t/(1-\alpha)$ are mapped onto $0$, and $0$ 
itself onto the tip at $t\alpha^{\alpha}/(1-\alpha)^{\alpha} ~ 
\e^{i\pi\alpha}$; this function is normalised
such  that $g_t(z) \longrightarrow z$ when $z$ tends to infinity. 

Under such a coordinate transformation 
the conformal class -- and hence the complex structure -- of a
locally conformally flat metric changes according to $|\d z|^2 \longrightarrow |\d z + \mu \d \bar{z}|^2$. The thereto associated  Beltrami
differential $\mu(z,\bar{z})$ can be used to detect where the
transformation ceases to be conformal, and by what amount. Given
the transformation $f:=g_t^{-1}$ on $\H$ it can be readily
calculated from the defining equation $\partial f = \mu\,
\bar{\partial} f$. The Beltrami differential associated to $g_t$
given in Eq.~(\ref{HL}) can be argued to be a 
distribution of the form
\be
\mu_t(z,\bar{z}) &=& -2\pi i ~ \frac{\alpha}{1-\alpha} (z-t)^2
\delta^{(2)}(z-t) ~, \label{lemBeltrami}
\ee
where the parameter $t$ is related to the length of the slit. 
The distribution $\delta({\gamma_0})$ in this case is 
\be
\delta({\gamma_0}) &=& \frac{\alpha}{1-\alpha} (z-t)^2 \delta^{(2)}(z-t)  ~. 
\ee
In view of the regularisation procedure that will follow, we choose first to change the parameterisation of the path $\Gamma$ from $t $ to $t'$ where
\be
\frac{\alpha}{1-\alpha} &=& \frac{\d t'}{\d t}~.  
\ee 
This is precisely the definition of the parameter on the path $\Gamma \subset {\cal M}$ alluded to below Eq.~(\ref{formalMu}).

This distribution is zero when integrated with regular test functions; in our case it will appear with functions with second order poles precisely at $z=t$ 
\be
\langle T(z) \phi_h(t,\bar{t}) \rangle &\sim& \frac{h}{(z-t)^2}  
\phi_h(z,\bar{z}) + \frac{1}{z-t} \partial_z \phi_h(z,\bar{z}) ~,
\ee 
so that it picks the coefficient of the leading pole in the operator product expansion, provided the operator $\phi_h$ is inserted precisely at $z=t$. It is therefore useful to define the operator $\hat{T}$ that does just that, namely picks the leading term in the Laurent expansion of the stress-energy tensor with operator insertions at the specified point, for instance
\be
\hat{T}(z) \cdot \phi_{h}(t)  &:=& h~ \phi_h(t,\bar{t}) ~,
\ee 
for $z=t$, otherwise $\hat{T}(z) \cdot \phi_{h}(t) =0$.  
One could represent this operator e.g.~in terms of contour integration of the standard operator product. 
All other operator insertions provide a trivial result. Note also that the contribution coming from the operator insertion at $z=t$ will now be finite, and its precise value does indeed depend on the regularisation or other details of the distribution $\delta({\gamma_0})$.  

We may now express the change of the correlator through 
\be
\Big\langle (\mu_{t}, ~ {T} ) \cdot \phi_{h}(t) \Big\rangle_{\text{regularised}} &=&  \langle 
\d t' ~ \hat{T}(t) \cdot \phi_{h}(t) \rangle
\ee
which is now to be looked upon as a one-form in  $T^*_X {\cal M}$. Note that the argument of $\phi_{h}$ is a point on the boundary of the Riemann surface and the differential $\d t'$ refers to a parameterisation of the path $\Gamma \subset {\cal M}$.

The operator insertions we consider $\phi_{h}$ are precisely where we want the boundary conditions in CFT to change -- at the intersection of the considered path $\gamma_{t}$ and the boundary $\partial X$. After having performed the above iterative step, we should, therefore, translate the operator insertion from the original point $z=t$ to $z=0$ where the rest of the path  intersects the boundary. This is to be seen as a part of what we mean by L{\oe}wner procedure, but boils down to choosing to evaluate the field $\phi_{h}(z,\bar z)$ precisely at $z=0$; the difference is indeed negligible as long as the infinitesimal path $\gamma_{0}$ is small enough. This is also reminiscent of the definition of stochastic integrals, where one chooses to evaluate the integrand on the left point of each interval.  
 
Inserting these results in  Eq.~(\ref{varX}) and not forgetting the anti-holomorphic sector produces now 
\be
\langle \delta \Xope \rangle 
&=& -  \d t' ~ \Big[ \langle \hat{T}(\gamma_0(t)) \Xope \rangle + \langle \hat{\bar{T}}(\gamma_0(t)) \Xope \rangle \Big] ~,
    \label{varB}
\ee 
with the understanding that the right-hand side involves an insertion of $\Xope$ in the beginning of the path and the left-hand side in the end of the path. 
In summary, the concrete analysis has produced the following results: 
\begin{itemize}
\item[1)]
With an operator insertion at the intersection of the path $\gamma$ and the boundary $\partial X$, correlators transform 
by an infinitesimal but nontrivial amount;  
\item[2)] 
The transformation can be expressed in terms of a (finite) line integral of the stress-energy tensor along the path $\gamma \subset X$; and, 
\item[3)] 
The integration parameter is determined in regularising Eq.~(\ref{msures}); the integral is invariant under reparameterisations as long as we change the volume measure on the Riemann surface $X$ on the right-hand side of this equation or the regularisation of the distribution on the left-hand side as well.  
\end{itemize}

Let us now return to the general discussion, assuming we have regularised the distribution $\delta(\gamma_{0})$, introduced operators $\hat{T}$, and a parameterisation $t$ of the path $\Gamma \subset {\cal M}$: 

Iterating this procedure for all infinitesimal paths $\gamma_{i}$ 
is tantamount to exponentiating the infinitesimal operation: the procedure leads indeed to (essentially) the standard path-ordered exponential that appears in parallel transports, which  can in this case be defined in a regularised form, in notation that will be explained below,  as  
\be
\Pexp  - \int_{\gamma} {T}(\gamma_t) \d t  &:=& \prod_{i} \Tset( \gamma_i) \Big( \unit  - \int_{\gamma_{i}}\d t  ~ \hat{T}(p_{i}^{0})   \Big)  ~. \label{regu}
\ee  
Here the stress-energy tensor is evaluated in the starting point $p_{i}^{0} \in \partial \gamma_{i}$ of each infinitesimal path $\gamma_i$, and the integral itself reduces to the (Lebesgue or Hausdorff) measure of the infinitesimal path $\int_{\gamma_{i}} \d t$. Now, in the limit where the paths are indeed taken arbitrarily small, the operator product expansion of the factors in this formula with any operator insertion $\phi_{h}(x)$ on the boundary $x \in \partial X$ are never more singular than with insertions at the beginning of each path $p_{i}^{0} \in \partial\gamma_{i}$ for the simple reason that this is the only place where the infinitesimal path intersects the boundary: Therefore, all operator product 
expansions are dominated by whatever contribution arises from the starting points $x=p_{i}^{0}$ and, as was shown above, these contributions are
finite. We have stipulated a specific way of doing this 
expansion, by determining that the operator insertion $\phi_{h}$, if any, be translated  from the beginning of the path $p_{i}^{0}$ to the end $p_{i}^{1}$
by inserting explicitly the translation operators
\be
\Tset(\gamma_{i}) \cdot \phi_{h}(p_{i}^{0}) & := & 
\phi_{h}(p_{i}^{1})  ~. 
\ee 

What all of this amounts to is a specific regularisation of the 
formal, a priori perhaps rather singular operator 
\be
\Pexp  \Big( - \int_{\gamma} {T}(\gamma_t) \d t \Big) \cdot \phi_{h}  ~. 
\ee 
There might be other ways of regularising this object -- essentially a parallel transport of operators along the path $\gamma$. It would be interesting to investigate further how different regularisations affect the present discussion or under what conditions this formal product should indeed converge. In particular, one should consider insertions of boundary operators in terms of insertions of bulk operators together with their mirror image on the other half of the Schottky double. 
We leave these issues, however, to later study.

\subsection{Parallel transport and conditioning correlators}

Let us consider now the correlation function
\be
{\langle \Xope(y_i) \rangle}_{\gamma[0,t]} &:=&  \Big\langle \Pexp -
\int_{\gamma[0,t]} \d s \Big( T(\gamma(s))
     + \bar{T}(\gamma(s))\Big) \cdot \Xope(y_i)  \Big\rangle
     \label{varE} ~,
\ee
associated to the path $\gamma[0,t]$ that follows the path $\gamma$ from time $0$ until time $t$. The definition of the path-ordered exponential used here was the subject of Sec.~\ref{moduli}; the regularised expression we intend to use here was explicitly constructed in Eq.~(\ref{regu}). In particular, the exponential is to be expanded  
in a product of exponentials of infinitesimal contributions 
along the path acting successively on $\Xope(y_i)$, and the operator insertion should be transported along the path in the process of performing the integration. The operator $\Xope$ was defined in Eq.~(\ref{xope}) in terms of chiral primaries. In what follows we shall choose to restrict to operator insertions on the boundary $z_i = y_i \in \partial X$, and choose the operators themselves from the pertinent BCFT. These operators are, by construction, invariant under complex conjugation and the associated correlators are thus real. As was observed after Eq.~(\ref{realanalytic}) in Sec.~\ref{moduli}, they transform as elements of $|T^{*}_{y_{i}}\partial X|^{\otimes h_{i}}$
under conformal transformations. 

Using the explicit Beltrami differential (\ref{lemBeltrami}), it
was possible in fact to show in the previous Sec.~\ref{moduli} that the L{\oe}wner procedure along the path $\gamma$ maps the correlator ${\langle \Xope \rangle}$ to the above correlator ${\langle \Xope \rangle}_{\gamma[0,t]}$. What we see here is, therefore, its parallel transport with respect to the operator valued connection $T$ \cite{RSZ}. Physically one can think of this as the partition function in the presence of energy density, or a current, distributed along the path $\gamma$.

The argument to this effect 
made use of the explicit form of the
Beltrami differential $\mu$ on $\H$ given in Eq.~(\ref{lemBeltrami}),
and the fact that it 
always arises together with the correlator of the
stress-energy tensor $T$ 
and the explicit operator insertion $\Xope$: The 
operator product expansion has precisely the quadratic
pole needed to produce a nontrivial result. 
The infinitesimal change in the correlator can be thought of as a
suitably normalised differential on the moduli space; cutting out parts of
the path repeatedly leads to the path-ordered exponential integrated along
a finite path in the moduli space. 

This can be generalised to arbitrary Riemann surfaces with boundary components. 

We can now define a probability density associated to any parametrised path in 
the space of paths on the surface $X$,
$\Pi(X,p;t)$, namely let
\be
{\cal P}_X : \Pi(X,p;t) \longrightarrow \R^*_+ ~ ; ~ \gamma
\mapsto  \frac{{\langle \Xope \rangle}_{\gamma[0,t]}}{{\langle \Xope
\rangle}} \label{main}
\ee
for each path in $\Pi(X,p;t)$ that starts from a fixed point 
$p \in \partial X$ and goes on until the final parameter value 
$t$. This density has, first of all, the property that it
is real and normalised ${\cal P}_X \in (0,1]$ such that ${\cal
P}_X(p) = 1$. Reality follows from the facts that the operator
insertions on the boundary are, by construction, real $\Xope^* =
\Xope$ and that the conformal mapping $f : X\setminus \gamma
\longrightarrow X$ preserves the boundary on the real axis. The
fact that it is non-negative follows from the assumption that
$\langle \Xope \rangle \neq 0$ as a physical 
partition function, and the fact that $\langle \Xope \rangle_{\gamma[0,t]}$ 
was constructed from it by exponentiation.  It is bounded by its value ${\cal
P}_X(p) = 1$ if we assume that the classical weak energy condition
 ${T}+{\bar{T}}>0$ translates to a positivity 
condition on the spectrum of the corresponding operator in CFT. 
Whether these (mild) assumptions are satisfies 
depends on the details of the pertinent CFT model.

The above observations amount to the statement that we can
consider ${\cal P}_X(\gamma)$ to be a probability associated to a
path $\gamma$. It is, however, {\em not} the probability of the
path occurring among all the paths of $\Pi(X,p)$ 
but rather the probability of finding a path in a
hull of $\gamma$, perhaps, whose width is related to the structure
of the CFT, and the central charge i.e.~the diffusion coefficient
in particular. 

Suppose now that the simple curve $\gamma[0,t]$ is an
SLE${}_\kappa$ process $\gamma_t$ on a Riemann surface $X$ 
with $0 \leq \kappa < 4$. We can then likewise associate to this path
the correlator
\be {\langle \Xope \rangle}_{\gamma[0,t]} &:=&  \Big\langle \Pexp -
\int_{\gamma[0,t]} \d \sigma_H(s) \Big( T(\gamma(s))
     + \bar{T}(\gamma(s))\Big) \Xope \Big\rangle
     \label{varEE} ~,
\ee
where $\sigma_H$ is the Hausdorff measure of dimension \cite{Be3}
\be
\dim_H(\gamma) &=& \min \left(2, 1 + \frac{\kappa}{8} \right)~.
\label{Bdim}
\ee
This extends the definition (\ref{varE}) to the case of fractal
curves. It shows, again, how the correlation functions of 
observables ${\langle \Xope \rangle}$ transform under L{\oe}wner
processes, and it continues to be normalised as suggested above.

A local conformal transformation $\rho$ induces a transformation
$R(\rho)$ of the pertinent CFT (Hilbert) space. This corresponds to a representation of the group $\Aut({\cal O})\cong\{a_1 t+ a_2 t^2+\cdots,\; a_1\neq0\}$ of changes of local (formal) coordinates $t$,  on the pertinent Hilbert space, constructed essentially  by exponentiating the positive part of Virasoro algebra $\Vir_{\geq 0}$, cf.~e.g.~Lemma 5.2.2 in Ref.~\cite{FbZ}. The operator insertions of primary fields transform homogeneously, whereas the stress-energy operator changes inhomogeneously
\be
\Xope (y_i) &=& R(\rho) \Xope \Big(\rho(y_i)\Big)R(\rho)^{-1} ~
\prod_i \Big( \rho'(y_i) \Big)^{h_i} \\
T(z) &=& R(\rho)  T\Big(\rho(z)\Big)  R(\rho)^{-1} + \frac{c}{12}
\Big\{ \rho(z), z \Big\} ~ \unit
\ee
where $\{ ~ ,~ \}$ is the Schwarzian derivative (\ref{schwarzian}).
The correlator ${\langle \Xope \rangle}_{\gamma[0,t]}$ transforms
therefore as
\be
\rho^*{\langle \Xope  (y_i) \rangle}_{\gamma[0,t]} 
&=& \prod_i \Big( \rho'(y_i)
\Big)^{h_i} ~ \exp -\frac{c}{6} \int_{\gamma[0,t]} 
\d s ~ \Re\Big\{ \rho(\gamma(s)), \gamma(s) \Big\} \nonumber \\
  & & \qquad  \qquad  \cdot {\langle \Xope  (y_i) 
\rangle}_{\gamma[0,t]} \label{transfR} ~. 
\ee

In order to condition the probability to paths that do not enter 
a given simply connected domain $A \subset X$ that touches the
boundary, we need to investigate the behaviour of the density 
${\cal P}_X$ under the pertinent diffeomorphism $\rho : X \setminus A
\longrightarrow X$. In the simple case that this diffeomorphism happens 
to be a conformal mapping or that the pertinent moduli space is a point, as in the case of the upper half-plane $\H$, we find 
\be
\frac{ {\cal P}_{X \setminus A} }{ {\cal P}_X } &=& \prod_i \Big(
\rho'(y_i) \Big)^{h_i} ~ \exp -\frac{c}{6} 
\int_{\gamma[0,t]} \d s ~
\Re \Big\{ \rho(\gamma(s)), \gamma(s) \Big\} ~.
\ee

Comparing to the martingale $Y_t$ 
(\ref{martingaleY}), we see that ${\cal P}_X$ and $P$ have precisely the
same behaviour under conditioning. In
comparing this expression to $Y_t$ in stochastic analysis, one
needs to take into account the fact that there it was convenient
to keep the origin of the complex plane fixed and let the
intersection $W_t = \gamma_t \cap
\partial \H$ move, whereas in the CFT analysis one kept the
intersection fixed at the origin and allowed the original zero to
move.

In the general case where the diffeomorphism  $\rho : X \setminus A
\longrightarrow X$ changes the conformal structure of the Riemann surface, 
it is difficult to give an explicit formula for the transformation. 
For infinitesimal such deplacements this is nevertheless possible, and 
reduces clearly to insertions of the stress-energy tensor 
integrated with the Beltrami 
differential associated to $\rho$ in the correlators.

As we will explain later, the correlator $\langle \Xope \rangle_{\gamma[0,t]}$ can be recognised as
a section of a certain bundle $\Obdle_h$ over
the moduli space of Riemann surfaces. In this context it follows
that ${\cal P}_X(\gamma)$ is  a Wilson
line, of this section when parallel transported from the fibre at
$X$ to the fibre over ${X \setminus \gamma}$ with respect to the
connection  $\d + T$.
\section{CFT I}
In this section we collect facts from Conformal Field Theory (CFT). It should serve as a preparation to ease the understanding of  the discussion in the text and to introduce some of the  common expressions from the vocabulary of high-energy physics. There exist several references one might consult for details~\cite{BPZ1, BPZ2, Ca1, Cardy:ir, Ca2, Ca3, CK, DiFrancesco:nk, Friedan:1986ua, G1, G2, J, KNTY, Ne, P, RSZ, Segal, TUY}. 

\subsection{General Field Theories}
We start by explaining  what a field theory might be. Here our summary follows parts of the treatment in~\cite{CK}, Appendix B. We restrict to examples of the kind of theories, that might occur. Further we will stay more or less with the path integral (Lagrangian)  description, although there are cases where it would not apply, as for example in the case of percolation. But this will have no impact on the general outline.

Let us consider first  a classical theory on a space time $\M$. Typical examples of $\M$ can be 4-dimensional Minkowski space or the world sheet of a string (i.e. a Riemann surface). Then we have the following

\begin{df} A {\bf classical field} $\phi$ on $\M$ is either a function, a differential form, a section of  a bundle on $\M$, or a connection on a bundle.
\end{df}
The physical dynamics is determined by the {\bf Lagrangian density} $\LL$, which is a functional of the fields and their derivatives. The choice of a Lagrangian is greatly constrained by symmetries, and is determined by the particular physics that the field theory is supposed to model. 

The basic principle of dynamics is that the fields must be minima of the {\bf action}, which is an integral of the Lagrangian density over volumes $V\subset\M$. The classical field equations can then be determined from the {\bf Euler-Lagrange equations} for the action integral, assuming the values of the fields at different points are treated as independent dynamical variables. This will need to be modified when the Lagrangian has symmetries.

The next step is to proceed to a quantum field, which is not well-defined mathematically in the desired generality. The idea is that we seek a Hilbert space $\HH$ of states, and that classical fields $\phi(x)$ get replaced by {\bf quantum fields}, denoted $\Phi(x)$, which are operators on $\HH$ depending locally on $x$ and behave as distributions in the variable $x$. The classical values of the fields are reproduced as eigenvalues of the operators. For example, if there is a state $|\psi\rangle\in\HH$ for which $\Phi(x)|\psi\rangle=f(x)|\psi\rangle$ for some function $f(x)$, then
we think of this quantum state as corresponding to the classical field with value $\phi(x)=f(x)$. The inner product is used to determine relative probabilities  as in ordinary quantum mechanics.

So far this was the Hamiltonian formulation of quantum field theory. As mentioned earlier, there is also a Lagrangian formulation, which involves the notion of a {\bf path integral}. Suppose that at times $t< t'$, the classical fields have values $\phi_0({\bf x},t)$ and $\phi_1({\bf x},t')$. We will let $|\phi_0({\bf x},t)\rangle$ and $|\phi_1({\bf x},t')\rangle$ denote the quantum states with respective eigenvalues $\phi_0({\bf x},t)$, $\phi_1({\bf x},t')$ under the operator valued distribution $\Phi({\bf x},t)$. In other words, we have
\begin{equation}
\label{operdist}
\Phi({\bf x},t)|\phi_0({\bf x},t)\rangle=\phi_0({\bf x},t)|\phi_0({\bf x},t)\rangle
\end{equation}
with a similar expression for $\phi_1$. The left hand side of (\ref{operdist}) involves the action of an operator and the right hand side involves multiplication by a distribution.

The Hilbert inner product of these two states is $\langle\phi_1({\bf x},t')|\phi_0({\bf x},t)\rangle$. Its physical meaning is that its squared norm gives the probability density for the state $|\phi_0({\bf x},t)\rangle$  to propagate into $|\phi_1({\bf x},t')\rangle$. Formal calculations suggest that if $S[\phi]$ is the action, then 
\begin{displaymath}
\langle\phi_1({\bf x},t')|\phi_0({\bf x},t)\rangle={\cal N} \int[D\phi]\;e^{iS[\phi]}
\end{displaymath}
where on the right hand side, the path integral is over the space of all paths within the space of fields with initial point $\phi_0$ and with terminal point $\phi_1$. The factor $\cal N$ is a suitable normalisation factor, and $[D\phi]$ is an appropriate measure on the space of paths. 

The path integral also allows for the computation of certain physical {\bf correlation functions}, sometimes called {\bf $n$-point functions}. Pick $n$ points $x_1,...,x_n\in\M$. The $n$-point function is formally defined to be
\begin{equation}
\label{B15}
\langle\Phi(x_1)\cdots\Phi(x_n)\rangle:=\frac{1}{Z}\int[D\phi]
\phi(x_1)\cdots\phi(x_n)\;e^{iS[\phi]}
\end{equation}
These are functions of the points $x_1,\cdots, x_n$ and of the types of fields associated to these points, but not the particular values of the fields, as these are integrated over. In our example of a scalar field, there is only one type of field. These quantities are of intrinsic physical interest. For example, the two-point functions are just the Green's functions of the theory. One can argue that the $n$-point functions contain all of the physical predictions of the theory.

While path integrals such as (\ref{B15}) cannot be formulated rigorously, there are some accepted heuristics for their calculation that do not have partial justifications. 

Here is a subtle point one should mention. In the path integral (\ref{B15}), the terms $\phi(x_i)$ are distributions, hence commute. On the other hand, the {\bf operators} $\Phi(x_i)$ do not commute in general, so there appears to be an inconsistency. To resolve this one defines the {\bf time-ordered product} $T(\Phi(x_1)\cdots\Phi(x_n))$ to be the operator obtained by applying the $\Phi(x_i)$ in chronological order. Then we have
\begin{displaymath}
\langle\Omega|T(\Phi(x_1)\cdots\Phi(x_n))|\Omega\rangle=\frac{1}{Z}\int[D\phi]
\phi(x_1)\cdots\phi(x_n)\;e^{iS[\phi]}
\end{displaymath}
where $|\Omega\rangle$  denotes the {\bf vacuum} state.

Some  understanding of a {\bf gauge theory} is  helpful. Here, the action is invariant under a continuous group of local transformations on $\M$. This group is usually infinite-dimensional, the typical example being Yang-Mills theory, where the gauge group is the space of maps to a finite-dimensional Lie group. We explain the role of  what is called {\bf gauge fixing} in the context of our discussion of quantum field theory, as well as the notation of a conserved quantity.

The simplest example to consider is the classical theory of electricity and magnetism in the absence of charged matter. The field strength $F=d\phi$ and hence the associated Lagrangian $\LL$ are unchanged by the addition of an exact form to $\phi=A_j\, dx^j$, i.e. by the substitution 
\begin{equation}
\label{EMgauge}
A_j\mapsto A_j+\frac{\partial\Lambda(x)}{\partial x^j},
\end{equation}
where $\Lambda(x)$ is an arbitrary real function on $\M$. This substitution is associated to the {\bf gauge group of local transformations} $e^{i\Lambda(x)}$; in fact, if we identify $\phi=\sqrt{-1}\;A_j\, dx^j$ with a connection form on a principal $U(1)$ bundle over $\M$, then the gauge transformation $e^{i\Lambda(x)}$ on the $U(1)$ bundle induces (\ref{EMgauge}). It is therefore sufficient to choose a slice of the parameter space of the $\phi=A_j\, dx^j$ which has the property that every possible $\phi$ is equivalent to one of the $\phi$ in the slice via a gauge transformation (\ref{EMgauge}). Such a slice is called a gauge choice.
We come back to this matters in the next section in more specific terms. 

It is an important fact  (Noether's theorem) that continuous groups of symmetries give rise to conserved quantities. Letting the fields transform infinitesimally as $\phi\mapsto\phi+\epsilon\psi$ under an infinitesimal gauge transformation (where $\epsilon$ is an infinitesimal parameter), the quantity
\begin{equation}
\label{conservedcurrent}
\frac{\delta\LL}{\delta\partial_j\phi}\psi
\end{equation}
has zero divergence, and is called a {\bf conserved current}. As a consequence, the spatial (fixed time) integral of a conserved current is constant in time, and is called a {\bf conserved charge}.

The quantum field theory discussed so far applies best to {\bf bosons}, which are particles of integer spin. Several classical (i.e., non-quantum) bosonic fields can be combined by multiplying the fields, and this multiplication is commutative. But in order to describe field theories for {\bf fermions} (particles of half-integer spin), one starts with classical field theories which multiply in an anticommutative fashion. 

\subsection{Conformal Field Theory}

A conformal theory (CFT) is a theory of fields on Riemann surfaces where conformal transformations of surfaces play a distinguished role. The connection to {\bf string theory} is that the Riemann surfaces occur as the surface swept out by the string as it propagates in time. Hence this Riemann surface is referred to as the {\bf world sheet}. For details cf.~\cite{AGMV, BPZ1, BPZ2, CK, DiFrancesco:nk, Friedan:1986ua, G1, G2, J, KR, KNTY, Ne, P, RSZ, Segal, TUY}.

There exist several attempts to axiomatise CFT. The best known is what is called Segal's axioms~\cite{Segal}, or equally, a modular functor. But so-far none of this frameworks is completely satisfactory. Nevertheless, we shall recall some of the main features that have partial justification through the study of specific models.

Probably one of the reasons why the axiomatic treatment still causes difficulties, is that in general physicists tend to make statements that are too general, i.e. they do not specify the class, for that it should hold. This applies e.g. for the renormalisation group, the path integral etc.  The task is therefore to make ``just the right" restriction, so that one has a working implementation, but with keeping the main features of the original idea. One could compare this with the relation of herbal medicine to a pharmacist who  actually determines the active ingredient.

Depending on whether the Riemann surfaces are allowed to have boundaries or not, one distinguishes between ``ordinary" CFT or boundary CFT (BCFT). 

In this section we describe a two-dimensional conformally invariant quantum field theory by some basic concepts and postulates. We will assume the Euclidean signature (+,+) on $\R^2$ or on surfaces because of the close connection of conformal field theory to statistical mechanics.

Let us  now introduce some of the necessary data for a CFT on (general) Riemann surfaces.

Every Riemannian metric on a real two-dimensional surfaces induces a complex structure. This gives a Riemann surface where the conformal mappings are self-mappings preserving the angles and the orientation, and they are holomorphic. After choosing complex coordinates $z=x+iy$ one could want to investigate conformal mappings locally.  But there is no group of local conformal transformations, because neighbourhoods are not preserved and so mappings cannot be composed. However the conformal algebra, generated by the  infinitesimal conformal transformations, induces a real vector field on the underlying real surface. Locally, in holomorphic coordinates these can be written as
\begin{displaymath}
f(z)\frac{\partial}{\partial z}+\bar{f}(\bar{z})\frac{\partial}{\partial\bar{z}}
\end{displaymath} 
where $f(z)$ is purely holomorphic. Usually one separates holomorphic and antiholomorphic parts (light-cone coordinates, left and right movers) by choosing bases
\begin{displaymath}
\ell_n:= -z^{n+1}\partial_z\quad\quad\quad \bar{\ell}_n:=-\bar{z}^{n+1}\partial_{\bar{z}}
\end{displaymath} 
which satisfy the Witt algebra relations:
\begin{displaymath}
{[\ell_n, \ell_m]}=(n-m)\ell_{n+m},\quad\quad\quad [\bar{\ell}_n,\bar{\ell}_m]=(n-m)\bar{\ell}_{n+m}.
\end{displaymath}
The conformal algebra can be recovered from the Witt algebra as the subalgebra generated by the $\ell_n+\bar{\ell}_n$ and $i(\ell_n-\bar{\ell}_n).$ 

Now we come to  the fields of interest in a conformal field theory.  First, a conformal field theory is a quantum field theory but with a class of fields singled out by having  a number of special properties:
\begin{df}
A {\bf primary field} of {\bf weight} $(h,\bar{h})$ is a field $\Phi(z,\bar{z})$ which transforms as 
\begin{displaymath}
\Phi(z,\bar{z})\mapsto\left(\frac{\partial f}{\partial z}\right)^h\left(\frac{\partial f}{\partial\bar{z}}\right)^{\bar{h}} \Phi(f(z),\bar{f}(z))
\end{displaymath}
under a conformal transformation $z\mapsto f(z)$.
\end{df}
In other words, the expression $\Phi(z,\bar{z}) dz^h\otimes d\bar{z}^{\bar h}$ is invariant when $\Phi(z,\bar{z})$ is a primary field. An important example of a primary field is the  $\bf {1}$ field.

The gauge group of CFT is the local conformal group and  the gauge choice is determined by fixing the metric on the surface to be diagonal. If a classical action is given that is conformally invariant, it might happen that in the process of quantisation the invariance is lost. To maintain it, one has to introduce certain additional fields, called {\bf ghosts}.  The relevant example in this context is the Polyakov action which we briefly introduce. So we fix a closed, oriented and smooth reference surface $M$ of genus $p$ (number of handles) and consider the set of arbitrary smooth embeddings of $M$ in $\R^d$
\begin{equation}
\label{ }
x\equiv(x^1,...,x^d) : M\rightarrow\R^d 
\end{equation}
and further an arbitrary Riemannian metric $(g_{ij})_{i,j=1,2}$ on $M$ with $(g^{ij}):=(g_{ij})^{-1}$, $\det g=g_{11}g_{22}-g^2_{12}$. The image $x(M)$ is the so-called vacuum-to-vacuum world-sheet and the random metric $(g_{ij})$ is an extra dynamical variable, which is also summed over and has nothing to do with the induced metric on $M$ via the embedding into $\R^d$. Then the {\bf Polyakov action}~\cite{P} for the world-sheet and the assigned metric is
\begin{equation}
\label{Polyakov}
S_P[g,x]:=\int_Mg^{ij}\frac{\partial x^{\alpha}}{\partial z^i}\frac{\partial x^{\alpha}}{\partial z^j} \sqrt{\det g}\, d z^1d z^2.
\end{equation}
where $z=(z^1, z^2)$ is a local coordinate. 
A solution to the problem
\begin{displaymath}
S_P[g,x]\longrightarrow\min
\end{displaymath}
is then harmonic w.r.t. the metric $g$:
\begin{equation}
\label{ }
\frac{1}{\partial z^j}\left(g^{ij}\sqrt{\det g}\,\frac{\partial}{\partial z^j}x^{\alpha}\right)=0,\quad\alpha=1,...,d
\end{equation}
and conformal:
\begin{equation}
\label{EMTensor}
T_{ij}:=\frac{\partial x^{\alpha}}{\partial z^i}\frac{\partial x^{\alpha}}{\partial z^j}-\frac{1}{2}g_{ij}g^{kl}\frac{\partial x^{\alpha}}{\partial z^k}\frac{\partial x^{\alpha}}{\partial z^l}=0.
\end{equation}
The fundamental quantity $(T_{ij})$, which is symmetric, is called the {\bf energy-momentum tensor}. 
We note, that the action $S_P[g,x]$ has three types of symmetries (or invariances)
\begin{itemize}
  \item {\bf Invariance under  isometries of $\R^d$}\quad If $t\in\R^d\rtimes O(d)$ i.e. a Euclidian motion, hence an element of the semi-direct product of the group of rotations and translations, then 
  \begin{displaymath}
S_P[g,x]=S[g, t\circ x]
\end{displaymath}
    \item {\bf Invariance under reparametrisations}\quad For every    diffeomorphism $f: M\rightarrow M$:
  \begin{displaymath}
S_P[g,x]=S[f^*g,f^* x]
\end{displaymath}
    \item {\bf Invariance under Weyl rescaling or conformal invariance}\quad For all $\varphi\in C^{\infty}(M,\R)$:
  \begin{displaymath}
S_P[g,x]=S[e^{\varphi}\cdot g,x]
\end{displaymath}
\end{itemize}
It follows from the conformal or Weyl invariance that the energy-momentum tensor is {\bf traceless}
\begin{equation}
\label{traceless}
g^{ij} T_{ij}=T^i_{\;\;i}=0.
\end{equation}
This means that the metric $g_{ij}$ is only determined up to a conformal factor, which allows to choose the conformal gauge, i.e. a conformally flat metric $e^{\varphi}\delta_{ij}$ for some real function $\varphi$.
In complex notation the vanishing of the trace is expressed as
\begin{displaymath}
T_{z\bar{z}}=T_{\bar{z}z}=T_{00}+T_{11}=0.
\end{displaymath}
One can then infer, considering the currents, that
\begin{equation}
\label{stress2}
T(z):=T_{zz}=T_{00}-i T_{01}\quad\quad\quad\overline{T}(\bar{z}):=T_{\bar{z}\bar{z}}=T_{00}+i T_{01}
\end{equation}
i.e. $T$ is a holomorphic, resp. $\overline{T}$ is an antiholomorphic (classical) field.  In the quantum case, as we will see, there will be some modifications. 

\begin{figure}[ht]
\begin{center}
\includegraphics[scale=0.4]{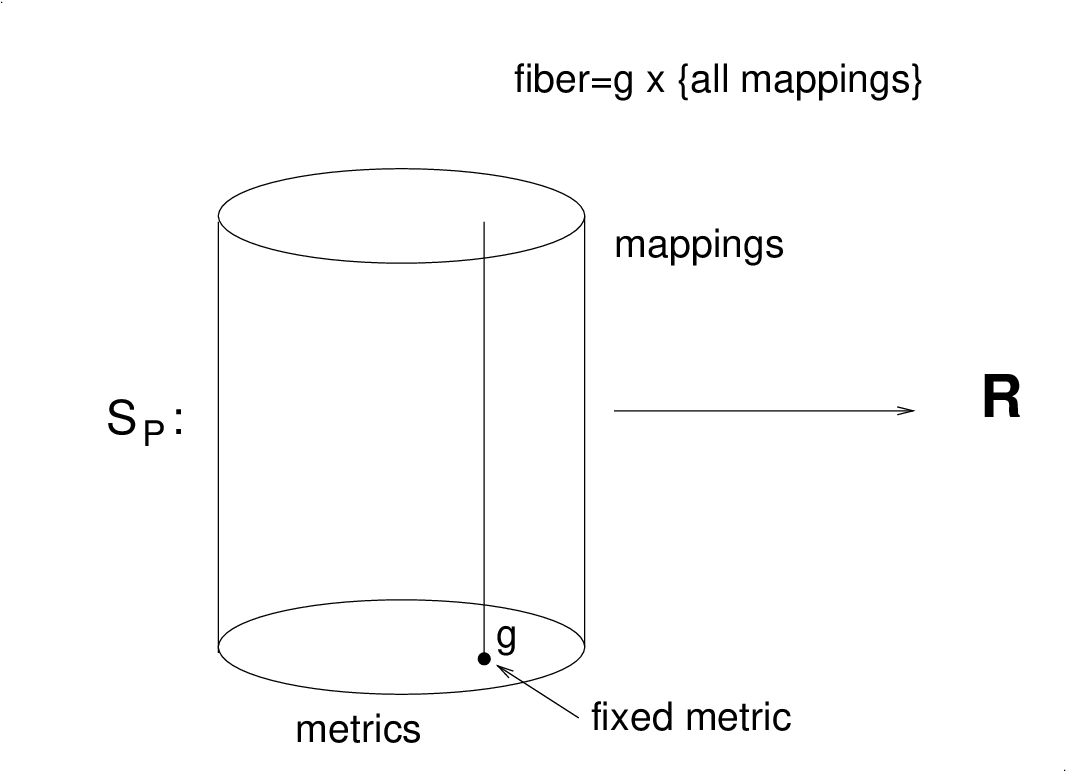}
\caption{Coarse space of integration, for the action $S_P$} 
\label{cd}
\end{center}
\end{figure}
The quantisation of the classical system, i.e. the path-integral, is now defined as
\begin{equation}
\label{Polyakov1}
Z:=\int[D x][D g]\;e^{-S_P[g, x]}
\end{equation}
where the integral is over 
\begin{displaymath}
\map(M)\times\met(M)
\end{displaymath}
i.e. the space of all mappings from $M$ times all possible metrics on the base surface $M$. This space  is  ``coarse", since it possesses an infinite number of symmetries or redundancies. To get rid of the massive over-counting, one has to reduce the space over which one would like to integrate. 

Now, the method that allows to eliminate the integration over orbits of an infinite dimensional symmetry group from the partition function or correlation function, is called the {\bf Faddeev-Popov procedure}, cf. e.g.~\cite{J}. 

So, let us assume that we consider an integral of the form
\begin{equation}
\label{FP121}
Z:=\int e^{-S[A]}dA
\end{equation}
where $S[A]$ is some action, and $A$ varies over a configuration space. We suppose that the action $S[A]$ and the metric on the configuration space underlying the measure $dA$ are invariant under the action of a group $G$, $A\rightarrow g.A$ for $g\in G$. Then the elements of $G$, if they act locally (cf. the example below), are called {\bf gauge transformations}. Because of this invariance under the action of $G$, we try to eliminate the integration over the orbits of $G$ from the definition of $Z$, as $Z$ should only count {\bf physically inequivalent situations}. To achieve this, we have first to choose a {\bf slice} in the $A$-space which is transversal to the orbits of $G$. This is called {\bf gauge fixing}. We can then write 
\begin{equation}
\label{FP122}
A=(A_1, A_2),
\end{equation}
where $A_2=0$ corresponds to the slice chosen while each fixed $A_1, A_2$ varies on the orbit through $(A_1,0)$. This change of variables yields a functional determinant via
\begin{equation}
\label{FP123}
dA=\det\left(\frac{\partial A}{\partial(A_1, A_2)}\right)dA_1 dA_2;
\end{equation}
where
\begin{displaymath}
\det\left(\frac{\partial A}{\partial(A_1, A_2)}\right)
\end{displaymath}
is called the {\bf Faddeev-Popov determinant}. One finally redefines $Z$ as 
\begin{equation}
\label{ }
Z:=\int e^{-S[A]}\det\left(\frac{\partial A}{\partial(A_1, A_2)}\right)dA_1~,
\end{equation}
i.e. one discards the integration over the orbits of $G$, because each point on an orbit is already represented by a physically equivalent point on the gauge slice. This way, $Z$ becomes independent of the choice of local slice and of the parameter $A_2$ on the orbits of the symmetry group.

However in applications, as e.g. in string theory, it may be the case that the classical action $S[A]$ is invariant under the action of the group $G$, the measure $dA_1$ on the slice is not. This is the case for the expression (\ref{Polyakov1}), i.e. which is not conformally invariant anymore, because the formal measures $[Dx]$ and $[Dg]$ are not. This is the ``famous" {\bf conformal anomaly}; and in general, when going from the classical to the quantised theory, this phenomenon is just called {\bf anomaly}.

In our case, if we denote by $M$ an oriented and connected surface (two-dimensional real manifold) of genus $h>1$ and by
\begin{itemize}
  \item $\Diff_+$ the group of diffeomorphisms of $M$ that preserve the orientation, 
  \item $\Diff_0$ the sub-group of diffeomorphisms that are homotopic to the  identity, i.e. for every $f\in\Diff_0$ there exists a differentiable map $F:[0,1]\times M\rightarrow M$ such that $F(0,\cdot)\equiv\id$ and $F(1,\cdot)\equiv f$,
  \item $\Weyl:=C^{\infty}(M,\R)$ the additive group of globally defined real functions,
\end{itemize}
then the group $\Diff_+$ has a natural action from the right on $\Weyl$:
\begin{equation}
\label{DiffactiononWeyl}
\varphi.f:=f^*\varphi=\varphi\circ f,\qquad\varphi\in\Weyl, f\in\Diff_+.
\end{equation}
Further, this action preserves the additive group structure of $\Weyl$ and therefore we can define the semi-direct product $\Weyl\rtimes\Diff_+$
and its sub-group $G:=\Weyl\rtimes\Diff_0$. Let 
$\met=\met(M)$ denote the space of $C^{\infty}$-Riemannian metrics on $M$.
Then the semi-direct product $\Weyl\rtimes\Diff_+$, and hence also its sub-group $G$, act from the right on $\met$:
\begin{equation}
\label{actiononmetric}
g.(\varphi,f):=f^*(e^{\varphi}g),\qquad g\in\met(M).
\end{equation}

Therefore, we have to divide the space of all metrics $\met(M)$ by $\Weyl\rtimes\Diff_+$ which yields a finite dimensional space, the moduli space. The usual gauge slice consists in choosing for genus $h>1$ the hyperbolic metrics of constant curvature $-1$.

Now  for a fixed metric $g_0$ if we integrate out the maps, i.e. determine the marginals of the path integral, will give us the ``marginal" partition function
\begin{equation}
\label{Marginal_part_funct}
Z[g]:=\int[D x]\;e^{-S_P[g,x]}
\end{equation} 

After this more general part which applied to string theory and as well to CFT, we return to CFT specific notions.

As in any quantum field theory the essential parameters of the theory, which are also important since they connect the theory with experimental data, are the correlation functions or also called {\bf Green's functions}. Particularly interesting are those of the primary fields. These Green's functions are defined, e.g. via the path integral, for non-coinciding points, are symmetric in pairs of arguments and assumed to be real analytic. 
Further they will depend on the Riemannian metric $g$ on the surface. 
Time-ordering means in this context that after having chosen a local coordinate, the equal time curves are the concentric  circles, whose radii are the curves with fixed spatial coordinate. This is called {\bf radial quantisation}. However, for surfaces the distinction between space and time is not invariantly defined. The correlation function of the ${\bf 1}$ field, with respect to the metric $g$, is simply the partition function and denoted as $Z[g]$, as already introduced. In essence the CFT is now specified by the partition function and the  set of primary fields $\{\Phi_i\}$ where the correlation functions of primary fields are assumed to satisfy some basic symmetry relations:
\begin{description}
  \item[Diffeomorphism covariance] 
  \begin{eqnarray}
Z[g]& = & Z[f^*g] \\
\langle\Phi_1(f(p_1))\cdots\Phi_n(f(p_n))\rangle_g& = & \langle\Phi_1(p_1)\cdots\Phi_n(p_n)\rangle_{f^*g}
\end{eqnarray}
  \item[Weyl covariance]
  \begin{eqnarray}
  \label{Weyl_anomaly}
Z[e^{\sigma}g] & = & e^{\frac{ic}{48\pi}\int_M (\partial\sigma\wedge\bar{\partial}\sigma+2\sigma R_g)}Z[g] \\
\langle\Phi_1(p_1)\cdots\Phi_n(p_n)\rangle_{e^{\sigma}g} & = & \prod^n_{l=1} e^{-\Delta_l\sigma(p_l)}\langle\Phi_1(p_1)\cdots\Phi_n(p_n)\rangle_{g} 
\end{eqnarray}
\end{description}
We note that (\ref{Weyl_anomaly}) is a manifestation of the so-called conformal anomaly as it shows up for the Polyakov action. Further it is relevant in SLE, as it is connected with the restriction property. 

Since correlation functions of CFT should reflect the infinite fluctuations of a quantum field taken at a precise position, they have singularities on the  diagonal, i.e. when the positions of two or more fields coincide. The fundamental {\bf operator product expansion} OPE captures the behaviour at these poles and allows to write the correlation function as a sum of its principal parts and an entire function (Mittag-Leffler decomposition).

The  OPE of the energy-momentum tensor, a quantum field now, with primary fields is written usually by removing the brackets $\langle...\rangle$, being understood that the OPE is meaningful only within correlation functions. For a single primary field $\Phi$ of conformal dimension $h$, we have
\begin{equation}
\label{Stress-OPE}
T(z) \Phi(w,\bar{w})\sim\frac{h}{(z-w)^2}\Phi(w,\bar{w})+\frac{1}{z-w}\partial_w\Phi(w,\bar{w})+\mbox{regular terms}
\end{equation} 

The  OPE for the energy-momentum tensor with itself is:
\begin{equation}
\label{Stress-Stress-OPE}
T(z)T(w)\sim\frac{c/2}{(z-w)^4}+\frac{2T(w)}{(z-w)^2}+\frac{\partial_w T(w)}{(z-w)}+\mbox{regular terms}
\end{equation}
where the constant $c$ depends on the specific model under study and is called the {\bf central charge}. Except for this anomalous term, the OPE (\ref{Stress-Stress-OPE}) means that $T$ is a quasi-primary field (which usually means invariance under the action of $SL(2,\C)$\,) with conformal dimension $h=2$. 
If we expand 
\begin{displaymath}
T(z)=\sum_{n\in\Z} L_n z^{-(n+2)}
\end{displaymath}
and substitute it into (\ref{Stress-Stress-OPE}), we get by contour integration
\begin{equation}
\label{Vir_Relation}
[L_n, L_m]=(n-m)L_{n+m}+\frac{c}{12}(n^3-n)\delta_{n,-m}{\bf 1}.
\end{equation}
This implies that in the quantum theory we have a representation of the  centrally extended Witt algebra, which is the Virasoro algebra of central charge $c$. 
\subsection{Boundary CFT}
\label{BCFT}
We stressed earlier the point that  the choice of proper boundary conditions was crucial to get in a stable way the random
chordal lines for the discrete models. More generally, one can study CFT in the presence of boundaries and how the theories depend on the various choices of boundary conditions. 
The corresponding formalism was initiated and developed by J. Cardy~\cite{Ca1,Cardy:ir}. Nevertheless BCFT is still (very much) in development (cf. also~\cite{DiFrancesco:nk, G2, G3}).

The prototype of a boundary CFT is that of a theory defined in the upper half-plane $\H$. Because of the presence of a boundary, i.e. the real axis, the set of conformal transformations is restricted to those that preserve it. So for an infinitesimal conformal transformation  $z\mapsto z+\epsilon(z)$ the $\epsilon(z)$ must be real for $z\in\R$, i.e. if $\epsilon(z)=\sum_n a_n z^{n+1}$ where $z$ is real then $a_n\in\R$. This has as consequence that there is only one set of corresponding Virasoro generators $L_n$, instead of the pair $(L_n, \bar{L}_n)$ as in the usual case.   Further  from the physical point of view, it is natural to demand, that there should be no flow of conformal currents across the boundary. Since the charge generates the conformal transformations in the quantum theory this is a reasonable requirement. So we have the additional  condition that
\begin{equation}
\label{realonboundary}
T(x)=\overline{T}(x)\quad\quad x\in\R.
\end{equation}
This condition ensures that the conformal symmetry is unbroken, but the translational symmetry  perpendicular to the boundary is lost.

One common way to relate the theory defined on a bordered surface to the one on a closed surface is first by taking the Schottky double and then by reflecting the system defined on one half to the other half. This method is similar to the known from classical electrodynamics, where it is called the ``doubling trick" or ``method of images".   In particular it yields
\begin{displaymath}
\overline{T}(z)=T(\bar{z}),
\end{displaymath}
what is consistent due to (\ref{realonboundary}). We note that because the  symmetry axis, i.e. the former boundary, still  plays  a special role, the holomorphic and antiholomorphic part are no longer independent. Nevertheless, the theory can so be developed  in analogy with the  boundary-less case.

In the path integral formalism we have the following requirements for the boundary conditions. First they should be  invariant under conformal transformations which requires them to be homogenous. So, e.g. for a scaling field $\Phi$ on $\H$ that extends  to the boundary, they would be as follows: 
\begin{displaymath}
\Phi\big|_{\R}=0,\quad\quad\Phi\big|_{\R}=\infty,\quad\quad\frac{\partial\Phi}{\partial\vec{n}}\Big|_{\R}=0.
\end{displaymath}

So far we have put the same homogenous conditions on the whole boundary which ensured the existence of the Virasoro algebra.  

But next we consider (as a toy model) the case, when we assign  from the set of  allowable  boundary conditions one to the left of $0$, say $(\alpha)$ and another to the right, say $(\beta)$, on $\R$ such that they change  discontinuously.  

In the operator formalism, in which we shall proceed now, this implies for radial  quantisation that the vacuum is no longer invariant under translations, i.e. it is no longer annihilated by $\hat{L}_{-1}$. This state might be considered as equivalent to the action of a {\bf boundary operator} $\Phi_{\alpha\beta}(0)$ acting on the true vacuum $|0\rangle$.  It is a highest weight state with weight $h_{\alpha\beta}$. In general boundary operators are  scaling fields  living on the boundary and when inserted at a point $x$ on the real axis (boundary), they  change the boundary condition from $(\alpha)$ to $(\beta)$.

The introduction of boundary operators allows further to relate the partition function of a system with discontinuously changing boundary conditions to a correlator of boundary operators.
Suppose we require the boundary conditions $|\alpha_i\rangle$ to apply on intervals labelled by $i=1,\ldots,n$. This can be done technically by inserting the
corresponding boundary operators $\Phi_i :=
\Phi_{\alpha_i\alpha_{i+1}}$ that change the state
$|\alpha_i\rangle$ to the state $|\alpha_{i+1}\rangle$
at the endpoints of the intervals $\{ x_1,\ldots,x_n\}$. Then 
partition function of the constrained theory can be expressed in terms of the BCFT correlation function as
\be
Z_{\alpha_1\ldots\alpha_n} &=& Z_f\cdot \langle \prod_{i=1}^n
\Phi_i(x_i) \rangle~.
\ee

We close this section by discussing the example of the Ising model on $\overline{\H}$. There are three distinct conformally invariant boundary conditions. Either all the boundary spins are in the $\sigma=-1$ state, or all are in the $\sigma=+1$ state or finally, all are free. We denote them by $(-)$, $(+)$ and $(f)$. Then we have
\begin{eqnarray*}
(++)\;\;\text{or}\;\;(--) & : & \quad h=0 \\
(ff) & : & \quad h=0,\,\frac{1}{2}\\
(-+) &:&\quad h=\frac{1}{2}\\
(-f)\;\;\text{or}\;\;(+f) & : & \quad h=\frac{1}{16}
\end{eqnarray*} 
Further we note, that a boundary operator and its bulk counterpart have, in general, different scaling dimensions. With free boundary conditions the spin operator has boundary scaling dimension of $1/2$, whereas in the bulk the respective values are $(1/16, 1/16)$. Additionally, an operator which is primary for the bulk Virasoro algebra may have no primary analogue in the set of boundary operators. So the energy operator is primary in the bulk, but a descendant of the identity, according to the classification of the irreducible representation of the boundary Virasoro algebra.


\section{Moduli spaces of curves and CFT II}
\subsection{Moduli spaces}

In this section, we recall basic definitions and facts about the objects needed in the sequel, in particular about the moduli spaces of curves. We will use both notations $C$ and $X$ to denote a ``curve", resp. objects build out of it.  For details cf.~\cite{AGMV, ADKP, BK, FbZ, Friedan:1986ua, J, KNTY, Ko, K, ShU, TUY}.

A surface is a real two-dimensional topological manifold, that may carry a differentiable structure. Topologically any compact, oriented surface is uniquely determined by its genus up to diffeomorphisms or homeomorphisms. A closed Riemann surface, also called a {\bf complex curve}, is a two-dimensional differentiable real manifold with an atlas such that the  transitions between charts are biholomorphic functions.  A bordered (Riemann) surface generalises the idea of a domain with (smooth) boundaries. 
\begin{df} A connected topological Hausdorff space $M$ is a {\bf Riemann surface with boundary} if every point $p\in M$ has an open neighbourhood $U$ which is homeomorphic to a relatively  open set in the closed upper half-plane $\overline{\H}$ such that, the transition functions are again conformal. The homeomorphism $z:U\rightarrow z(U)$  is called a {\bf local variable} at the point $p\in U$.
\end{df}
We call a homeomorphism $f: A\rightarrow f(A)$ of open sets of the closed upper half-plane  {\bf holomorphic} or {\bf analytic} if it is holomorphic in the usual sense in $A\cap\H$.
Points $p\in M$ for which all the local charts $(U,z)$, $p\in U$, satisfy $z(p)\in\R$ are called {\bf boundary points} of $M$ and they form the {\bf boundary} $\partial M$ of $M$.  Surfaces that are allowed to have a boundary are referred to as {\bf surfaces with boundary} or {\bf bordered}.

Now, if a surfaces carries a Riemannian metric then we have the following
\begin{df} Let $M$ be an oriented , closed and differentiable real 2-manifold. Two Riemannian metrics $g_1, g_2$ on $M$ are called {\bf conformally equivalent}, if there exists a function $f\in C^{\infty}(M,\R)$, such that $g_1=e^f\cdot g_2$. The corresponding equivalence classes of Riemannian metrics on $M$ are called {\bf conformal structures}.
\end{df} 
A conformal structure permits to measure (oriented) angles but no more lengths. By a {\bf complex structure $\SS$} on surface $M$ we shall denote the choice of a maximal holomorphic atlas on it. 
By the Theorem of Newlander-Nierenberg, the two notions are the same for two-dimensional manifolds. Therefore a Riemann surface is equivalently  described by a complex or conformal structure.

The next step is to classify the complex or conformal structures. 
\begin{df} 
\label{mdf}
Let ${\SS}_1, {\SS}_2$ be two complex structures on a fixed compact surface $M$ of genus $g$. We define (as an equivalence relation)
\begin{displaymath}
\SS_1\sim_{\MM}\SS_2
\end{displaymath}
if there exists a biholomorphic mapping $\SS_1\rightarrow\SS_2$. The set of  the equivalence classes ${\MM}_g$ is called the {\bf moduli space}.
\end{df}
Geometrically, ${\MM}_g$ is an {\bf orbifold}, or a {\bf V-variety} because of the presence of automorphisms of the surface . To overcome the thereto related problems, there exists another identification.
\begin{df}
\label{tdf}
Let $\SS_1, \SS_2$ be two complex structures on a fixed compact surface $M$ of genus $g$. We define an equivalence relation
\begin{displaymath}
\SS_1\sim_{T}\SS_2
\end{displaymath}
if there exists a biholomorphic map $\SS_1\rightarrow\SS_2$, that is isotopic to the identity on $M$. The set of equivalence classes is called 
the {\bf Teichm\"uller space}.
\end{df}
We note that Definitions~\ref{mdf} and \ref{tdf} are independent of the choice of the underlying surface $M$.
The relation between the moduli space and the Teichm\"uller space is given by
\begin{prop}
The moduli space ${\MM}_g$ is isomorphic to $T_g\big/\Gamma_g$, where $\Gamma_g$ is the {\bf mapping class group} of a surface of genus $g$.
\end{prop}
The next, classical, result describes the dimension of the set of equivalence classes and states that it carries an additional structure. .
\begin{thm}
The set $T_g$ has a natural structure of a complex analytic manifold such that the action of $\Gamma_g$ on $T_g$ is holomorphic, which gives also ${\MM}_g$ the structure of a complex analytic variety.\\
As a {\bf real analytic} manifold, $T_g$ is isomorphic to $\R^{6g-6}$ for $g>1$, but as a complex analytic manifold we have
\begin{displaymath}
T_g\neq\C^{3g-3}
\end{displaymath} 
\end{thm} 
One can generalise the above notions to surfaces with marked points (punctures) or with boundaries. First we have
\begin{df}
A {\bf pointed curve} is a complex curve $C$ with an ordered set of marked points $p_1,...,p_n\in C$ and with a non-zero tangent vector $v_i$ given at each point. We say that $C$ is of type $(g,n)$ if it has genus $g$ and $n$ marked points.
\end{df} 
The isomorphisms of pointed curves are biholomorphic maps that  map marked points onto marked points, thereby respecting the ordering and the tangent vectors as well. We then have (in analogy with Def.~\ref{mdf}\,)
\begin{equation}
\label{pmdf}
\widetilde{\MM}_{g,n}:=\{\text{isomorphism classes of pointed curves of genus}\, g\, \text{with}\, n\, \text{marked points} \}.
\end{equation} 
and we call $\widetilde{\MM}_{g,n}$ the {\bf moduli space of pointed curves} of type 
$(g,n)$, which is also a complex variety.

Although from the physical point of view bordered surfaces are more natural  to have, e.g. open strings or finite samples,  for the mathematical treatment it is often  desirable to work with closed ones, e.g.  closed strings. A standard technique to pass from open to closed surfaces is by taking the double.

Let $(M,\Sigma)$ be a bordered Riemann surface (we shall simply write  $M$) and let $(M,\overline{\Sigma})$ be the mirror image of $M$ (we write $\overline{M}$), which is obtained by replacing each local chart $\varphi\in\Sigma$ by $\varphi^*: p\mapsto-\overline{\varphi}(p)$. Then the  new structure is conformal, since $\varphi_1^*\circ{\varphi_2^*}^{-1}$ has the explicit expression $z\mapsto -\overline{\phi}_{12}(-\bar{z})$, where $\phi_{12}:=\varphi_1\circ\varphi_2^{-1}$,  and this is a conformal mapping. So the {\bf (Schottky) double} $M^d$ of $M$ is the topological sum modulo the identification of the border points by means of the identity mapping, i.e. $M\sqcup\overline{M}\large/{}_{\sim}$. Hence  if $M$ is a surface of genus $g$ with $k$ boundary curves then we get an oriented surface $M^d$ without boundary of genus
\begin{displaymath}
\tilde{g}=2g+k-1,
\end{displaymath}
which carries an antiholomorphic  {\bf involution}
\begin{displaymath}
\iota : M^d\rightarrow M^d\qquad\text{i.e.}\quad\iota^2=\text{id}_{M^d}
\end{displaymath}
that interchanges $M$ and $\overline{M}$ and has the boundary of $M$ as its fixed point set.  We have  
\begin{thm}
The Teichm\"uller space $T_{g,k}$ of Riemann surfaces of genus $g$ with $k$ boundary curves, is a totally real submanifold of the Teichm\"uller space $T_{2g+k-1}$ of closed Riemann surfaces of genus $2g+k-1$.
\end{thm}

Before proceeding further with the general discussion of moduli spaces, let us fix some (standard) notation.
For a variety $S$, we denote by ${\OO}_S$ the {\bf structure sheaf} of $S$, i.e. the sheaf of analytic functions on $S$, and for an open set $U\subset S$ and a sheaf $\cal F$ on $S$, we denote by ${\cal F}(U)$  the {\bf vector space of sections} of $\cal F$ over $U$.
If $s$ is a section such that $s(x)\neq 0$ $\forall x\in U$, where $U$ is an open domain, we call $s$ a frame over $U$. A $C^{\infty}$-frame of a line bundle $L$ on an open subset $U$ is also called a {\bf trivialisation} of $L$ on $U$. For trivial line bundles over $X$, its $C^{\infty}$-sections may be identified with $C^{\infty}$-functions.
If $D$ denotes a divisor, $C$ a compact curve and ${\cal F}$ an ${\OO}$-module over $C$, then we shall denote by ${\cal F}(C-D)$ the space of {\bf meromorphic sections} of $\cal F$ that are regular outside of $D$.

For a point $p\in S$ we denote by ${\OO}_{S,p}$ the local ring at $p$, i.e. the ring of germs of analytic functions at $p$, and by ${\mathfrak m}_p$ the maximal ideal of this ring, which consists of functions vanishing at $p$, i.e. $\forall f\in{\mathfrak m}_p\Rightarrow f(p)=0$. We denote by $\widehat{\OO}_{S,p}$ the completion of the local ring ${\OO}_{S,p}$ with respect to the topology given by the powers of the maximal ideal ${\mathfrak m}_p$.

Now we fix the notation concerning universal coordinates by setting:
\begin{eqnarray*}
{\cal O}:=\C[[z]] & := & \{\sum^{\infty}_{n=0} a_n\;z^n\, |\, a_n\in\C\}:\quad\mbox{the ring of formal power series}, \\ 
{\cal K}:=\C((z)) &: = & \{\sum^{\infty}_{k\geq m} b_k\;z^k\, |\, b_k\in\C ,\, m\in\Z \}:\quad\mbox{the field of formal Laurent series}.
\end{eqnarray*}

If in particular $C$ is a Riemann surface and $p$ a regular point then $\widehat{\OO}_{C,p}$ is {\bf non-canonically} isomorphic to $\C[[z]]$. To specify such an isomorphism one has to choose a formal coordinate  $z$ at $p$. 

\begin{df}
Let $C$ be a curve and $q$ a non-singular point on $C$. An $n$-th {\bf infinitesimal neighbourhood} $s^{(n)}$ of $C$ at the point $q$ is a $\C$-algebra isomorphism 
\begin{equation}
\label{n_neighbourhood}
s^{(n)}:{\OO}_{C,q}/{\mathfrak m}_q^{n+1}\simeq\C[[z]]/(z^{n+1})
\end{equation}
where ${\mathfrak m}_q$ is the maximal ideal of ${\OO}_{C,q}$ consisting of germs of holomorphic functions vanishing at $q$ and $(z^{\bullet})$ denotes the ideal generated by the element $z^{\bullet}$.
\end{df}
Taking the limit $n\rightarrow\infty$ in the isomorphism~(\ref{n_neighbourhood}), we have an isomorphism 
\begin{equation}
\label{formal_neighbourhood}
s:\widehat{\OO}_{C,q}\simeq\C[[z]].
\end{equation}
The isomorphism $s$ is called a {\bf formal neighbourhood} of $C$ at $q$ and its inverse $s^{-1}(z)$ a {\bf formal coordinate}.
\begin{df}
\label{npointed}
The data $(C; q_1, q_2,..., q_N; s_1, s_2,...,s_N)$ is called an $N$-{\bf pointed stable curve of genus} $g$ with {\bf formal neighbourhoods}, if
\begin{enumerate}
  \item $(C;q_1, q_2...,q_N)$ is an $N$-pointed stable curve of genus $g$.
  \item $s_j$ is a formal neighbourhood of $C$ at $q_j$.
\end{enumerate}
Similarly one defines an $N$-pointed stable curve with $n$-th {\bf infinitesimal neighbourhoods}.
\end{df}

We are now going to introduce an {\bf infinite dimensional variety} $\widehat{\MM}_{g,1}$ that  parametrises triples $(C,p,z)$ where $C$ is a smooth curve of genus $g$, $p$ a point on $C$, and $z$ a formal coordinate defined and vanishing at $p$. 

Here we shall assume, that we did not fix a tangent vector. This is a fine point and it changes things a little, but not drastically. The correct way would be to say, that we are dealing with the moduli stack. 

But first let us consider the case when we have a fixed curve $C$ with a marked point $p$, and $S$ is the set of all possible choices of a formal local parameter $z$ at $p$.  

This set has a natural structure of a projective limit of the smooth manifolds
\begin{displaymath}
S^{(N)}:=\{N-\text{jets of local parameters at}\;\, p\}.
\end{displaymath}
Then we have a tautological family of curves $C_S:=S\times C$ over $S$, with the same marked point $p$ and with the formal local parameter determined by $s\in S$.

The {\bf pro-Lie group} $\Aut({\OO})$ (i.e., the  projective limit of Lie groups)  of continuous automorphisms of ${\OO}$ (changes of local parameter) acts naturally on $S$. 
The action is free and transitive so that  $S$ is a principal homogenous space  for $\Aut({\OO})$, i.e. an $\Aut({\OO})$-{\bf torsor}.

Such an automorphism is completely determined by its action on the topological generator $z$ of $\C[[z]]$. Therefore, any element $\rho$ of $\aut({\OO})$ can be represented by $\rho(z)\in{\OO}$, and $\rho$ is a continuous automorphism of $\OO$ iff $\rho(z)$ is a formal power series in $z$ of the form
\begin{displaymath}
a_1 z+a_2 z^2+\dots\qquad\text{with}\;a_1\in\C^*.
\end{displaymath}
The  group structure is given by composition, i.e. for $\rho_1, \rho_2\in\Aut({\OO})$ we have
\begin{displaymath}
\rho_1*\rho_2(z):=\rho_2(\rho_1(z)).
\end{displaymath}
Further these automorphisms take $z$ to all possible topological generators of $\OO$.

We have the following Lie groups and Lie algebras:
\begin{eqnarray*}
\Aut_+({\OO})=\{z+a_2 z^2+\cdots\} &  &\qquad\der_+({\OO})=z^2\C[[z]]\partial_z \\
\cap\qquad & & \qquad\qquad\cap \\
\Aut({\OO})=\{a_1z+a_2 z^2\cdots,\,a_1\neq0\} & &\qquad\der_0({\OO})=z\C[[z]]\partial_z \\
& & \qquad\qquad\cap \\
& & \qquad\der({\OO})=\C[[z]]\partial_z
\end{eqnarray*}

Again, there exists the notion of an equivalence of such triples.
\begin{df}
Two triples $(C, p, z)$ and $(C', p', \xi)$ are called {\bf (rigged) equivalent} iff there is a biholomorphic map $\varphi: C\rightarrow C'$ such that $\varphi(p)=p'$ and $z=\xi\circ\varphi$. 
\end{df}
The  moduli space of rigged surfaces carries the structure of a complex analytic  manifold, obtained as the projective limit of finite-dimensional complex varieties, namely of pointed curves with $n$-th infinitesimal neighbourhoods.
Moreover it  has a bundle structure, with base ${\MM}_{g,1}$ and as fiber model the infinite-dimensional space $\aut({\OO})$. Therefore locally it looks like $\C^{3g-1}\times\{a_1z+a_2 z^2+...,\; a_1\neq0\}$. 

Let us remark, that if we include a tangent vector at the marked point, the fiber model changes to $\Aut_+({\OO})$. This is the relevant group for SLE on the upper half-plane.

We discuss now, in what sense $\widetilde{\MM}_{g,n}$ is the moduli space of curves, i.e. the space that parameterises the possible complex structures. 
For a complex manifold $M$ and  a point $q\in M$, we denote by $T_qM$ the {\bf holomorphic tangent space} at the point $q$. A holomorphic mapping $\pi: X\rightarrow B$ from an $m+n$-dimensional complex manifold $X$ to an $m$-dimensional complex manifold $B$, is called a {\bf smooth family of compact complex manifolds} or a {\bf complex analytic family of compact complex manifolds} over the complex base manifold $B$, if it satisfies the following conditions:
\begin{enumerate}
  \item $\pi$ is a proper mapping, i.e. for any compact set $K\subset B$ the inverse image $\pi^{-1}(K)$ is compact.
  \item $\pi$ is a smooth holomorphic mapping, that is, for any point $p\in X$ the linear mapping $d\pi_p: T_pX\rightarrow T_{\pi(p)}B$ of the holomorphic tangent spaces is surjective.  
  \item For any point $w\in B$ the fiber $\pi^{-1}(w)$ is connected.
\end{enumerate}
Similarly, a {\bf family of pointed curves} is a family of curves $X_B$ together with $n$ non-intersecting sections $p_i: B\rightarrow X$ and a non-vanishing vertical field $v_i$ on $p_i(B)$ (vertical means that $\pi_*(v_i)=0$).

Now by the conditions (1.) and (2.) the fiber $X_w:=\pi^{-1}(w)$ of each point $w\in B$ is a compact complex manifold. 
\begin{df}
For a point $w_0\in B$ we call $X_w$, $w\in B$ a {\bf deformation} of the compact complex manifold $X_{w_0}$.
\end{df}
The idea~\cite{Ko, SchS} behind a deformation is, that the base manifold $B$ parametrises a set of other varieties, and therefore it allows to relate the different objects, i.e. the structures of the fibers to each other.  The most familiar example  of a compact complex manifold depending on a parameter, that is  already built into its definition, is that of a torus or elliptic curve. Let  $L:=\{m\tau+n\; | \;m,n\in\Z\,\,\,\mbox{and}\,\,\Im(\tau)>0\}$  be a lattice and define $T_{\tau}:=\C/L$. Then this gives a family of tori depending on the parameter $\tau$. 

We shall now study a complex analytic family $\pi: X\rightarrow B$ by using local coordinates, cf.~\cite{Ko, ShU}. Let us choose a coordinate neighbourhood $U$ of a point $0_p\in B$ and local coordinates $\{w^1,\dots, w^m\}$ of $U$ vanishing at $0_p$. Let $\pi^{-1}(U)$ be covered by open sets $\{U_{\lambda}\}_{\lambda\in\Lambda}\,:\,\bigcup_{\lambda\in\Lambda} U_{\lambda}=\pi^{-1}(U)$. 
\begin{figure}[ht]
\begin{center}
\includegraphics[scale=0.5]{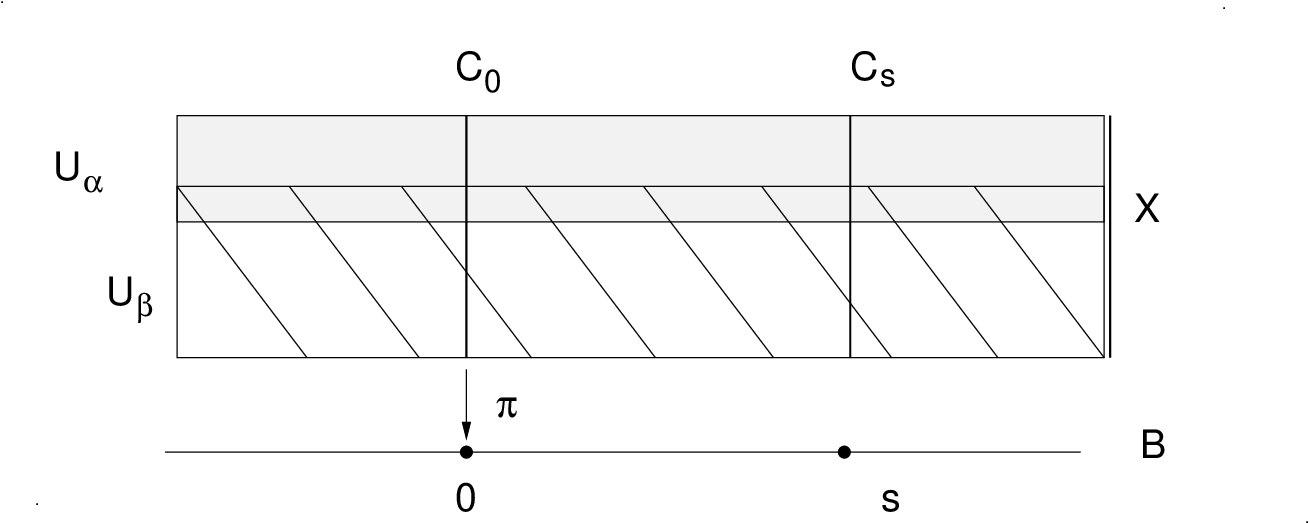}
\caption{The covering of the deformation space by horizontal cylindrical sets.}
\label{covering_fig}
\end{center}
\end{figure}
Since $\pi$ is a smooth holomorphic mapping, we can choose local coordinates of $U_{\lambda}$ as $(w^1,\dots, w^m, z^1_{\lambda},\dots, z^n_{\lambda})$, i.e. $w^1,\dots, w^m$ is chosen as a part of the local coordinates of $U_{\lambda}$ for all $\lambda$.
If $U_{\lambda}\cap U_{\mu}\neq\emptyset$, then between the two local coordinates we have the following relation:
\begin{equation}
\label{U11}
z^i_{\lambda}= f^i_{\lambda\mu}(w^1,\dots, w^m, z^1_{\mu},\dots, z^n_{\mu}),\qquad i=1,2,\dots,n.
\end{equation}
Here, $f^i_{\lambda\nu}$ is holomorphic on $U_{\lambda}\cap U_{\mu}$. The manifold $\pi^{-1}(U)$ can be regarded as obtained by patching $\{U_{\lambda}\}_{\lambda\in\Lambda}$ together by the relation~(\ref{U11}) and $(w^1,\dots, w^m)$ are regarded as parameters of changing the patching. That is, for $(w^1,\dots, w^m)=(0_p,\dots,0_p)$ the manifold $\pi^{-1}(0_p,\dots, 0_p)$ is a complex manifold $X_{0_p}$ which is obtained by gluing according to 
\begin{displaymath}
z^i_{\lambda}=f^i_{\lambda\mu}(0_p,\dots,0_p, z^1_{\mu},\dots, z^n_{\mu})
\end{displaymath}
Now for $w=(w^1,\dots, w^m)$ we obtain $X_w:=\pi^{-1}(w)$ by changing the patching of $X_{0_p}$ slightly by $w$. Therefore, the  term of the Taylor expansion of~(\ref{U11}) with respect to the variables $w^1,\dots, w^m$ gives the first order approximation of deformations, which are usually called {\bf infinitesimal deformations} of the complex manifold $X_{0_p}$. In particular, this means that if $U_{\lambda}\cap U_{\mu}\neq\emptyset$ then 
\begin{displaymath}
\left\{\frac{\partial f^i_{\lambda\mu}}{\partial w^k}(0_p,\dots, 0_p, z^1_{\mu},\dots, z^n_{\mu})\right\},\qquad 1\leq i\leq n,\; 1\leq k\leq m,
\end{displaymath}
give information on the deformation of $X_{0_p}$.

Let us consider for $U_{\lambda}\cap U_{\mu}\neq\emptyset$ a holomorphic tangent vector field
\begin{equation}
\label{U12}
\theta_{\lambda\mu}^{(k)}:=\sum_{i=1}^n \frac{\partial f^i_{\lambda\mu}}{\partial w^k}(0_p,\dots, 0_p, z^1_{\mu},\dots, z^n_{\mu})\frac{\partial}{\partial z^i_{\lambda}}
\end{equation}
on $X_{0_p}\cap U_{\lambda}\cap U_{\mu}$. In the above definition of the vector field $\theta_{\lambda\mu}^{(k)}$ we use the vector field $\frac{\partial}{\partial z^i_{\lambda}}$ on $U_{\lambda}$, but not $\frac{\partial}{\partial z^i_{\mu}}$. 

If
\begin{displaymath}
U_{\lambda}\cap U_{\mu}\cap U_{\nu}\neq\emptyset,
\end{displaymath}
then by~(\ref{U11}) we have
\begin{eqnarray*}
z^i_{\lambda} & = & f^i_{\lambda\mu}(w^1,\dots, w^m, f^1_{\mu\nu}(z_{\nu}, w),\dots,f^n_{\mu\nu}(z_{\nu}, w)) \\
 & = & f^i_{\lambda\nu} (w^1,\dots, w^m, z^1_{\nu},\dots, z^n_{\nu})~.
\end{eqnarray*}

Therefore, on the triple intersection $U_{\lambda}\cap U_{\mu}\cap U_{\nu}$, as an explicite  calculation shows, we have the following identity
\begin{displaymath}
\theta^{(k)}_{\lambda\nu}=\theta^{(k)}_{\mu\nu}+\theta^{(k)}_{\lambda\mu}
\end{displaymath}

That is, $\left[\theta_{\lambda\mu}^{(k)}\right]$ is a {\bf \v{C}ech one-cocycle} with coefficients in holomorphic tangent vector fields. The cohomology class of $H^1(X_{0_p}, \Theta)$ defined by the above one-cocycle is denoted by the same symbol $\left[\theta_{\lambda\mu}^{(k)}\right]$, where $\Theta$ denotes the sheaf of holomorphic vector fields on $X_{0_p}$. The cohomology class $\left[\theta_{\lambda\mu}^{(k)}\right]$ is uniquely determined for fixed local coordinates $(w^1,\dots, w^m)$, and is independent of the choice of open covering $\{U_{\lambda}\}_{\lambda\in\Lambda}$ of $\pi^{-1}(U)$, of coordinate neighbourhoods and local coordinates $(w^1,\dots, w^m, z^1_{\lambda},\dots, z^n_{\lambda})$, as follows from a direct calculation.

Hence, we can define a linear mapping from the holomorphic tangent space $T_{0_p} B$ at $0_p\in B$ to $H^1(X_{0_p},\Theta)$:
\begin{equation}
\label{KS}
KS_0: T_{0_p} B\rightarrow H^1(X_{0_p},\Theta),\qquad \sum a_k\frac{\partial}{\partial w^k}\mapsto \sum a_k\left[\theta_{\lambda\mu}^{(k)}\right].
\end{equation}
The linear map~(\ref{KS}) is called the {\bf Kodaira-Spencer mapping}~\cite{Ko}, and cf.~\cite{KNTY, ShU, TUY}. We note, that there exists a sheaf version of the $KS$-mapping as well.  

\subsection{From Schiffer Variation to Virasoro uniformisation}
\label{SchVVU}
We already mentioned the method of boundary variation as first considered by Hadamard. To overcome the difficulties in the case of non-smooth boundaries, Schiffer devised the method of interior variation, which is a special, but important, aspect of Kodaira-Spencer deformation theory, for details cf.~\cite{ AGMV, ADKP, BK, BS, FbZ, KNTY, Ko, K, PS, SchS, TUY}.

We are now going to introduces parts of this concepts and show, how they relate to physics. The basic idea of Schiffer variation is to remove a ``small" disc from the Riemann surface and then to sew in a deformed disc, depending on a complex parameter.

\begin{figure}[ht]
\begin{center}
\includegraphics[scale=0.5]{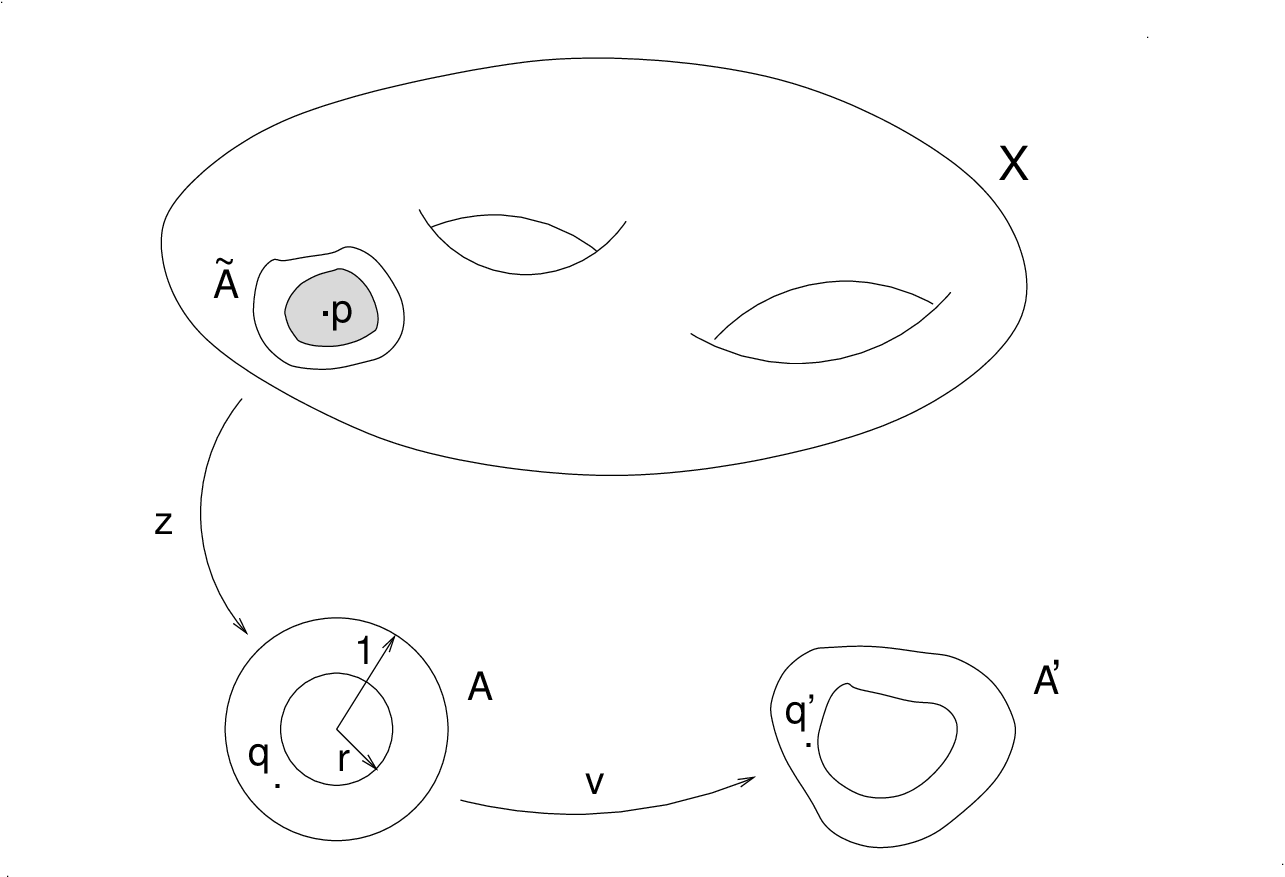}
\caption{The Riemann surface $X$ with a puncture at $p$} 
\label{Schiffer_fig}
\end{center}
\end{figure}

So let us consider a Riemann surface $X$ with a marked point (puncture) at $p$ and a coordinate $z$ vanishing at $p$, such that a closed disc of radius $>1$ is contained in the image of $z$.  Denote by $\tilde{A}:=\{y\in X : 0\leq r<|z(y)|<1\}$ the annular region  on $X$ and let $U:=z^{-1}({\D})$. The Riemann surface $X$ can be obtained from $X_1:=X\setminus\{y\in X: |z(y)|\leq r\}$ by attaching $U$ and identifiying the annulus $\tilde{A}$ with the corresponding points  (cf. Fig.~\ref{Schiffer_fig}). Now let us consider a meromorphic vector field $v(z)\frac{\partial}{\partial z}$ which is holomorphic in $U\setminus\{p\}$ and is allowed to have a pole only at $z(p)=0$. We write
\begin{displaymath}
v(z)=\sum_{m\leq n} a_{n-1} z^n\qquad\text{for}\quad m\in\Z.
\end{displaymath}
The vector field $v$ permits to construct  a new surface by first  deforming the  annulus  $A\rightarrow A'$ infinitesimally, i.e. by
\begin{equation}
\label{AlVafa41}
z\mapsto z+\varepsilon\cdot v(z)\qquad\text{for}\quad z\in A,\; \varepsilon\in\C,
\end{equation}
and then by filling the inside of the new annulus $A'$ to obtain a new ``disc" $D'$. We then glue this disc $D'$ to $X_1$ by identifying the new annulus $A'$ with the previous collar on the Riemann surface. So, if $q'\in A'$ is the image of $q\in A$ under (\ref{AlVafa41}), and $q$ corresponds to a point on the surface $X_1$, then $q'$ is ``glued" to $q$.

This way we get a new, topologically equivalent, Riemann surface $X'$. The  question is, what is the relation of the new surface $X'$ to the  old surface $X$. There are basically three possibilities:
\begin{enumerate}
  \item $v$ extends to a holomorphic vector field on $U$ and $v(0)=0$. This corresponds to a change of the local coordinate. If however $v(0)\neq0$, then the marked point $p$ is shifted (infinitesimally). 
  \item $v$ extends holomorphically to $X_1$, i.e. to the rest of the surface. In this case the vector field induces an infinitesimal conformal isometry of the rest of the surface and the triple of data gives the same point in the infinite moduli space.
  \item $v$ does not extend holomorphically to the disc $U$ or to the rest of the surface. Using the Riemann-Roch gives, that the vector fields with this property are (the insertion not being at a Weierstrass point).
\begin{equation}
\label{Alvarez51}
\left\{\frac{1}{z},\frac{1}{z^2},...,\frac{1}{z^{3g-3}}\right\}
\end{equation}
and they represent the tangent space to moduli space ${\MM}_g$ at $X$.
\end{enumerate}
Therefore we observe, that with these basic operations we can either change the moduli of the surface, change the position of the marked point or just change the local coordinate. In other words, the space of meromorphic vector fields maps onto the tangent space of the moduli space of punctured Riemann surfaces with a local coordinate at the puncture. The construction generalises to the case of $n$-punctured Riemann surfaces, thereby giving a mapping onto the tangent space of the moduli space of $n$-times punctured surfaces with coordinates at the punctures.

If the meromorphic vector field $v(z)$ is of the form $1/z$, and hence the infinitesimal transformation reads
\begin{displaymath}
z\mapsto z+\frac{\varepsilon}{z}, \quad \varepsilon\in\C,
\end{displaymath}
we obtain the classical {\bf Schiffer variation}~\cite{SchS}.

Let us note that the deformation (\ref{AlVafa41}) can be exponentiated by a finite amount. This corresponds to the action of 
\begin{displaymath}
\exp\left(a\cdot v(z)\frac{\partial}{\partial z}\right)
\end{displaymath}
on $\tilde A$ which is well defined for small $a$.

As we will explain later in a geometric setting (and in more detail), in conformal field theory one associates a {\bf ray} to the collection of data, i.e.
\begin{displaymath}
\text{(surface, marked point, local coordinate)}\longrightarrow |\phi\rangle,
\end{displaymath}
where the original motivation stems from the path-integral treatment of open strings with parametrised boundaries. The assignment of the rays has to satisfy some requirements on how they depend on the underlying data; i.e. they should form a holomorphic vector bundle (conformal blocks) over the moduli space of the above triples.

But as we just discussed, every vector field induces a variation of the data (point, moduli, coordinate), and therefore the natural question arises, what is the corresponding variation for $|\phi\rangle$? If we demand that this variation is to be given by some operator, represented as  $\mathfrak{O}(v)$, then in CFT it has naturally to be the energy-momentum tensor. Locally on the surface it is has a representation as (a quadratic differential) 
\begin{displaymath}
T(z)=\sum_{n\in\Z} L_n z^{-(n+2)}\,(dz)^2
\end{displaymath}
where the $L_n$'s are the Virasoro generators. So for a vector field $v(z)$ in the neighbourhood of $p$, we consider
\begin{equation}
\label{ }
T(v)=\frac{1}{2\pi i}\oint_p T(t)\,v(t)=\sum_{n\in\Z} a_{n} L_n.
\end{equation}

Hence the explicit change of $|\phi\rangle$ caused by the vector field  $v$ is given by
\begin{equation}
\label{ }
\delta_v|\phi\rangle=(T(v)+\bar{T}(\bar{v}))|\phi\rangle,
\end{equation}
which indicates that $T(v)$ gives a {\bf projective connection} over $\widehat{\MM}_{g,1}$.

If $v$ extends holomorphically off $p$, what should $T(v)|\phi\rangle$ be? Since $v$ does not change the point in the infinite moduli space, all we can expect is that $T(v)|\phi\rangle\propto|\phi\rangle$. The anomalous commutator 
\begin{equation}
\label{ }
[T(v_1),T(v_2)]=T([v_1, v_2])+\frac{c}{12}\oint_p v_1 v'''_2 
\end{equation}
indicates, that if we take a {\bf state} $|\phi\rangle$, i.e. a normalised vector, instead of the ray, then what we get is 
\begin{equation}
\label{ }
\frac{c}{12}\{w(z),z\}\id|\varphi\rangle
\end{equation}
where $c$ denotes the central charge, $\id$ the identity operator and $\{~,~\}$ the Schwarzian derivative. 

We can now summarise the variation of  a state on moduli space in the case of surfaces with a single puncture (not a Weierstrass point) as:
\begin{itemize}
  \item $L_n$,\; changes the coordinate for $n\geq 0$,
  \item $L_{-1}$,\; moves the puncture,
  \item $L_{-k}$,\; changes the moduli for $k=2,\dots , 3g-3+1$,
\end{itemize}
and the rest of the $L_n$'s can be written as linear combinations of the above.

The above discussion can be rephrased  in rigorous terms, known as {\bf Virasoro uniformisation (VU)}.

Fix $g>1$ and let ${\MM}_{g,1}$ denote the moduli stack of smooth pointed curves of genus $g$ and  $\widehat{\MM}_{g,1}$  the moduli stack of triples $(X, p, z)$, where $(X,p)\in{\MM}_{g,1}$ and $z$ is a formal coordinate at $p$. Then $\widehat{\MM}_{g,1}$ is fibered over  ${\MM}_{g,1}$, i.e. it has the structure of an $\aut({\OO})$-bundle. Further  the moduli space ${\MM}_{g,1}$ naturally projects onto ${\MM}_g$ by forgetting the marked point; therefore it can be identified as the {\bf universal curve} ${\mathfrak X}_g$ over ${\MM}_{g}$. Schematically we have the following sequence of spaces
\begin{displaymath}
\widehat{\MM}_{g,1}\longrightarrow{\MM}_{g,1}\longrightarrow {\MM}_{g}~.
\end{displaymath}

Before giving the first key result in VU, we recall the notion of an infinitesimal action of a Lie algebra on a manifold.
\begin{df}
Let $M$ be a (possibly infinite dimensional) complex analytic manifold. A complex Lie algebra $\mathfrak{g}$ acts complex analytically on $M$ if there is a homomorphism of Lie algebras
\begin{displaymath}
\mu:{\mathfrak g}\rightarrow\vect(M),
\end{displaymath}
called the {\bf action} where $\vect(M)$ denotes the complex Lie algebra of complex analytic vector fields on $M$.
\end{df}
The kernel of the composition of the evaluation map $\text{ev}_p$ at a point $p\in M$ and the action $\mu$
\begin{displaymath}
\text{ev}_p\circ\mu:\mathfrak{g}\rightarrow\vect(M)\rightarrow T_pM
\end{displaymath}
is denoted by $\mathfrak{g}_p$ and called the {\bf isotropy subalgebra} or  {\bf stabiliser} of $\mathfrak{g}$ at the point $p\in M$.
The action $\mu$ is called {\bf homogenous} or  {\bf transitive} if the composite map $\text{ev}_p\circ\mu$ is surjective at each point $p\in M$. 

The action we are going to describe now is a formal version of the action of the Lie algebra of complex-valued vector fields on the unit circle $S^1$ on the moduli space of surfaces with complex structure and one boundary component, parameterised by $S^1$. If we consider $\widehat{\MM}_{g,1}$ as an infinite-dimensional {\bf real} manifold, then the holomorphic action of the Lie algebra $\der{\cal K}$ (recall that ${\cal K}:=\C((z))$\,) gives rise to a homomorphism of the direct sum of a ``left" and a ``right" moving complex Lie algebra, i.e. of
\begin{displaymath}
\C((z))\frac{\partial}{\partial z}\oplus\C((\bar{z}))\frac{\partial}{\partial\bar{z}}
\end{displaymath}
to the complex Lie algebra of complex valued smooth vector fields on 
$\widehat{\MM}_{g,1}$.

We have
\begin{thm}[Virasoro uniformisation \cite{ADKP, BS, K, TUY}]
\label{VU}
The moduli space $\widehat{\MM}_{g,1}$ carries a transitive action of the Lie algebra $\der({\cal K})$ compatible with the $\aut({\OO})$-action  along the fibres of the map $\widehat{\MM}_{g,1}\rightarrow {\MM}_{g,1}$, i.e. we have the following exact sequence:
\begin{equation}
\label{virunif}
0\rightarrow\vect(X\setminus\{p\})\hookrightarrow\der({\cal K})\rightarrow T_{(X,p,z)}\widehat{\MM}_{g,1}\rightarrow0.
\end{equation}
\end{thm}
The above Thm.~\ref{VU} is the ``elaborate" or algebraic  version of the previously introduced gluing construction. We are now going to sketch the proof (following the presentation in \cite{FbZ}).
\begin{proof}
In order to define an action of the Lie algebra $\der({\cal K})$ on $\widehat{\MM}_{g,1}$, we construct a (right) action of the corresponding group $\aut({\cal K})$ on $\widehat{\MM}_{g,1}$.

If $(X, p, z)$ is a point of $\widehat{\MM}_{g,1}$, and $\rho\in\Aut({\cal K})$, we construct a new point $(X_{\rho}, p_{\rho}, z_{\rho})$ of $\widehat{\MM}_{g,1}$ by ``gluing"  the formal neighbourhood of $p$ in $X$ and $X\setminus\{p\}$ with a ``twist" by $\rho$.

As a topological space, $X_{\rho}=X$, but the structure sheaf ${\OO}_{X_{\rho}}$ is changed as follows.  Let $U\subset X$ be Zariski open in $X$. If $p\notin U$, then ${\OO}_{X_{\rho}}:={\OO}_X(U)$. If $p\in U$, then ${\OO}_{X_{\rho}}(U)$ is defined to be the subring of ${\OO}_X(U\setminus\{p\})$ consisting of functions $f$ whose expansion $f_p(z)\in C((z))$ at $p$ in the coordinate $z$ satisfies
\begin{displaymath}
f_p(\rho^{-1}(z))\in \C[[z]].
\end{displaymath} 
Now  $X_{\rho}$ is an algebraic curve over $\C$. Note that if $p\in U$, then under the embedding ${\OO}_{X_{\rho}}(U\setminus\{p\})\rightarrow \C((z))$ defined by the formula $f\rightarrow f_p(\rho^{-1}(z))$, the subspace ${\OO}_{X_{\rho}}(U)\subset {\OO}_{X_{\rho}}(U\setminus\{p\})$ embeds into $\C[[z]]\subset \C((z))$. 

Next, we define the point $p_{\rho}$ of $X_{\rho}$. Choose an open subset $U$ of $X$ containing $p$. Then $p_{\rho}$ is defined as corresponding to the ideal in ${\OO}_{X_{\rho}}(U)$ equal to the intersection of $z \C[[z]]$ with the image of ${\OO}_{X_{\rho}}(U)$ in $\C[[z]]$ under the above embedding ${\OO}_{X_{\rho}}\rightarrow\C[[z]]$. Then $(X_{\rho}, p_{\rho}, z)$ is a point of $\widehat{\MM}_{g,1}$.

Now $\rho: (X,p,z)\rightarrow (X_{\rho}, p_{\rho},z)$ defines a (right) action of $\aut({\cal K})$ on $\widehat{\MM}_{g,1}$. This action extends the right action of $\aut({\OO})$ on $\widehat{\MM}_{g,1}$ by changes of coordinate $z$, because if $\rho\in\aut({\cal O})$, then $(X_{\rho}, p_{\rho}, z)\simeq(X, p, \rho(z))$.

The corresponding action of $\der({\cal K})$ is transitive.
\end{proof}

The above construction shows that the stabiliser of $(X,p,z)\in \widehat{\MM}_{g,1}$ under the action of $\aut({\cal K})$ is identifiable with the automorphisms of $\cal K$ which preserve ${\OO}(X\setminus\{p\})\subset{\KK}$. 
This gives us a presentation of the tangent space of $\widehat{\MM}_{g,1}$ as a homogenous space.
\begin{cor}
\begin{equation}
\label{Virtangentspace}
T_{(X,p,z)}\widehat{\MM}_{g,1}=\der({\cal K})\big/\vect(X\setminus\{p\}).
\end{equation}
where it is understood, that we take the image of the vector fields (cf.~(\ref{virunif})\,).
\end{cor}

Let us remark, that in physical terminology, we are in  the ``central charge zero" case, since the action of $\der({\cal K})$ is the action of a completion of the Witt algebra (the classical symmetry algebra). Further, to obtain the previously mentioned rays, we have to introduce line bundles,  more specifically, determinant line bundles.  This we shall describe next. 

\subsection{The partition function and real determinant line bundles}
\label{partdetbundel}
We have already discussed in section~\ref{moduli} the nature of the partition function in general CFT. The main features were, that the partition function in not a function, but rather a section of a line bundle over moduli space and that the energy-momentum tensor (of string theories) can be interpreted as a (projectivley) flat connection of the partition function line bundle on the moduli space of Riemann surfaces. There was also the important fact, that the partition sum depended in an essential way on the metric, i.e. on the concrete representative within a conformal class; and hence on a choice of  renormalisation scheme. This is the picture as it was  first given by D.~Friedan and S.~Shenker~\cite{Friedan:1986ua} and then formalised in the language of (twisted)~${\cal D}$-modules, or axiomatised by G.~Segal~\cite{Segal} and M.~Kontsevich~\cite{Ne} in the language of modular functors.
However, the first rigorous construction was done by A.~Tsuchiya, K.~Ueno and Y.~Yamada~\cite{TUY}, (cf. the exposition in~\cite{BK,FbZ}).

In general, for a finite-dimensional vector space $V$, one defines the 1-dimensional vector space $\det(V)$ as the highest exterior power of $V$:
\begin{equation}
\label{ }
\det(V):=\bigwedge^{\dim V}(V).
\end{equation} 
More generally, for a finite complex of finite-dimensional vector spaces $V^{\bullet}=(\cdots\rightarrow V_{i-1}\rightarrow V_i\rightarrow L_{i+1}\rightarrow\cdots)$, one writes
\begin{displaymath}
\det(V^{\bullet}):=\bigotimes(\det(V_i))^{(-1)^i},
\end{displaymath}
with the convention, that for a 1-dimensional vector space $K$, we let $K^1=K$ and $K^{-1}:=K^*$. E.g., if $K$ denotes the canonical bundle on a Riemann surface, i.e. the cotangent bundle, then $K^{-1}$ is the holomorphic tangent bundle.

For families, one has
\begin{df}
\label{BK674}
Let $C_S$ be a family of pointed curves over $S$. We define the {\bf determinant line bundle} $Q_S$ by
\begin{equation}
\label{BKdetline}
Q_S:=\det R^{\bullet}\pi_*{\OO}_{C_S}=\bigotimes(\det R^i\pi_*{\OO}_{C_S})^{(-1)^i}~,
\end{equation}
where $\pi$ is the projection $C_S\rightarrow S$, and $R^i\pi_*$ are the {\bf right derived functors} of $\pi_*$.
\end{df}
An equivalent definition, without the notion of higher direct images, states that the fiber of $Q_S$ at a point $s\in S$ is
\begin{equation}
\label{BK673}
Q_s=\bigotimes(\det H^i(C_s, {\OO}_{C_s}))^{(-1)^i},
\end{equation}
where one should notice, that in~(\ref{BK673}) we did not use the marked points. 

This applies also to the case of the moduli space,  because the definition of $Q_S$ is functorial in $S$, and therefore it also yields a line bundle $Q_{{\MM}_{g,n}}$ over the moduli stack ${\MM}_{g,n}$. Further, it works also for singular curves, which is important in our context, so that $Q_{{\MM}_{g,n}}$ is well defined over the compactification of moduli space $\overline{\MM}_{g,n}$. 

We are now going to describe another new construction of a line bundle, which is based on physical considerations. 

Let $(M,\SS)$ be a Riemann surface, i.e. a closed surface $M$ with a conformal structure $\SS$, i.e. an equivalence class of conformally equivalent metrics.

We will define the {\bf determinant line} $|\det_{\SS}|$, an oriented one-dimensional real vector space, as follows.

Any smooth Riemannian metric $g$ on $M$, compatible with the conformal structure $\SS$, gives a positive point (base vector) in $|\det_{\SS}|$, denoted by $[g]$.

For two metrics $g_1, g_2$, the ratio of the corresponding vectors is defined by
\begin{displaymath}
[g_1]/[g_2]:=e^{S_{L}[g_1,g_2]}~,
\end{displaymath}
where the {\bf Liouville action} for two metrics, of  the same conformal class, is given by:
\begin{equation}
\label{KLiouville}
S_{L}[g_1,g_2]:=\frac{1}{48\pi i}\int_{\SS}(\phi_1-\phi_2)\partial\bar{\partial}(\phi_1+\phi_2)~, 
\end{equation}
where we use the representation of the metric $g_i$ with respect to a local coordinate $z$ on $\SS$ (which by construction is isothermal), i.e.  $g_i$ can be written as 
\begin{displaymath}
g_i=e^{\phi(z,\bar{z})}\cdot|dz|^2~.
\end{displaymath}
A calculation, using Green's identity, gives 
\begin{prop}
Let $g_1, g_2$ and $g_3$ be Riemannian metrics on a Riemann surface $M$ compatible with the underlying conformal structure $\SS$. Then the Liouville action (\ref{KLiouville}) satisfies the following cocycle identity:
\begin{displaymath}
S_L[g_1,g_3]=S_L[g_1,g_2]+S_L[g_2,g_3]~.
\end{displaymath}
\end{prop}

By deforming the conformal structure $\SS$ on $M$, we obtain an oriented, real line bundle $|\det|$ on the moduli stack ${\MM}$ of conformal structures on $M$.

Now, in the case of surfaces with boundaries, one can use the same formulae as above, but the metrics have additionally to be flat near the boundaries, such that they become geodesic.

Since the bundle is real, one can define for all $c\in\R$ the tensor power $|\det|^{\otimes c}$ of the line bundle $|\det|$.

In the case of holomorphic line bundles the situation is somewhat different. There, if the bundle is raised to an arbitrary power, the resulting object needs in general not to be a holomorphic line bundle any more, but is (rather) a projective bundle. 

The general motivation in physics to consider (holomorphic) line bundles on families of (pointed) curves, comes from the idea, as we already discussed in Sec.~\ref{SchVVU}, that the particles which enter a surface at different points, change the state of the surface, according to their nature. This is modeled, by assigning a ``conformal block", i.e. a finite dimensional vector space, and in its simplest case, it becomes a line. Now, one additionally requires that all these vector spaces form a vector bundle over the family of curves, i.e. the moduli space, with a projectively flat connection. The fact, that it is not flat, is a manifestation  of the central charge $c$, i.e. the conformal anomaly. This is modelled, by raising the line bundle to the power of the central charge, i.e. by defining a $c$-connection. 

Roughly speaking, the assignment of a vector bundle to the moduli space of punctured curves is called a {\bf modular functor}, and the simplest modular functor with central charge, is the determinant line bundle to the power $c$, which is the mathematical version of our previously encountered assignment in Sec.~\ref{SchVVU} of a ray, with the stress-energy tensor giving a projectively flat connection. 

We shall close this section, by giving the rigourous definitions, i.e. the mathematical translation of the objects just (and previously) described. Further it serves to deepen the understanding of the matters which will be discussed in the next section.

Let us consider a complex analytic vector bundle $E$ over a complex analytic manifold $M$. Then a {\bf connection} is a morphism of sheaves
\begin{equation}
\label{connectionBK}
\nabla:{\cal E}\rightarrow{\cal E}\otimes\Omega^1
\end{equation}
satisfying 
$\nabla(s f)=(\nabla s)f+s\otimes df$, with $s\in{\cal E},\, f\in{\OO}_M$, and 
where ${\cal E}$ is the sheaf of sections of $E$, and $\Omega^n$ is the sheaf of differential forms on $M$ of degree $n$.

For a fixed vector holomorphic field $X$ on $M$, (\ref{connectionBK}) induces a linear morphism 
\begin{equation}
\label{ }
\nabla_X:{\cal E}\rightarrow{\cal E}
\end{equation}
such that $\nabla_X(s f)=(\nabla_X s)f+s X(f)$. 
Then $\nabla$ is {\bf flat} iff $X\mapsto\nabla_X$ is a homomorphism of Lie algebras (representation on the space of sections), i.e., $[\nabla_X, \nabla_Y]=\nabla_{[X,Y]}$. In local coordinates $x_i$, $X_i:=\partial/\partial x_i$, $\nabla_i:=\nabla_{X_i}$, this means $[\nabla_i, \nabla_j]=0$.

This is the situation for a ``modular functor with zero central charge".

In order to include the central charge, we have to introduce the notion of a bundle that carries instead of a flat connection a {\bf projectively flat} one, i.e. a connection such that $[\nabla_X, \nabla_Y]-\nabla_{[X, Y]}$ is an operator of {\bf multiplication by a function}, depending on $X,Y$. 
Equivalently, we can say that the sheaf of sections carries a {\bf projective action} of the Lie algebra of vector fields.

To describe the failure of the connection to be flat, we have to introduce the notion of a central extension of the Lie algebra of vector fields.

\begin{df}
A {\bf central extension} of~$\Theta_S$ is an ${\OO}_S$ module ${\cal A}$ on $S$ with a structure of a Lie algebra on section $s\in{\cal A}$ and with two maps of ${\OO}_S$-modules giving a short exact sequence 
\begin{equation}
\label{ }
\begin{CD}
0@>>>{\OO}_S@>\psi>>{\cal A}@>\varepsilon>>\Theta_S@>>>0,
\end{CD}
\end{equation}
such that:
\begin{enumerate}
  \item the maps $\psi,\varepsilon$ preserve the Lie bracket (${\OO}_S$ is endowed with the zero Lie bracket)
  \item $\psi(1)$ is central in ${\cal A}$;
  \item for $a,b\in{\cal A}$, $f\in{\OO}_S$, we have $[a, fb]=f[a,b]+(\varepsilon(a)f)b$
\end{enumerate} 
\end{df}
Then one can choose locally a lifting, i.e. a morphism of ${\OO}_S$-modules: $a:\Theta_S\rightarrow{\cal A}$
such that $\varepsilon\circ a=\text{id}$. Therefore the bracket can be written as
\begin{equation}
\label{local_ coc}
[a(X), a(Y)]=a([X,Y])+c(X,Y),
\end{equation}
where $c(X,Y)\in\Theta_S$ is a  $2${\bf -cocycle} on $\Theta_S$.

Let us discuss the relevant example for us. So let ${\cal A}_{\OO}:={\OO}_S\oplus\Theta_S$ (direct sum as ${\OO}_S$-modules), with the bracket given by
\begin{equation}
\label{ex663}
[X+f, Y+g]:=[X,Y]+X(g)-Y(f),
\end{equation}
where $X,Y\in\Theta_S$, $f,g\in{\OO}_S$, and $[X,Y]$ it the usual bracket of vector fields. This is the trivial central extension. 

If $L$ is  now be a line bundle on $S$, and ${\cal L}$ the sheaf of sections of $L$ then define ${\cal A}_{\cal L}$ as the Lie algebra of first-order differential operators acting on ${\cal L}$. If we choose a local trivialisation of $L$, then sections of ${\cal A}_{\cal L}$ have the form 
\begin{displaymath}
\partial:=X+f,\qquad\text{for}\quad X\in\Theta_S,\, f\in{\OO}_S.
\end{displaymath}
Differently stated,  a choice of trivialisation ${\cal L}\big|_U\rightarrow{\OO}\big|_U$
defines an isomorphism 
${\cal A}_{{\cal L}|U}\rightarrow{\cal A}_{{\cal O}|U}$.

Let ${\cal A}$ be a central extension of $\Theta$, and $k\in\C^*$. Then we can define another central extension ${\cal A}^k$, that, 
as a sheaf of Lie algebras, coincides with ${\cal A}$ but the embedding ${\OO}_S\rightarrow{\cal A}^k$ is given by $\frac{\psi}{k}$ where $\psi$ is the embedding ${\OO}_S\rightarrow{\cal A}$. 

Equivalently, if we locally choose a lifting $\Theta_S\rightarrow{\cal A}$ so that the extension ${\cal A}$ is given by a $2$-cocycle $c(X,Y)$, then ${\cal A}^k$ is given by the $2$-cocycle 
\begin{displaymath}
k\cdot c(X,Y),
\end{displaymath} 
which also shows that ${\cal A}^k$ is well defined for $k=0$. 
For {\bf integer} $k$, one has 
\begin{equation}
\label{truetensor}
{\cal A}_{{\cal L}^k}=({\cal A}_{\cal L})^k.
\end{equation}
Using this, one can define for any $k\in\C$ the {\bf sheaf of first-order differential operators in} ${\cal L}^k$ by
\begin{equation}
\label{faketensor}
{\cal A}_{{\cal L}^k}:=({\cal A}_{\cal L})^k.
\end{equation}

Therefore we see that every projectively flat connection $\nabla$ in a vector bundle $E$ defines a central extension ${\cal A}$ of $\Theta$ such that $\nabla$ defines a true action of ${\cal A}$ by first-order differential operators in $E$. In other words, the failure of a projectively flat connection to be flat can be described by a central extension ${\cal A}$ of the Lie algebra of vector fields.

We note, that in the case of our real line bundles, the expression in~(\ref{truetensor}) applies, i.e. one has an equality.

\subsection{The Fock space and non-zero central charge}
\label{Fockneq0}
This section is key, as it contains the essence of what has been alluded to at several occasions. To understand the connection of SLE  with CFT in its most general form, one has to refer to the material presented here. 

We already discussed the concept of Virasoro uniformisation, we should rather say Witt uniformisation, determinant line bundles etc., but now we are going to present the unifying framework, that is needed for the quantum treatment.

Basically, we will introduce the primary space for the representation of the Virasoro algebra, the Fock space. There are two explicit descriptions of this space, either as the semi-infinite exterior algebra, i.e. the  {\bf fermionic} Fock space, or as a sum of symmetric algebras, i.e. the  {\bf bosonic} Fock space. In the later each of the  algebras involved, is isomorphic to the ring of universal symmetric polynomials. 

We remark, that  the Fock space showed already up in the treatment of SLE, as e.g. in the construction of the representation in Sec.~\ref{HWR}.

For the presentation of the material in this section, we shall closely follow the careful exposition in~\cite{BK}, which is based on a host of ``unpublished manuscripts" and some of the founding articles, as well. Let us mention, that so far almost no attempts have been made to give a ``nice" and ``streamlined" exposition of many of the important contributions to algebro-geometric CFT, although two decades went by; a notable  exception is~\cite{BK} (or~\cite{FbZ, TUY}). See also~\cite{ADKP,BS,KR,KNTY,K,PS}.

The first basic concept we shall need, is that of a {\bf polarised} topological vector space $V$, i.e. a super-space. Then, a polarisation of $V$ is a class of allowable decompositions $V=V^+\oplus V^-$ which are fairly close to each other. We are now going to formalise this.

Let $V$ be a vector space, and $V_1, V_2$ be its subspaces. 
\begin{df}
\begin{enumerate}
  \item $V_1\prec V_2$ if there exists a finite-dimensional subspace $W\subset V$ such that $V_1\subset V_2+W$.
  \item $V_1, V_2$ are called {\bf commensurable} (notation: $V_1\sim V_2$)~if $V_1\prec V_2$ and $V_2\prec V_1$. In other words, $V_1\sim V_2$ iff\, $V_1/(V_1\cap V_2)$ and $V_2/(V_1\cap V_2)$ are finite-dimensional.
\end{enumerate}
\end{df}
Informally speaking, $V_1\sim V_2$ if they differ only by a finite dimensional subspace of $V$, i.e. they are ``fairly close" to each other.
\begin{df}
A {\bf Tate vector space} is a vector space $V$ with a collection of subspaces $V_{\alpha}$ such that:
\begin{enumerate}
  \item $\forall\,\alpha, \beta$,\quad $V_{\alpha}\sim V_{\beta}$;
  \item $\bigcap_{\alpha} V_{\alpha}=\{0\},\quad\bigcup_{\alpha} V_{\alpha}=V$;
  \item $\forall\, \alpha, \beta$,\, $\exists \gamma$ such that\, $V_{\alpha}\cap V_{\beta}\supset V_{\gamma}$.
\end{enumerate}
\end{df}
The fundamental  example of a Tate space for us is $V:=\C((z))$, with the collection of subspaces given by 
\begin{displaymath}
V_k:=z^k\,\C[[z]],\;k\in\Z.
\end{displaymath}
More generally, we can consider the direct sum $V:=\bigoplus^n_{i=1}\C((z_i))$ with $V_{k_1,...,k_n}:=z^{k_1}_1\,\C[[z_1]]\oplus\cdots\oplus z^{k_n}_n\,\C[[z_n]]$.

We shall define a topology on $V$ through the  collection of subspaces $V_{\alpha}$, such that $V$ becomes a topological vector space. The $V_{\alpha}$ yield a basis of neighbourhoods of zero, and henceforth we assume this topology to be given. As one checks, the $V_{\alpha}$ are {\bf closed} and a subspace $X$ is {\bf discrete} iff  $(X\cap V_{\alpha})\sim0$, independently of the choice of $V_{\alpha}$. Finally, $V$ is {\bf compact} iff $V_{\alpha}$ has finite codimension in $V$.

The spaces $V_{\alpha}$ and in general, any vector subspace $X\subset V$, such that $X\sim V_{\alpha}$, are called {\bf lattices}.

Let us now define the following spaces of operators:
\begin{eqnarray}
V^* & := & \{l: V\rightarrow \C~ |~ l\;\text{is linear and continuous}~\}, \\\nonumber
\mathfrak{gl}(V) & := & \{\varphi: V\rightarrow V~|~\varphi\;\text{is linear and continuous}~\},
\end{eqnarray}
where $\C$ is considered with the discrete topology. 

The main goal  is to define the central extension $\widehat{\mathfrak{gl}}(V)$, by using the space of semi-infinite forms, which is a module over the Clifford algebra $C(V)=Cl(V\oplus V^*)$.
\begin{df}
\label{Cl_algebra}
The {\bf Clifford algebra} $C(V)$ is an associative algebra with unit, generated by elements $v\in V$, $v^*\in V^*$ with relations
\begin{eqnarray}
\{v,w\} & = & 0\, = \{v^*, w^*\},\qquad v,w\in V,\quad v^*,w^*\in V^*, \\\nonumber
\{v,v^*\} & = & (v,v^*), 
\end{eqnarray}
where $\{x,y\}:=xy+yx$ is the anti-commutator and $(~,~)$ is the pairing $V\otimes V^*\rightarrow\C$.
\end{df}

Now we define the {\bf Fock space} as the semi-infinite wedge space, which is originally due to P.~Dirac. It is the vector space spanned by formal monomials of the form 
\begin{displaymath}
w:=e_{i_1}\wedge e_{i_2}\wedge\cdots,
\end{displaymath}
where $i_1, i_2,\dots$ is an infinite integer sequence such that $i_1< i_2<\cdots$ and $i_{k+1}=i_k+1$ for $k\gg0$. Further the vectors $\{e_i\}_{i\in\Z}$  are assumed to form a topological basis of $V$. In physics, they could be interpreted as eigenvectors of a self-adjoint operator $D: V\rightarrow V$.

A typical example of such a monomial, is 
\begin{equation}
\label{ }
w_n:=e_n\wedge e_{n+1}\wedge\cdots .
\end{equation}
The elements of the Fock space can be multiplied by vectors $v\in V$ and by 1-forms (dual vectors). In Quantum Field Theory, the action of an element of $V$ is called a {\bf creation operator}, and that of an element of the dual space $V^*$ an {\bf annihilation operator}; and the two actions fit together to form the action of the Clifford algebra $Cl(V\oplus V^*)$. The action can be defined as follows:
\begin{eqnarray*}
v.w & := & v\wedge w,  \\
v^*.(e_{i_1}\wedge e_{i_2}\wedge\dots) & := & \sum_k e_{i_1}\wedge\dots\wedge e_{i_{k-1}}\wedge(v^*,e_{i_{k}})e_{i_{k+1}}\wedge\cdots~.
\end{eqnarray*}
Let us provide the rigorous 
\begin{df}
For every lattice $V_{\alpha}\subset V$, we denote by $\bigwedge_{\alpha}^{\infty/2}(V_{\alpha})$ the space of {\bf semi-infinite forms}, i.e. the Clifford module which is generated by one vector $w_{\alpha}$ with relations
\begin{eqnarray}
v.w_{\alpha} & = & 0,\qquad v\in V_{\alpha}, \\\nonumber
v^*.w_{\alpha} & = & 0, \qquad v^*\in V^{\bot}_{\alpha}, 
\end{eqnarray}
where $V^{\bot}_{\alpha}:=\{v^*\in V^*~|~v^*(V_{\alpha})=0~\}$.
\end{df}  
We remark, that if we let $V_-\subset V$ be any subspace complementary to $V_{\alpha}$, i.e. $V=V_-\oplus V_{\alpha}$, (which then has to be discrete and thus closed), then as a module over $\bigwedge^{\bullet}(V_-\oplus V^*_{\alpha})\subset C(V)$; \;$\bigwedge^{\infty/2}_{\alpha}(V)$ is free of rank one, i.e.:
\begin{equation}
\label{EXE15}
\textstyle{\bigwedge}^{\infty/2}_{\alpha}(V)=\textstyle{\bigwedge}^{\bullet}(V_-\oplus V^*_{\alpha})w_{\alpha}.
\end{equation}
Now the Fock space, i.e. the space of semi-infinite forms, can be characterised as an irreducible $C(V)$-module, and any two lattices $V_{\alpha}, V_{\beta}$, give isomorphic $C(V)$-modules  $\bigwedge_{\alpha}^{\infty/2}(V_{\alpha})$ and $\bigwedge_{\beta}^{\infty/2}(V_{\beta})$, unique up to a constant.
Therefore we will use the notation $\bigwedge^{\infty/2}(V)$ for any of the modules $\bigwedge_{\alpha}^{\infty/2}(V_{\alpha})$. 

Therefore we can characterise the central extension  $\widehat{\mathfrak{gl}}(V)$, by the requirement that it acts on $\bigwedge^{\infty/2}(V)$, as follows.
\begin{df} 
\label{tateextension}
For any Tate space $V$, we let
\begin{eqnarray}
\label{centralext}
\nonumber
\widehat{\mathfrak{gl}}(V) &:=&\{(\hat{\varphi}, \varphi)~|~\varphi\in{\mathfrak {gl}}(V),\;\;\hat{\varphi}\in\End_{\C} ({\textstyle\bigwedge^{\infty/2}}(V)),\\
& &
\;\; [\hat{\varphi}, v]=\varphi(v),\;\; [\hat{\varphi}, v^*]=-\varphi^*(v^*)\quad \forall\, v\in V, \;v^*\in V^*\}~.
\end{eqnarray}
\end{df}
In the above definition, we do not distinguish the notation for a vector $v$ and for the corresponding operator $v_{\bullet}(\cdot)$, e.g. $\varphi(v)$ is considered as an element in $\End_{\C}({\bigwedge^{\infty/2}}(V))$.

One has a natural map $\widehat{\mathfrak{gl}}(V)\rightarrow{\mathfrak{gl}}(V)$,  where the kernel of this projection is 1-dimensional, because  ${\bigwedge^{\infty/2}}(V)$ is irreducible. Further the projection is surjective, and therefore $\widehat{\mathfrak{gl}}(V)$ is a central extension of ${\mathfrak{gl}}(V)$. 

Let us consider the case $V=\C((z))$ in detail (cf. \cite{K}).  We define a lifting 
\begin{displaymath}
{\mathfrak{gl}}(V)\rightarrow\widehat{\mathfrak{gl}}(V),\qquad\varphi\mapsto(s(\varphi),\varphi),
\end{displaymath}
as follows. Choose a polarisation $V=V_+\oplus V_-$, where $V_+=V_{+n}$ for some $n$. Then every operator $\varphi\in{\mathfrak{gl}}(V)$ can be uniquely written as a sum
\begin{displaymath}
\varphi=\varphi_{++}+\varphi_{+-}+\varphi_{-+}+\varphi_{--},
\end{displaymath} 
where $\varphi_{+-}: V_-\rightarrow V_{+}$, etc. Define $s(\varphi_{-+})$, $s(\varphi_{+-})$, $s(\varphi_{--})$ by the Leibniz formula, which will work, because only a finite number of terms in the sum are non-zero. As for $\varphi_{++}$, define $s(\varphi_{++})$ by the condition $s(\varphi_{++})w_n=0$ and use (\ref{centralext}) to extend it to all of $\bigwedge^{\infty/2}(V)$. One can now check, that this is well defined, and thus we get a map $s : {\mathfrak{gl}}(V)\rightarrow\widehat{\mathfrak{gl}}(V)$

This lifting $s$ is not a Lie algebra homomorphism, since
\begin{displaymath}
[s(\varphi),s(\psi)]=s([\varphi,\psi])+c(\varphi,\psi),
\end{displaymath}
where the {\bf 2-cocycle} $c$ is given by
\begin{equation}
\label{2cocycle}
c(\varphi, \psi)=\tr_V(\varphi_{+-}\psi_{-+}-\psi_{+-}\varphi_{-+}).
\end{equation} 
The lifting $s$ and the 2-cocycle $c$ depend on the choice of $V_+, V_-$. 

Now for $f,g\in\C((z))$,\, $V_+:=\C[[z]]$ and $V_-:=z^{-1}\C[z^{-1}]$,  we have, as a calculation shows,
\begin{eqnarray}
\label{Virresid}\nonumber
c(f\partial_z, g\partial_z) & = &\frac{1}{6}\cdot\text{Res}\, f'' dg~;\qquad  \text{and in particular,}\\
c(-z^{n+1}\partial_z, -z^{m+1}\partial_z) & = &-\delta_{n,-m}\cdot\frac{n^3-n}{6},  
\end{eqnarray}
which is the cocycle of the Virasoro algebra.

We note, that there exists a functional analytic version of the above construction, introduced by Segal and Wilson. In that context the Tate space is called the {\bf Sato Grassmannian}~(cf.~\cite{KNTY,TUY}) or the {\bf Segal-Wilson Grassmannian}~(cf.~\cite{PS}).  It is  applied to Virasoro uniformisation in the general case, i.e. with non-vanishing central charge, in ~\cite{ADKP}. 

Let us proceed with the general theory, where we need the following
\begin{df}
A {\bf colattice} $X\subset V$ is a subspace satisfying the conditions
\begin{displaymath}
X\cap V_{\alpha}\sim0,\qquad\text{and}\qquad X+V_{\alpha}\sim V.
\end{displaymath}
\end{df}
We note, that every colattice $X$ is discrete and thus, closed. 

The basic example for us is the space of meromorphic functions on a compact curve $C$, considered as a subspace of $\C((z))$ via series expansion in a local parameter $z$, in a neighbourhood of a point $p\in C$.

Let us define the Lie algebra
\begin{displaymath}
{\mathfrak{gl}}_X(V):=\{\varphi\in{\mathfrak{gl}}(V)~|~\varphi(X)\subset X~\};
\end{displaymath}
which naturally projects to ${\mathfrak{gl}}(X)$ and ${\mathfrak{gl}}(V/X)$. By $\widehat{\mathfrak{gl}}_X(V)$ we shall denote the restriction of the central extension $\widehat{\mathfrak{gl}}(V)$ to ${\mathfrak{gl}}_X(V)$. This central extension is trivial and therefore it admits a canonical splitting. To define it, we will select a lattice $V_{\alpha}$ and define the vector space
\begin{equation}
\label{lambdabuendel}
\lambda_{\alpha}:=\textstyle{\bigwedge}^{\infty/2}_{\alpha}(V)/(X\oplus X^{\bot})\textstyle{\bigwedge}^{\infty/2}_{\alpha}(V),
\end{equation}
where we consider $X, X^{\bot}$ as subspaces of the Clifford algebra $C(V)$.
\begin{lem}
The space $\lambda_{\alpha}$ is 1-dimensional and it is canonically isomorphic to 
\begin{displaymath}
\det(V_{\alpha}\cap X)^*\otimes\det(V/(V_{\alpha}+X))
\end{displaymath}
\end{lem}
We have now the important 
\begin{prop}
\label{BK71021}
There exists a natural action $\mu$ of the central extension $\widehat{\mathfrak{gl}}_X(V)$ on the vector space $\lambda_{\alpha}$ so that the central element acts by $1\cdot\id$. This gives a splitting $\widehat{\mathfrak{gl}}_X(V)=\C\oplus\ker\mu$, where $\ker\mu\simeq{\mathfrak{gl}}_X(V)$, in particular this shows that the central extension $\widehat{\mathfrak{gl}}_X(V)$ is trivial.
\end{prop}
\begin{proof} (following~\cite{BK})
The central extension $\widehat{\mathfrak{gl}}_X(V)$ acts on $\bigwedge^{\infty/2}_{\alpha}(V)$ and this action normalises the subalgebra $\bigwedge^{\bullet}(X\oplus X^{\bot})\subset C(V)$. Therefore, this action descends to the space $\lambda_{\alpha}$.
\end{proof}

We shall now apply the above, general discussion, to the situation relevant for us, as it describes the Virasoro uniformisation in the case of non-trivial central charge (cf.~\cite{ADKP,BS,K,TUY}).

Let $C$ be a non-singular compact complex curve with marked points $\vec{p}=(p_1,\dots,p_n)$, such that there is at least one marked point on  each of the connected components of $C$. Let ${\KK}_{p_i}$ be the {\bf field of fractions} of the completed local ring at $p_i$. The choice of a local coordinate $z_i$ at $p_i$ gives the identification ${\KK}_{p_i}\simeq\C((z_i))$. Then $\vec{\KK}:=\bigoplus{\KK}_{p_i}$ is a Tate space and its structure does not depend on the choice of local coordinate $z_i$. 

Let $\der(\vec{\KK}):=\bigoplus\der({\KK}_i)$ be the Lie algebra of $\C$-linear derivations of $\vec{\KK}$. Let us recall that we could equally call it the Witt algebra.  Again by the choice of a local coordinate $z_i$ we get the isomorphism $\der(\vec{\KK})\simeq\oplus\C((z_i))\partial_{z_i}$.

This Lie algebra acts in an obvious way on $\vec{\KK}$, such that the action is continuous in the Tate topology. 

Hence, $\der(\vec{{\KK}})\subset\mathfrak{gl}(\vec{\KK})$, and by restricting the central extension $\widehat{\mathfrak{gl}}(\vec{\cal K})$ to $\der(\vec{\cal K})$, we get a central extension $\widehat{\der(\vec{{\KK}})}$ of $\der(\vec{{\KK}})$, which is, up to a factor of $-2$, a sum of Virasoro extensions (cf. the discussion around Eq.~(\ref{Virresid})~).

Therefore, a module over $\widehat{\der(\vec{{\KK}})}$ with central element acting by $a\cdot\id$, is the same as a module over the direct sum $\vir\oplus\dots\oplus\vir$, with the central element of each of the Virasoro algebras acting by $-2a\cdot\id$. This way we get a coordinate-free definition of the Virasoro central extension.

The following construction is usually called the {\bf Krichever construction}. It allows to embed the various varieties into a linear space, where one can do analysis much easier. Here, in our case, the linear space is the Tate space, i.e. the infinite Grassmannian.

Let us denote by $X:={\OO}(C-\vec{p})$ the space of meromorphic functions on $C$  that are holomorphic on $C\setminus\vec{p}$. Since every meromorphic function is uniquely determined by its Laurent series near $p_i$, the space $X$ can be considered as a subspace of $\vec{\KK}$. 
From the Riemann-Roch theorem it follows that $X$ is a colattice, i.e. we have
\begin{lem}
\label{lemma612}
Let $\vec{\KK}_0\subset\vec{\KK}$ be the completed local ring of regular functions (in local coordinates $z_i$, $\vec{\KK}_0\simeq\bigoplus\C[[z_i]]$~). Then there exist canonical isomorphisms
\begin{eqnarray}
H^0(C,{\OO}) & \simeq & {\cal O}(C-\vec{p})\cap\vec{\KK}_0 \\ \nonumber
H^1(C,{\OO}) & \simeq & \vec{\KK}/({\OO}(C-\vec{p})+\vec{\KK}_0) 
\end{eqnarray}
where both of these vector spaces are finite-dimensional. 
\end{lem}
Let us also consider the Lie algebra $\Theta(C-\vec{p})$ of global meromorphic vector fields on $C$ that are holomorphic outside of $\vec{p}$. Expanding such a vector field near each of the points $p_i$ gives an embedding
\begin{displaymath}
\Theta(C-\vec{p})\hookrightarrow\der(\vec{\KK})\subset{\mathfrak {gl}}(\vec{\KK}).
\end{displaymath}
Moreover, the action of $\Theta(C-\vec{p})$ preserves the subspace $X$ of meromorphic functions. In combination with Prop.~\ref{BK71021}, we get the following
\begin{prop}
The restriction $\widehat{\Theta}(C-\vec{p})$ of the Virasoro central extension of $\der(\vec{\KK})$ to $\Theta(C-\vec{p})$ is trivial, i.e. there exists a canonical way to define an action $\mu$ of this central extension on the 1-dimensional vector space
\begin{equation}
\label{ }
\lambda_C:=\det(H^0(C,{\OO}))^*\otimes\det H^1(C,{\OO}),
\end{equation} 
such that the central element of Virasoro acts by $-2\cdot\id$; this gives a splitting
\begin{displaymath}
\widehat{\Theta}(C-\vec{p})\simeq\C\oplus\ker\mu,\qquad\ker\mu\simeq\Theta(C-\vec{p})~.
\end{displaymath}
\end{prop}

We can do the above construction in families, as well. So let us assume that we are given a non-singular family $\pi: C_S\rightarrow S$ of compact connected curves over a smooth base $S$, with marked points $p_i : S\rightarrow C_S$. Then we have an ${\OO}_S$-module ${\cal K}_S$ of Tate vector spaces defined as before ; a choice of a local coordinate $z_i$ near $p_i(S)$ gives an isomorphism ${\cal K}_S\simeq\,\oplus\,{\OO}_S((z_i))$. One can now define sheaf versions of the perviously introduced objects, i.e. of the Clifford algebra $C({\cal K}_S)$, the space of semi-infinite forms $\bigwedge^{\infty/2}({\cal K}_S)$ etc.; and all of them are free ${\OO}_S$-modules. Further, there is a sheaf version of the Lie algebra ${\mathfrak{gl}}({\cal K}_S)$ and of its central extension $\widehat{\mathfrak{gl}}({\cal K}_S)$. We note, that everything is ${\OO}_S$-linear.

Let ${\cal N}^0$ be the sheaf of meromorphic {\bf vertical} vector fields, (vector fields $\theta$ such that $\pi_*\theta=0$), that  are holomorphic on $C_S\setminus\vec{p}(S)$.  This sheaf is locally free over ${\OO}_S$, and its fiber at a point $s$ is the  Lie algebra $\Theta(C_s-\vec{p}(s))$ discussed before. By the same construction as before, we get a canonical central extension $\widehat{\cal N}^0$, which acts (${\OO}_S$-linearly) on the line bundle $\lambda_S$ and thus splits:
\begin{equation}
\label{s140}
\widehat{\cal N}^0\simeq{\OO}_S\oplus\ker\mu,\qquad\ker\mu\simeq{\cal N}^0~.
\end{equation}
We are now ready to make the last step, allowing for non-vertical vector fields. Define the sheaf ${\cal N}$ on $S$ by
\begin{displaymath}
{\cal N}(U):=\{\theta,\tilde{\theta}\},
\end{displaymath}
where $U$ is an open subset of $S$, $\theta$ is a vector field on $U$ and $\tilde{\theta}$ is a lifting of $\theta$ to a meromorphic vector field on $\pi^{-1}(U)$ such that $\widetilde{\theta}$ is holomorphic outside of $\vec{p}(U)$. The sheaf ${\cal N}$ has a natural projection $\pi_*:{\cal N}\rightarrow\Theta_S$ to the sheaf of vector fields on $S$, which is surjective, i.e. every $\theta$ admits a lifting. 

The kernel of the projection is exactly the sheaf ${\cal N}^0$ of vertical vector fields. Further, ${\cal N}$ has the structure of an ${\cal O}_S$-module and a Lie bracket.  The collection of all this structures, which give rise to a reasonably new one, can be formalised  in the following 
\begin{df}
A {\bf Lie algebroid} on $S$ is an ${\OO}_S$-module $\cal A$, with a Lie bracket on sections and a morphism of ${\OO}_S$-modules $\pi_*:{\cal A}\rightarrow\Theta_S$ such that 
\begin{enumerate}
  \item $\pi_*$ preserves the Lie bracket;
  \item For $f\in{\cal O}_S$, \, $a_1, a_2\in{\cal A}$, one has
  \begin{equation}
\label{ }
[a_1, f a_2]=f[a_1, a_2]+(\pi_*(a_1)f)a_2.
\end{equation}
\end{enumerate}
\end{df}

An {\bf action} of a Lie algebroid $\cal A$ on an ${\OO}_S$-module $\cal E$ is a map $\cal A\otimes_{\C}{\cal E}\rightarrow{\cal E}$ such that:
\begin{eqnarray*}
[a_1, a_2]e & = & a_1(a_2 e)-a_2(a_1 e),\qquad a_1\in{\cal A},\, e\in{\cal E}, \\
(fa)e & = & f(ae),\qquad\qquad\qquad\quad a\in{\cal A},\, f\in{\cal O}_S, \\
a(fe) &=& f(ae)+(\pi_*(a)f)e.
\end{eqnarray*}
The important example in the context of Virasoro uniformisation with central charge is the sheaf ${\cal A}_{\cal L}$ of first-order differential operators in ${\cal L}$, where $L$ is a line bundle. 

With the above Def. we can now state that the sheaf ${\cal N}$ is a Lie algebroid, that acts on the sheaf ${\cal K}_S$ and with an adjoint action on the sheaf ${\cal K}^*_S$. All in all, these actions can uniquely be extended to an action of ${\cal N}$ by derivations on the Clifford algebra $C({\cal K}_S)$.

We can again define a central extension $\widehat{\cal N}$ of ${\cal N}$ as a sheaf whose sections are pairs $(\hat{\theta},\theta)$, where $\theta$ is a section of $\cal N$ and $\hat{\theta}\in\End_{\C}(\bigwedge^{\infty/2}({\cal K}_S))$ satisfying the following condition (cf. Def.~\ref{tateextension})
\begin{equation}
\label{ }
[\hat{\theta}, x]=\theta(x), \qquad x\in C({\cal K}_S).
\end{equation}

As before, it follows from the irreducibility of $\bigwedge^{\infty/2}({\cal K}_S))$ as a $C(V_S)$-module that $\widehat{\cal N}$ is a central extension of ${\cal N}$, i.e. we have the following short exact sequence
\begin{equation}
\label{ }
0\rightarrow{\OO}_S\rightarrow\widehat{\cal N}\rightarrow{\cal N}\rightarrow0
\end{equation}
that preserves the Lie bracket and the structure of an ${\OO}_S$-module. Further we can define the projection map $\pi_*:\widehat{\cal N}\rightarrow\Theta_S$ as the composition $\widehat{\cal N}\rightarrow{\cal N}\rightarrow\Theta_S$.

However, it is not true that the central extension $\widehat{\cal N}$ splits. Instead, we have following result. Take $\vec{\cal K}_{\alpha}:=\vec{\cal K}_0\simeq\oplus{\cal O}_S((z_i))$. In this case by Lem.~\ref{lemma612}, the line bundle $\lambda$ is given by $\det(R^0\,\pi_*{\cal O}_{C_S})^*\otimes\det R^1\,\pi_*{\cal O}_{C_S} $, i.e. it coincides with the {\bf inverse of the determinant line bundle}. 

\begin{prop}
The Lie algebroid $\widehat{\cal N}$ acts canonically on the sections of the line bundle $\lambda$, such that the central element acts by $1\cdot\id$.
\end{prop}
For the proof one should compare with that of Prop.~\ref{BK71021}. We have, as it follows from Eq.~(\ref{s140}),
\begin{cor}
One has a short exact sequence of Lie algebroids 
\begin{displaymath}
0\rightarrow{\cal N}^0\rightarrow\widehat{\cal N}\rightarrow{\cal A}_{\lambda}\rightarrow0,
\end{displaymath}
where ${\cal A}_{\lambda}$ is the algebroid of first-order differential operators in $\lambda$ (cf. example at~(\ref{ex663})~).
\end{cor}

Let us now summarise the various relations among all the Lie algebroids discussed so far, in the following commutative diagram, where all rows and columns are exact.
\[
\begin{CD}
{}@.{}@.0@.0\\
@.@.@VVV@VVV\\
{}@.{}@.{\cal N}^0@={\cal N}^0\\
@.@.@VVV@VVV\\
0@>>> {\OO}_S@>>>{\widehat{\cal N}}@>>>{\cal N}@>>>0\\
@. @|@VVV@VVV\\
0 @>>>{\OO}_S@>>>{\cal A}_{\lambda}@>>>\Theta_S@>>>0\\
@.@.@VVV@VVV\\
{}@.{}@.0@.0
\end{CD}
\]

\subsection{Glimpses of Axiomatic (Boundary) CFT}
\label{axiomatic}
We will reproduce here a description of CFT that is axiomatic in spirit. So far there is no fully and rigorously established version of CFT, although there has been (big) progress. Nevertheless  it is hard to implement in (all) generality the axiomatic approach as proposed by G. Segal~\cite{Segal} and M. Kontsevich around 1987. There are other approaches, e.g. the one developed by D. Friedan and S. Shenker~\cite{Friedan:1986ua} or that by G. Moore and N. Seiberg. Common to all of them is, that they are  geometric, as opposed to the purely functional analytic ones.

The situation is somewhat different in boundary CFT; there is still a lot that remains to be done. We are tempted to say, that rigorous (boundary) CFT is so far a ``locally consistent theory". For a recent introduction and aspects of BCFT on surfaces we recommend ~\cite{G2,G3}.

Nevertheless, let us give a partial description, at least of the data involved.  

The first building block is the list of data contained and a list of axioms that tell how the various parts are related to each other.

The basic classification of any CFT is given by a real number $c$, called the central charge and as in usual quantum field theory, a countable complex vector space ${\cal H}$,  whose elements $\psi$ are called {\bf states} or {\bf fields}. The fundamental symmetry algebra is the Virasoro algebra with the same central charge $c$. Two commuting copies, a ``holomorphic" and ``anti-holomorphic", act on the vector space, i.e. 
\begin{displaymath}
L_n, \bar{L}_n:{\cal H}\rightarrow{\cal H},\quad n\in\Z,
\end{displaymath}
\begin{displaymath}
[L_n, L_m]  = (n-m) L_{n+m}+c\, \frac{n^3-n} {12}\,\delta_{n+m,0}\cdot\mbox{Id}_{\cal H}\quad\mbox{the same for}\;\;\bar{L}_n,
\end{displaymath}
\begin{displaymath}
[L_n, \bar{L}_m]=0.
\end{displaymath}
The original idea of particle physics, namely  the description of scattering experiments of elementary particles, is translated  in CFT on surfaces into the language of correlators as follows. The positions where the total of $n$ particles enter or leave the surface are described by $n$ punctures or marked points and $n$ formal coordinates vanishing at these points. Then the correlations are given by a mapping from the space of states into the space of sections (Fock space) of the pull-back of the standard real determinant line bundle to the rigged moduli space $\widehat{\MM}_{g,n}$,  namely
\begin{displaymath}
\langle\,\rangle_{g,n}: {\cal H}^{\otimes n}\rightarrow\Gamma(\widehat{{\MM}}_{g,n},\hat{\pi}^*|\det|^{\otimes c}),\quad g,n\geq 0.
\end{displaymath} 

The main property that has to be satisfied is that $\langle\,\rangle_{g,n}$ is equivariant with respect to the action of $n$ copies of $\Vir$ and $\overline{\Vir}$, and also of the symmetric group ${\mathfrak S}_n$.

In the case $n=0$, if no particles are around, we get a section of the bundle $|\det|^{\otimes c}$ on ${\MM}_g$. This section is called the {\bf partition function} in CFT and its value at a point of the moduli space, represented by a complex structure $\SS$ on an underlying surface $M$, is denoted by  $Z[\SS]$ or $Z_{\SS}$.

The value~/~vector $Z_{\SS}$ is ideally related to the asymptotics of the lattice model partition function in the thermodynamic limit, as coming from a triangulation of the surface. For probabilistic purposes, as we already encountered in section~\ref{moduli} and we will see later, one has to normalise by the number ${Z}_{\SS}$.

The complete set of axioms should also include the operator product expansions (OPE's) (cf. expr.~(\ref{Stress-OPE})\,) and further (technical) constraints which are known for so-called {\bf unitary CFT} (which exist only for $c\geq0$), and are less clear in the  general situation.

In a CFT defined  on a domain or a surface with boundary, one has not only the central charge $c$ and the vector space ${\cal H}$ of states as before, but also a set BC of {\bf boundary conditions} that are required to be local and diffeomorphism invariant. Further for any ordered pair of its elements $(\alpha,\beta)\in\text{BC}$ a representation ${\cal H}_{(\alpha,\beta)}$ of just {\bf one} copy of the Virasoro algebra with central charge $c$ (the space of boundary fields). Correlators are defined for collections 
\begin{displaymath}
(\Sigma, p_1,...,p_n, q_1,...,q_m; z_1,...,z_n, w_1,...,w_m;bc)
\end{displaymath}
where $\SS$ is an oriented surface with conformal structure and possibly with boundary, $(p_i)_{i=1,...,n}$ is a collection of pairwise distinct points in the interior $\interior({\SS})$, $(q_j)_{j=1,...,m}$ is a collection of pairwise distinct points on the boundary $\partial\Sigma$, $(z_i)$ is a collection of formal local holomorphic complex coordinates on $\SS$ at the points $p_i$, whereas $(w_j)$ are formal {\bf real positively oriented local coordinates} on $\partial\SS$ at the points $q_j$, and 
\begin{displaymath}
bc:\partial{\SS}\setminus\{q_1,...,q_m\}\rightarrow\mbox{BC}
\end{displaymath}
is a locally constant map. The correlators for a given collection as specified above is a linear functional on the tensor product
\begin{displaymath}
{\cal H}^{\otimes n}\otimes\left(\bigotimes_{j=1}^{m}{\cal H}_{\alpha_j^{\mbox{\tiny left}}\alpha_j^{\mbox{\tiny right}}}\right)
\end{displaymath}
where $\alpha_j^{\mbox{\tiny left}}$ (resp. $\alpha_j^{\mbox{\tiny right}}$) are boundary conditions on the left (resp. on the right) of the point $q_j$. 

In general, if a field $\psi$ in the vector space of states $\cal H$, is a {\bf highest-weight vector}, then the correlator depends only on the first derivative of the coordinate $z_i$ at the marked point $p_i$. 
Therefore the correlator can be represented by a {\bf tensor of some weight} $h$ at the point $p_i$.

The way to extend the Virasoro uniformisation (VU) to surfaces with boundaries, is by taking the (appropriate) double, i.e. to consider complex algebraic curves with an anti-holomorphic involution.

Then the axiom should say again that the correlator is given by an equivariant map to the space of sections of the real line bundle $|\det|^{\otimes c}$. (cf. the discussion in section~\ref{moduli})

There are some models, where the complete set of BC is explicitly known and also the corresponding BCFT, as well as the correlators; e.g. the Ising model at criticality (cf. the discussion in section~\ref{BCFT}).

\section{Relations of domain walls on surfaces with SLE}
In this final part of the present work, we are going to describe a construction that relates the chordal domain walls on surfaces to correlators of some boundary field. The phase separating intervalls start and end at different points of the same  boundary component. 

To do so, we have to make extensive use of the material previously introduced (cf. sections~\ref{moduli}, \ref{SchVVU} and~\ref{Fockneq0}). As quite often, when one generalises a concept,  there are several directions one could go, and none of them is a priori canonic. However, we shall relay here on methods from conformal field theory. 

The main prediction is that the shape of chordal domain walls on surfaces is governed by a pair of distinguished boundary fields $|\psi_1\rangle$, resp. $|\psi_2\rangle$,  which have to be degenerate highest-weight vectors of weight $h$, such that the parameters $h$ and $c$ are related to  $\kappa$, by (cf. Sec.~\ref{HWR}, in particular (\ref{ckappa}) and (\ref{hkappa})~)
\begin{displaymath}
c=(3\kappa-8)(6-\kappa)/(2\kappa), \qquad\text{and}\qquad h=(6-\kappa)/(2\kappa)~.
\end{displaymath}

The precursor of this sort of techniques appeared first in the derivation of Cardy's formula~\cite{Ca2}.

\subsection{``L{\oe}wner process" on the determinant line bundle}
\label{lonmod}
Let us denote by ${\cal M}_{g,b,m}$ the moduli space of bordered Riemann surfaces of genus $g$ with $b$ boundary components and $m$ marked points. Accordingly, let $\widehat{\MM}_{g,b,m}$ denote the infinite moduli space, with additionally a formal coordinate at the marked points.

Then the  CFT partition functions on Riemann surfaces $X\equiv(M,{\SS})$ of genus $g>0$ with $b$ boundary components, can be seen as a section of the determinant bundle
\be
\langle \unit \rangle &\in& \Gamma\Big(  {\cal M}_{g,b}, 
{\Xdet}^{\otimes c} \Big)~,
\ee
as discussed in Sec.~\ref{moduli} and~\ref{axiomatic}.  
We recall that the fibre at
$X \in {\cal M}_g$ of the standard determinant bundle $\Det_j$
with $j \in \Z$ is
\be
\bigwedge^{\max}  H^0(X, \Omega_X^{\otimes j}) \otimes
\bigwedge^{\max} \Big( H^1(X, \Omega_X^{\otimes j}) \Big)^*~.
\ee
where $\Omega_X$ denotes the canonical bundle and also that the determinant bundle ${\det}_X$ associated to a surface $X$, is
the inverse of the Hodge bundle $\Det_1$, i.e. ${\det}_X =\Det_1^{-1}$. Further it is known, that the Hodge
bundle $\Det_1$ generates the Picard group of ${\cal
M}_g$. The relation $\Det_j \simeq \Det_1^{\otimes(6j^2-6j+1)}$, which plays an important role in String Theory, was proved by D. Mumford.  

Now by choosing a marked point $p\in X$  and a formal coordinate $z$, with values in the formal (half)-disc, the moduli space of triples
$(X,p,z)$, yields an
$\Aut({\cal O})$-bundle, i.e. 
\be
\label{proj152}
\pi : \widehat{{\cal M}}_{g,1}
\longrightarrow {\cal M}_{g,1} ~,
\ee
where the action of $\Aut({\cal O})$ induces changes of the formal
coordinate. The
fibres of this bundle represent the set of all choices of
formal coordinates around the marked point $p \in X$. The corresponding  Lie algebra of  $\Aut({\cal O})$ is $\Vir_{\geq 0} = z \C[[z]] \partial_z$. 
By projecting further down, i.e. by forgetting the marked point $p$, we can view $\widehat{{\cal M}}_{g,1}$ as the bundle 
\be
\label{projdown}
\hat\pi : \widehat{{\cal M}}_{g,1} \longrightarrow {\cal M}_g ~.
\ee 
But, now we have also to consider the shifts of
the marked point, generated by $\partial_z$. Changes of the
conformal structure of the surface are generated at the fixed
point by the singular vector fields of the form $z^{-n} \C[[z]]
\partial_z \subset \Vir_{<-1}$ for $n>0$. All these formal vector
fields included in $\Vir_{<-1}$, $\Vir_{\geq 0}$, and $\C
\partial_z$ form together the Witt algebra $\Der({\cal K})$. Then the 
actual Virasoro algebra $\Vir$ is the  central extension of this. Further,  
as we know from Sec.~\ref{SchVVU}, $\widehat{{\cal M}}_{g,1}$ carries a
transitive action of $\der({\cal K})$, compatible with the $\Aut(
{\cal O})$ action along the fibres. Now, as it follows from the results in Sec.~\ref{partdetbundel} and \ref{Fockneq0}, the transitive action of $\der({\cal K})$ can be lifted to an  epimorphism (onto)
\begin{eqnarray*}
\vir & \rightarrow & {\cal A}_{\hat{\pi}^*|\det|^{\otimes c/2}}, \\
{\bf c} & \mapsto & c\cdot\id,
\end{eqnarray*}
where ${\cal A}_{\hat{\pi}^*|\det|^{\otimes c/2}}$ denotes the Lie algebroid of infinitesimal symmetries of the line bundle $\hat{\pi}^*|\det|^{c/2}$, and $\hat{\pi}$ is the projection map defined in~(\ref{projdown}). This is the content of Virasoro uniformisation in the case of non-vanishing central charge. 

For one (spin-0) operator insertion at $p\in\partial X$ with conformal weight $h$ (where $h$ is the weight of the boundary field) the transformation rule (\ref{transfR}) implies ${\langle
\Xope \rangle}_{\gamma[0,t]} \in \Gamma({\Obdle}_h)$, where
\be
{\Obdle}_h &:=& {\Xpdet}^{\otimes c} \otimes |T_p^* \partial X|^{\otimes h} ~. \label{obdleDef}
\ee 
Here we have used the pull-back bundle 
$\Xpdet := \varpi^*\Xdet$ where $\varpi : {\cal M}_{g,1} 
\longrightarrow {\cal M}_{g}$. The fibre of ${\Xpdet}^{\otimes c}
\longrightarrow {\cal M}_{g,1}$ at the Riemann surface $X$ is
twisted by a tensor power of the modulus of the cotangent space of the boundary of the surface at the marked point $p \in \partial X$. The modulus appears, as was discussed at the end of Sec.~\ref{moduli}, because insertions of boundary operators can be described as insertions of bulk operators and their mirrors on the Schottky double; as the mirror transforms by the complex conjugate $\rho'(\bar p)^{*}$ of the transformation of bulk field
$\rho'(p)$, the total effect is a transformation by a positive function $|\rho'(p)|^{2}$.

As the L{\oe}wner process produces a nontrivial Beltrami
differential, it generates a motion in the pertinent moduli 
space ${\cal M}_{g,1}$. This involves 
deforming the surface, changing the complex structure, 
and displacing the marked point $p \in X$ on the surface. 
If we attach a formal disc with coordinate $z$, we 
see that the above operations induce actions of 
$\Vir_{\geq 0}$, $\Vir_{<-1}$,and $\Vir_{-1} \sim \C \partial_z$
on the disc. This means that the L{\oe}wner process acts 
on the disc by  $\Der({\cal K})$. 

In the special case of the upper half-plane we can simply identify 
the formal (half) disc with the Riemann surface $X\equiv\H$ itself. 
In  Sec.~\ref{OA} where we encountered the action of $\Der({\cal K})$
through operators ${\cal L}_n$, in particular the condition to be an SLE martingale~(\ref{singular_2}) was given by the second-order
differential operator 
\be
\label{OP155}
\frac{\kappa}{2} {\cal L}_{-1}^2 -2 {\cal
L}_{-2} ~.
\ee 
In the case of general Riemann surfaces 
this $\Der({\cal K})$ action then extends 
to a transitive action on the bundle  
$\widehat{{\cal M}}_{g,1} \longrightarrow {\cal M}_{g,1}$
compatible with the structure group, as was just recalled at the beginning of this section.

The CFT analysis leads us to consider sections 
of the line bundle $\Obdle_h$, which is a twisted 
version of the standard determinant bundle defined 
on the moduli space ${\cal M}_{g,1}$. 
By using the above defined projection $\pi$, we can 
construct the pull-back bundle  $\pi^*\Obdle_h$ on 
$\widehat{{\cal M}}_{g,1}$. This bundle carries now a transitive Virasoro 
action. 
In this way the $\Der({\cal K})$ action is lifted  
to a Virasoro action in the quantum theory. 
Then  the corresponding operator to~(\ref{OP155}) is a second-order element of the {\bf universal enveloping algebra}  ${\cal U}(\Vir)$ of the Virasoro algebra: 
\be
\label{HOP}
\hat{H}:=\frac{\kappa}{2} L_{-1}^2 -2 L_{-2}~,
\ee
which should be naturally seen as the map 
\be
\hat{H}:
\Gamma(\pi^* {\Obdle}_h) \longrightarrow \Gamma(\pi^* {\Obdle}_{h+2}) ~.
\ee  
The sections of these pull-back bundles differ from the correlators 
suggested by the CFT analysis only in that they also depend on the 
formal coordinate. This extra structure is just enough to enable us
to equip them with the appropriate Virasoro action. 
  
Recall that we could consider the family of correlators  $\omega_t := \langle \Xope \rangle_{\gamma[0,t]}$, associated to (random) paths $\gamma$ on the surface $X$. 
These objects are naturally defined as  sections of $\Obdle_h$. But when we pull 
these sections back into the bundle  $\pi^*\Obdle_h$ we need to specify
their dependence on the formal coordinate $z$. This can be 
done by requiring that the resulting ``process $\pi^* \omega_t$ 
is still a martingale", which means a tensor of some definite weight (cf. the discussion at the end of Sec.~\ref{axiomatic}), i.e. that it is annihilated by the action of 
$\hat{H}$.  In this way we are able to 
eliminate  the dependence on the formal coordinate $z$, but are nevertheless able to 
retain the Virasoro action on it. 

There is yet another characterisation.

If the correlator $\langle \Xope \rangle$ satisfies 
$\hat{H} \langle \Xope \rangle =0$, the operator creates a state in the 
Verma module $V_{2,1}$. This module is closed under Virasoro action, 
which in turn is generated by the stress-energy tensor $T$. Since 
the L{\oe}wner process involves only insertions of the stress-energy 
tensor in the correlator, the final correlator  
$\langle \Xope \rangle_{\gamma[0,t]}$ has to be that of an operator 
belonging to the same Verma module
and satisfying the same differential equation. This is true irrespective
of the moduli of the Riemann surface, and provides indeed an independent 
analytic characterisation of the correlators 
$\langle \Xope \rangle_{\gamma[0,t]}$ as those sections of  
$\Obdle_h$ that are annihilated by $\hat{H}$.

\subsection{Degenerate highest weight field $\psi$ and canonical differential operator} 
\label{DegHigh}
In this section we are going to describe now a natural way to construct a probability measure on path on the finite dimensional moduli space. These measure should be related with the measures supported by the phase boundaries on the surfaces itself, as obtained by a limiting procedure in the scaling limit. 

So let us fix $\kappa, c, h$ such that they satisfy the relations ({\ref{ckappa}) and (\ref{hkappa}). Then, by using the Virasoro uniformisation, i.e. by interpreting the moduli space of conformal structures as a double coset for the Virasoro algebra, we can associate with the second-order element (\ref{HOP})
\begin{displaymath}
\hat{H}:=\frac{\kappa}{2}L^{2}_{-1}-2 L_2,
\end{displaymath}
a canonical second-order differential operator
\begin{equation}
\label{2OD}
\Delta_{\kappa}:\Gamma({\MM}_{g,1,1}, |\det|^{\otimes c}\otimes |T^*_p\partial X|^{\otimes h})\rightarrow\Gamma({\MM}_{g,1,1}, |\det|^{\otimes c}\otimes |T^*_p\partial X|^{\otimes (h+2)})
\end{equation}
acting on sections of two different, oriented real line bundles on the finite moduli space ${\MM}_{g,1,1}$.

Namely, let us consider $\widehat{\MM}_{g,1,1}$, where we assume that the single marked point $p$ is always on the single boundary component, i.e. $p\in\partial X$ and therefore the (formal) coordinate at $p$ should always take its values in the (formal) half-disc, such that the part of the boundary around $p$ is mapped onto the real line, around $0$.

Then by Virasoro uniformisation, the Virasoro algebra acts on the space of sections, of the pull-back of the real bundle $|\det|^{\otimes c}$, i.e. on $\Gamma(\widehat{\MM}_{g,1,1},\hat{\pi}^*|\det|^{\otimes c})$\quad (cf.~(\ref{projdown})). 

Now, the space of highest-weight vectors, with weight $h$ (resp. $h+2$) is identified with the space of sections of a line bundle on the finite-dimensional moduli space ${\MM}_{g,1,1}$, and 
so the differential operator $\Delta_{\kappa}$ coming from the operator $\hat{H}$ in  ${\cal U}(\Vir)$, is acting on this representation.

Let us now  consider the moduli space ${\MM}_{g,k,m}$, where the  $m=2k$ marked points $(p_j)_{j=1,\dots,m}$ are on the  boundary of the underlying surfaces, i.e. on $\partial X$. Further let us assume that the boundary condition changes at all marked points. This is (physically) modeled by inserting a boundary condition changing operator $\psi(p_j)$, at the marked point 
$p_j$. 

Therefore the correlator
\begin{equation}
\label{PCor}
\langle\prod^m_{j=1}\psi(p_j)\rangle
\end{equation}
is a section of the complexified oriented line bundle $\tilde{\Obdle}_h\rightarrow{\MM}_{g,k,m}$ whose fibre is
\begin{displaymath}
\C\otimes|{\det}_{X}|^{\otimes c}\otimes\left(\bigotimes^m_{j=1}|T^*_{p_j}\partial X|^{\otimes h}\right)\equiv{\Obdle}_h\otimes\C.
\end{displaymath}
By our assumption regarding $\psi$ as a degenerate highest weight state at level two, this correlator is annihilated by $m$ second order operators $\Delta^{(j)}_{\kappa}$ acting from sections of $\tilde{\Obdle}_h$ to sections of $\tilde{\Obdle}_{h+2}:=\tilde{\Obdle}_h\otimes\left(\bigotimes_{j=1}^m|T_{p_i}^*\partial{X}|^2\right)$; where $\Delta^{(j)}_{\kappa}$ is constructed similarly to the operator $\Delta_{\kappa}$ in~(\ref{2OD})~.

On physical grounds, we expect, that the correlator~(\ref{PCor})
is {\bf real} and {\bf everywhere positive}. 

If we consider now the trivialisation of $\Obdle_h$, given by the correlator~(\ref{PCor}), and an {\bf arbitrary} trivialisation  of $\Obdle_{h+2}$, we obtain  a second-order differential operator acting on functions, vanishing on constants, and with (constant rank) non-negative symbol, i.e. an operator in H\"ormander form, also called of Kolmogorov type. Since such an operator defines a diffusion process, we obtain a random motion on the moduli space ${\MM}_{g,k,m}$, i.e. on a finite dimensional manifold. Further, the operators $\Delta^{(j)}_{\kappa}$  induce a probability measure on parameterised continuous paths, and  
if one changes the trivialisation of the second line bundle $\Obdle_{h+2}$, which results in a time change for the random process on $\MM$, i.e. the induced probability measure on the space of paths would be up to reparameterisation, the same. More precisely we have

\begin{df}
Let $M$ denote a manifold. For a vector field $A\in\Gamma(TM)$ let $A^2(f):=A(A(f))$, $f\in C^{\infty}(M)$. A mapping $L$ is called a {\bf partial differential operator (PDO) in H\"ormander form} if there exist vector fields $A_0, A_1,...A_r$ on $M$ such that $L$ can be written as
\begin{equation}
\label{HForm}
L=A_0+\sum_{i=1}^r A_i^2~.
\end{equation}
\end{df}
For example, if $M=\R^n$ and $A_i:=D_i=\frac{\partial}{\partial x^i}$ $(i=1,...,d)$, so $\Delta=\sum^d_{i=1} A^2_i$, is the Euclidean Laplace operator. 

In our case the number $r$ can be chosen to be equal 1. Further, for any such operator we have
\begin{df}
Let $M$ denote a manifold, $L$ a (PDO) in H\"ormander form on $M$, and $x\in M$. An adapted continuous process $X$ on a standard filtered probability space $(\Omega;{\cal F};\P;({\cal F}_t)_{t\in\R_+})$ with values in $M$ and $X_0=x$, is called the {\bf (flow) process associated to} $L$ ({\bf with starting point} $x$), if for every test function $f\in C^{\infty}_c(M)$ the process $N(f)$:
\begin{displaymath}
N(f)_t:=f\circ X_t-f\circ x-\int^t_0 Lf\circ X_r\, dr,\qquad t\in\R_+,
\end{displaymath}
defines a martingale, i.e. for all $s\leq t$ in $\R_+$ we have:
\begin{displaymath}
\E[N(f)_t-N(f)_s|{\cal F}_s]\equiv\E[f\circ X_t- f\circ X_s-\int^t_s Lf\circ X_r\,dr\, |{\cal F}_s]=0.
\end{displaymath} 
\end{df}
We allow, that the (flow) process $X$ has only a finite life-time $\zeta$. 

In our situation the underlying manifold $M$ is the moduli space $\MM$, which is a real-analytic manifold. The tangent vector fields on it, in which we express the differential operator in H\"ormander form, are Beltrami differentials. This way we get a ``continuous random walk" on $\MM$.

From the transitivity of the Virasoro uniformisation and from the fact that the elements $L_n,\;n\geq -2$ generate the whole Virasoro algebra, we can deduce, that the operators $\Delta^{(j)}_{\kappa}$ are {\bf hypo-elliptic}, since H\"ormander's condition~{\bf (H) } is satisfied. Let us recall it.

If $V$ and $W$ are two vector fields, one defines the Lie bracket of $V$ and $W$ by $[V,W]:=V\circ W-W\circ V$, 
where this expression is understood as acting on some test functions. 
If the vector fields $A_1,\dots,A_d$ are used to write the differential operator in H\"ormander form~(\ref{HForm}),  then the condition can be stated as follows:

{\bf (H)}\quad The vector space spanned by the vector fields
\begin{equation}
\label{ }
A_1,\dots, A_d,\qquad [A_i, A_j],\; 0\leq i,j\leq d,\qquad [A_i,[A_j, A_k]],\; 0\leq i,j,k\leq d,\,\dots
\end{equation}
at a point $p\in M$ is $T_p M$, i.e. they span the whole tangent space.

Now we are going to look somewhat closer to the global nature of such diffusive paths.

We recall from Sec.~\ref{moduli}, that by cutting the surface along a path, that starts at some point $p\in\partial X$ and is strictly transversal at the initial point to the boundary,  and then by shifting the marked point to the tip of the path, we got a unique and continuous path in the moduli space of punctured surfaces. 

The cutting by itself induces a boundary variation wich produces a new surface, that isn't conformally related to the original one, but only by a quasi-conformal (qc) mapping. Actually, one can write down an explicite qc-mapping, wich shows, that as we trace along the curve, we get a family of in-equivalent Riemann surfaces. Again, this is a special form of the Schiffer variation, called the ``buttonhole construction". It furnishes one coordinate direction around the initial point in moduli space. 
Now, by just moving the marked point, we can vary another coordinate direction, and both variations together span a cone at the point, as we shall see in the next.

Let us mention that we think that the cutting procedure can be extended also to fractal paths, as coming from phase boundaries, since conformal invariance is a strong condition.  In particular, if all the technicalities would be properly implemented, the Hausdorff dimension could be derived from the Liouville action, once the central charge $c$ is fixed; (cf.~\cite{FK}).

Let us fix $j$, i.e. the index of the marked point $p_j$ on the boundary, and further let us denote by ${ P}={P}_j$ the sub-bundle of the tangent bundle to $\MM$ generated by the vector fields $A_0, A_1$ from the representation in~(\ref{HForm}).  Then $P$ is an universal two-dimensional sub-bundle, independent of the choices made above, including the parameter $\kappa$. However the distribution of planes $ P$ is not holonomic. 

Further ${P}$ contains a sub-bundle of open half-planes,  
${P}_+:=\R\cdot A_1+\R^*_+\cdot A_0$ in which the traces of the above random walks lay.  

\begin{figure}[ht]
\begin{center}
\includegraphics[scale=0.5]{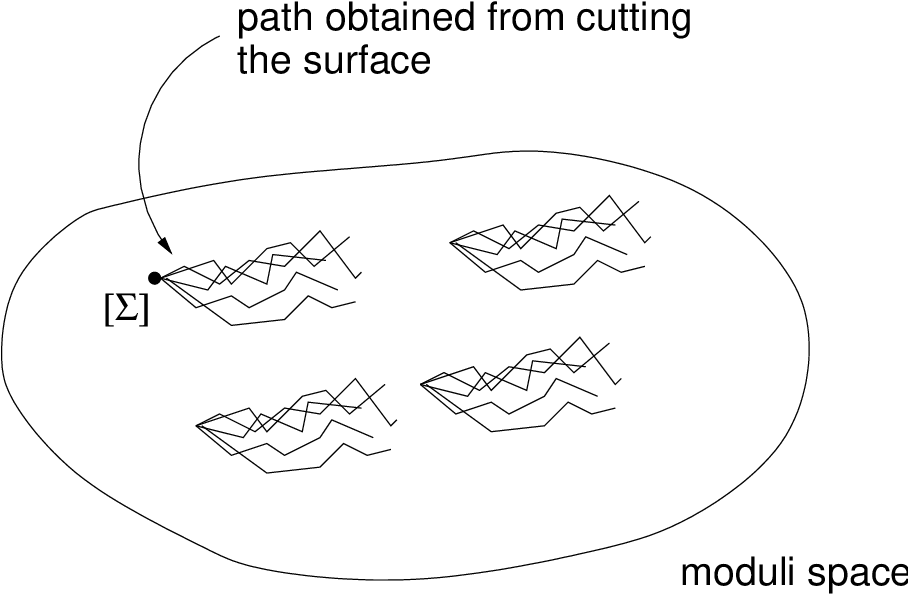}
\caption{From every fixed point  in moduli space emantes a family of simple path, obtained by cutting the underlying surface and thereby deforming the initial conformal structure $[\SS]$. } 
\label{cd}
\end{center}
\end{figure}

Now, the prediction is, that the self-avoiding random path on the surface $X$, associated with random walks on the moduli space ${\MM}$ coming from correlators of the highest-weight field $\psi$, are random phase boundaries. 

In the case of the disk with two marked points on the boundary, the moduli space is just one point. Therefore the correlator $\langle\psi(p_1)\psi(p_2)\rangle$, which only depends on the value of $\kappa$, is a priori known. The random path in this case, should coincide with the usual $\text{SLE}_{\kappa}$-path.

At this point it is instructive to compare with the construction of the Polyakov measure in string theory via the operator formalism (cf.~\cite{AGMV}).

In general, this last chapter of the present work, as well as Sec.~\ref{CFTresults}, show that the partition function carries genuine information about random quantities, which are well defined in the scaling limit, as e.g. the chordal domain walls on surfaces, coming from a statistical mechanics model, with proper boundary conditions.

So far, this has not been in the main focus of research in CFT, but the results presented in this work, which are inspired by SLE, show that there should exist a probabilistic description of CFT, which goes beyond the usual approaches. Further, it should reveal more concretely the stochastic nature of conformally invariant quantum fields. The price one has to pay is  that one has to overcome many technicalities, to properly interconnect the various pieces involved. On the other hand, the attempt to understand rigorously the fluctuating nature of two-dimensional critical systems, led to unexpected new connections of probability theory with other mathematical fields.

One of the striking examples is the existence of the connection between the central charge, and the variance of the Brownian motion in the SLE process, which is given by the representation theory of infinite Lie algebras. 

\subsubsection*{Acknowledgements}
We would like to thank very heartily  Jussi~Kalkkinen, Maxim~Kontsevich and Wendelin Werner for their support, energy, collaboration and help, which led to the basic material upon which this text is based. 

Further we would like to thank Vincent~Beffara, Klaus~Linde and Yuri~Suhov for the fruitful discussions we had with them. 

Finally we thank the Institut des Hautes \'Etudes Scientifiques~IHES for its support and hospitality, where the bulk of the work presented here was completed and the Institute for Advanced Study~IAS and Thomas~Spencer, where the additional  material is currently done. 
{}
\end{document}